\documentclass[sigconf, screen]{acmart}

\AtBeginDocument{}
\usepackage{hyperref}
\usepackage{url}
\usepackage{comment}
\usepackage{wrapfig}
\usepackage{amsmath,mathtools}
\usepackage{soul} 
\usepackage{graphicx}    
\usepackage{caption}     
\usepackage{subcaption}  
\usepackage{enumitem}
\usepackage{booktabs}    
\usepackage{multirow}    
\usepackage{tabularx}    
\usepackage{array}       
\newcolumntype{L}{>{\raggedright\arraybackslash}X} 
\usepackage{xcolor}
\usepackage{adjustbox}
\usepackage{algorithm}
\usepackage{algpseudocode}
\usepackage{xspace}
\usepackage[most]{tcolorbox}
\usepackage{balance}
\usepackage[utf8]{inputenc}

\newtcolorbox{findingsbox}{
  colback=gray!5,      
  colframe=black!50,   
  boxrule=0.8pt,       
  arc=3pt,             
  left=8pt, right=8pt, top=6pt, bottom=6pt,
  before skip=10pt, after skip=10pt,
}
\newif\ifcomments
\commentstrue              

\algrenewcommand\algorithmicrequire{\textbf{Input:}}
\algrenewcommand\algorithmicensure{\textbf{Output:}}
\algrenewcommand{\algorithmiccomment}[1]{\hspace{.5em}\(\triangleright\)\,#1}
\algtext*{EndIf}
\algtext*{EndFor}
\algtext*{EndWhile}
\newcommand{\graph}{\textsc{Graphectory}\xspace}
\newcommand{\lang}{\textsc{Langutory}\xspace}
\newcommand{\SA}{SWE-agent\xspace}
\newcommand{\TA}{Trae agent\xspace}
\newcommand{\OH}{OpenHands\xspace}
\newcommand{\MSA}{mini-SWE-agent\xspace}
\newcommand{\Rone}{DeepSeek-R1\xspace}
\newcommand{\Vthree}{DeepSeek-V3\xspace}
\newcommand{\devstral}{Devstral-small\xspace}
\newcommand{\gptf}{GPT-5 mini\xspace}
\newcommand{\swebv}{SWE-bench Verified\xspace}
\newcommand{\swebp}{SWE-bench Pro\xspace}

\newcommand{\cmt}[3]{\ifcomments\textcolor{#1}{\textbf{[#2:}~#3\textbf{]}}\fi}

\newcommand{\reyhan}[1]{\cmt{cyan}{Reyhan}{#1}}
\newcommand{\shuyang}[1]{\cmt{magenta}{Shuyang}{#1}}

\setcopyright{cc}
\setcctype{by}
\acmDOI{10.1145/3832783.3834400}
\acmYear{2026}
\copyrightyear{2026}
\acmISBN{979-8-4007-2882-2/2026/10}
\acmConference[ASE '26]{Proceedings of the 41st IEEE/ACM International Conference on Automated Software Engineering}{October 12--16, 2026}{Munich, Germany}
\acmBooktitle{Proceedings of the 41st IEEE/ACM International Conference on Automated Software Engineering (ASE '26), October 12--16, 2026, Munich, Germany}
\acmSubmissionID{ase26main-p1675-p}
\received{2026-03-26}
\received[accepted]{2026-06-18}

\begin{document}

\title{From Plan to Action: How Well Do Agents Follow the Plan?}

\author{Shuyang Liu}
\orcid{0009-0009-7264-268X}
\authornote{Both authors contributed equally to this work.}
\affiliation{%
  \institution{University of Illinois Urbana-Champaign}
  \city{Urbana}
  \country{USA}
}
\email{sl225@illinois.edu}

\author{Saman Dehghan}
\orcid{0009-0006-4388-6501}
\authornotemark[1]
\affiliation{%
  \institution{University of Illinois Urbana-Champaign}
  \city{Urbana}
  \country{USA}
}
\email{samand2@illinois.edu}

\author{Jatin Ganhotra}
\orcid{0000-0001-6212-0356}
\affiliation{%
  \institution{IBM}
  \city{New York}
  \country{USA}
}
\email{jatinganhotra@us.ibm.com}

\author{Martin Hirzel}
\orcid{0009-0006-8840-6065}
\affiliation{%
  \institution{IBM}
  \city{New York}
  \country{USA}
}
\email{hirzel@us.ibm.com}

\author{Reyhaneh Jabbarvand}
\correspondingauthor
\orcid{0000-0002-0668-8526}
\affiliation{%
  \institution{University of Illinois Urbana-Champaign}
  \city{Urbana}
  \country{USA}
}
\email{reyhaneh@illinois.edu}

\begin{abstract}

Agents 
are commonly instructed to follow a task-specific \emph{plan} for guidance.
However, it is unknown to what extent agents actually follow instructed plans. Without such an analysis---determining the extent agents \emph{comply} with a given plan---it is impossible to assess whether a solution was reached through correct strategic reasoning or through other means, e.g., data contamination or overfitting to a benchmark.
This paper presents the first extensive, systematic analysis of \emph{plan compliance} in programming agents, examining 21,120 trajectories from \SA across four LLMs on \swebv and \swebp under eight plan variations. Without an explicit plan, agents fall back on internalized workflows during training, which are often incomplete, overfit, or inconsistently applied. Providing the standard plan improves issue resolution, and we observe that periodic plan reminders can mitigate plan violations and improve task success. A subpar plan hurts performance even more than no plan at all. Surprisingly, inserting additional task-relevant phases in the early stage can degrade performance, particularly when these phases do not align with the model’s internal problem-solving strategy. These findings call for fine-tuning paradigms that teach models to follow instructed plans, rather than encoding task-specific plans in them, so that they
reason and act adaptively, rather than memorizing workflows.

\end{abstract}



\maketitle

\vspace{-12pt}
\section{Introduction}
\label{sec:introduction}

Agents have emerged as a promising paradigm for automating software engineering tasks, from code synthesis and translation to end-to-end issue resolution~\cite{SWE-agent,wu2024introducing,ibrahimzada2026recodeagent}. 
Central to these systems is the use of \emph{structured instructions}, a.k.a.\ a \emph{plan}, which decomposes a high-level objective of a given task into an ordered sequence of steps that the agent can follow to accomplish the task successfully. In theory, a plan can help reduce cognitive load for reasoning about future steps in the local reason-act-observe loop~\cite{yao2022react}. As a result, planning is commonly used in agentic frameworks, usually encoded as step-by-step instructions in the system prompt~\cite{swe-agent-plan,openhands-plan,trae-agent-plan,chen2026unlocking}. For example, a plan for fixing GitHub issues will instruct the agent to navigate to a potential bug location (based on the issue description), reproduce the bug to ensure correct localization, patch the bug, and validate the patch's correctness. 

In practice, the plan is only \emph{advisory} and 
without enforcement. 
At each trajectory step, the model performs local reasoning over its current context, and its actions may or may not align with the plan. As the trajectory grows and the context fills with error messages, file contents, and prior reasoning, the plan's influence may diminish, consistent with the known limitations of LLMs in attending to earlier context~\cite{liu2024lost}. Therefore, whether agents truly follow the instructed plan remains an open question. Evaluating plan compliance 
can reveal whether the agent accomplishes a task through correct strategic reasoning or through overfitting to benchmark trajectories or data contamination. 

This paper presents a large-scale
evaluation of plan compliance in programming agents. The analysis leverages a novel plan compliance metric, measured across three dimensions: Plan Phase Compliance, Plan Order Compliance, and Plan Phase Fidelity (\S \ref{sec:approach}). We evaluate 21,120 \SA trajectories, generated to resolve instances of two popular benchmarks (\swebv~\cite{chowdhury_et_al_2024} and \swebp~\cite{swebenchpro}), using four backbone LLMs (\gptf, \Vthree, \Rone, and \devstral), under eight plan settings: the standard navigate-reproduce-patch-validation plan, no specified plan, and six variations of the standard plan, obtained by removing, adding, 
and re-ordering
plan phases. Our study answers the following research questions:

\begin{itemize}[leftmargin=*]
  \item \textbf{RQ1: Standard Plan Compliance (\S \ref{sec:rq1-plan}).} To what extent do agents follow the instructed plan? What factors impact plan compliance and violations? Does plan compliance help agents resolve issues?
  \textbf{Findings.} Agents follow the standard plan, although with varying compliance rates. Some strictly follow the plan in the specified order, while others adaptively override the plan based on the trajectory, depending on the problem's difficulty. Following the plan positively helps all agents resolve more GitHub issues. The fine-tuning paradigm, context window pressure, data contamination, overfitting, and optimization for short-term reward are the most prevalent factors impacting plan compliance. 
  
  \item \textbf{RQ2: Behavior of Agents in the Absence of Plan (\S \ref{sec:rq2-noplan}).} How do agents operate in the absence of a plan? To what extent does removing the plan impact overall performance?
  \textbf{Findings.} Without a plan, agents follow their internalized prob\-lem-solving strategy, which overlaps with the standard plan to a varying degree. The success rate, however, drops in the absence of the standard plan. 

  \begin{figure*}[!tb]
\centering
\includegraphics[width=0.92\linewidth]{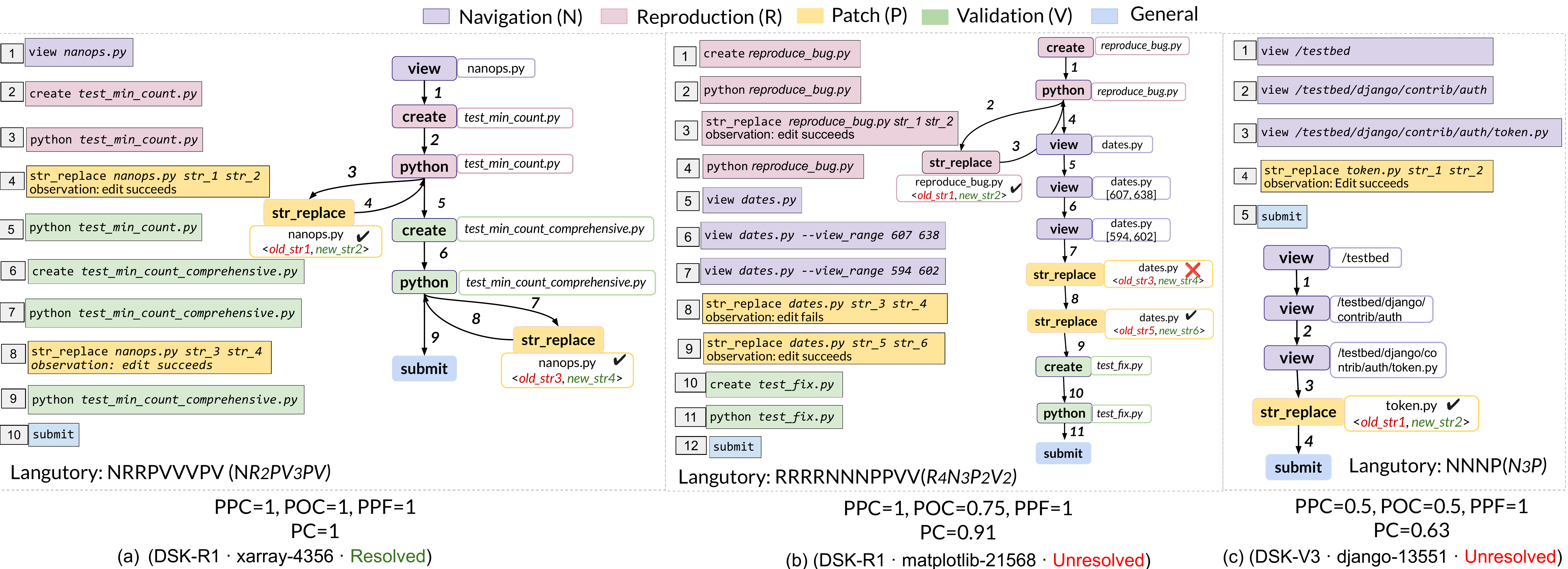}
\vspace{-10pt}
\caption{Illustrative examples of agent trajectories and their corresponding \graph and \lang representations. Plan: $\mathcal{L}^\star(\Phi)=\colorbox[HTML]{CCC7E6}{N}\colorbox[HTML]{EEC7D4}{R}\colorbox[HTML]{FFED99}{P}\colorbox[HTML]{D9EDCC}{V}$.}
\vspace{-10pt}
\label{fig:illustrative-examples}
\end{figure*}


  \item \textbf{RQ3--RQ5: Impact of Plan Variations (\S \ref{subsec-rq3-remove-plan}--\ref{subsec-rq5-reorderplan-repeat}).} Do agents heed removal, addition, and reordering of plan phases? To what extent does frequent reminding of the plan phases help compliance in the long-horizon task of program repair?
  \textbf{Findings.} Removing a standard plan phase, even if the agent usually ignores it under the standard plan setting, negatively impacts the agents' performance, confirming the overall impact of a global plan on local reasoning steps. The negative impact of a bad plan is greater than no plan at all. Surprisingly, augmenting plans with task-relevant phases inspired by best practices also negatively affects agents' performance when they are not aligned with the model's internal strategy. Periodic plan reminders reduce plan violations and improve performance.
  
  \item \textbf{RQ6: Generalization to Other Benchmarks (\S \ref{subsec-rq6-contamination}).} How much can the observations and conclusions about plan compliance generalize to another benchmark, i.e., \swebp~\cite{swebenchpro}?
  \textbf{Findings.} The plan compliance rate of the agents across all settings on \swebp drops by 13\%, on average, compared to \swebv. The agents exhibit different phase flow patterns, e.g., they give up on generating reproduction tests early and validate patches using existing regression tests rather than generating tests. This is likely because \swebp instances are more challenging and less contaminated, and the high-level standard plan is no longer effective at guiding the agents.
  
  \item \textbf{RQ7: Impact of Nondeterminism (\S \ref{subsec-rq7-nondeterminism}).} To what extent is plan compliance of agents under different plan settings attributed to nondeterminism?
  \textbf{Findings.} Nondeterminism exists but does not impact our findings. We account for nondeterminism by repeating experiments and comparing persistent behaviors across plan settings.
  
\end{itemize}

\vspace{-8pt}
We are the \emph{first} to (1) conduct a large-scale analysis of plan compliance by agents, (2) introduce novel plan compliance metrics, (3) speculate the root causes of plan violations, and (4) study how plan compliance relates to task success.
Our findings suggest that the effectiveness of the plan is tightly coupled to its alignment with the model's internalized workflow and the task's complexity. Therefore, future research should focus on fine-tuning paradigms that teach models to follow plans more effectively, rather than encoding task-specific plans into them.

\section{Experimental Design}
\label{sec:approach}

We aim to analyze whether and to what extent programming agents follow the specified, task-appropriate software engineering workflows. Given the popularity of programming agents for fixing real-world GitHub issues, this study will focus on program repair. The default practical workflow for this task involves localizing the bug, patching the code, and then validating whether the patch resolved the bug. Many existing scaffolds, e.g., \SA/\MSA, \TA, and \OH, \emph{explicitly instruct} the agent to follow a similar plan in their system prompt\footnote{The specified plan for some agents could be more verbose. Regardless, all existing agents follow similar high-level plans.}~\cite{swe-agent-plan,trae-agent-plan,openhands-plan}: 

\begin{itemize}[leftmargin=*]
    \item \textbf{Navigation (N)}. The agent searches for, opens, and reads files relevant to the issue description, building an understanding of the codebase and localizing the relevant components.
    
    \item \textbf{Reproduction (R)}. The agent generates \emph{new tests} to reproduce the bug, i.e., tests that fail on the buggy code.
    
    \item \textbf{Patch (P)}. The agent edits the application code to fix the bug.
    
    \item \textbf{Validation (V)}. The agent runs \emph{reproduction tests} and generates new tests to validate patch correctness.
\end{itemize}

\sloppy Assessing whether an agent follows the instructed plan for a task requires \emph{process-centric} analysis of trajectories. We build our process-centric analysis on top of \graph and \lang~\cite{graphectory}. \graph represents linear raw trajectories as enriched graph structures, where nodes are the agent's distinct actions and edges denote the chronological execution order. \lang is an abstract representation of the trajectory in the form of language. That is, by mapping the agent's actions across the sequence of $n$ trajectory steps $T = (s_1, \ldots, s_n)$ to an alphabet $\Phi=\{p_1,\ldots,p_m\}$ of $m$ letters\footnote{$m \ll n$ to show an overall strategy rather than detailed actions.}, \lang $\mathcal{L}(T,\Phi)$ explains the agent's problem-solving strategy as a sequence of letters. 

When the alphabet denotes plan phases, constructing \lang requires mapping each raw trajectory action to the phase it attempts to perform. We map file and directory inspection to Navigation, newly generated test creation and execution before any application-code edit to Reproduction, application-code edits to Patch, and newly generated test creation or execution after patching to Validation. Agent-generated tests are distinguished from existing repository tests: running existing tests is treated as regression testing, which is outside the default NRPV plan unless explicitly included in a plan variant. Actions that do not correspond to a plan phase, such as environment probing or dependency checks, are labeled as General. For dynamically executed Python code, such as heredocs, we parse the code to identify environment setup, file modifications, and test logic using signals such as assertions and test-specific imports (e.g., \texttt{pytest}, \texttt{unittest}, and \texttt{mock}). The mapping implementation is available in our artifact~\cite{website}.

Given a plan phase alphabet~$\Phi$ and expected phase sequence $\mathcal{L}^\star(\Phi)$, we say a \lang \emph{complies} with the instructed plan if it includes all and only specified plan phases in the specified order. We propose a novel process-centric metric, \emph{plan compliance}~($PC$), measured across three dimensions: \emph{plan phase compliance}~($PPC$), \emph{plan order compliance}~($POC$), and \emph{plan phase fidelity}~($PPF$). 

To illustrate the concept, Figure~\ref{fig:illustrative-examples} shows three trajectories generated by \SA along with their corresponding \graph and \lang. Figure~\ref{fig:illustrative-examples}a shows a \emph{compliant} and \emph{successful} execution. \emph{\SA\textsubscript{DSK-R1}} starts by navigating to the buggy file \texttt{\small{nanops.py}} (step 1), creates and executes a reproducing test (steps 2--3), edits the buggy file (step 4), validates the patch by creating and executing a more comprehensive test (steps 5--7), edits the file again to handle corner cases (step 8), and re-executes the test~(step 9) before submitting the patch. This yields a \lang of $NR_2PV_3PV$, which is compliant with the instructed plan $\mathcal{L}^\star(\Phi)=\colorbox[HTML]{CCC7E6}{N}\colorbox[HTML]{EEC7D4}{R}\colorbox[HTML]{FFED99}{P}\colorbox[HTML]{D9EDCC}{V}$. The execution in Figure~\ref{fig:illustrative-examples}b covers all plan phases in its trajectory, but violates the intended order, with excessive reproduction~\mbox{(steps 1--4)} preceding navigation (steps 5--7) and leading to an \emph{unresolved} patch. The execution in Figure~\ref{fig:illustrative-examples}c skips key phases, transitioning directly from navigation (steps 1--3) to patching (step~4) before submission, violating the plan. The consequence of plan violation is a low-quality patch that does not resolve the issue.

We will explain our novel process-centric plan compliance metrics using this illustrative example. $PPC$ measures whether \lang $\mathcal{L}(T,\Phi)$ covers the phases specified in the plan:

\vspace{-8pt}
\begin{equation}
\label{eq:PPC}
PPC =
\frac{\left|\Phi \cap \{\mathcal{L}(T,\Phi)_t \mid 1 \le t \le n\}\right|}
{|\Phi|}
\end{equation}

\begin{table*}[t]
\centering
\small
\caption{Summary of studied plan settings, their corresponding formulation, and type of plan variation.}
\vspace{-8pt}
\label{tab:plan_conditions_full}
\begin{tabular}{llll}
\toprule
\textbf{Plan Setting} & \textbf{Plan Formulation} & 
\textbf{Plan Variation} & 
\textbf{Plan Description} \\

\midrule
Standard (Default) Plan        & $\langle$$N$, $R$, $P$, $V$$\rangle$ & 
Baseline & 
Standard Navigation-Reproduction-Patch-Validation plan \\

No Plan             & --- & 
Reduction & 
Plan removed entirely from the system prompt\\

Default Plan - Reproduction     & $\langle$$N$, $\neg{\text{R}}$, $P$, $V$$\rangle$ & 
Reduction & 
Reproduction phase removed \\

Default Plan - Validation       & $\langle$$N$, $R$, $P$, $\neg{\text{V}}$$\rangle$ & 
Reduction & 
Validation (after patching) phase removed \\

Default Plan + Regression Test Execution          & $\langle$$R_G$, $N$, $R$ $P$, $V$, $V_G$$\rangle$ & 
Augmentation & 
Regression test execution phases added\\

Default Plan + Summary of Changes             & $\langle$$N$, $R$, $P$, $V$, $S$$\rangle$ & 
Augmentation & 
Summarizing changes before submission added\\

Reordered Default Plan          & $\langle$$N$, $P$, $R$, $V$$\rangle$ & 
Reordering & 
Patching moved before Reproduction \\

Periodic Plan Reminder           & $\langle$$N$, $R$, $P$, $V$$\rangle$ & 
Repeating & 
Default plan re-injected every \emph{five} trajectory steps \\
\bottomrule
\end{tabular}
\vspace{-9pt}
\end{table*}

\noindent Here, $\mathcal{L}(T,\Phi)_t$ denotes the phase label assigned to the t-th action in trajectory $T$ under the phase vocabulary $\Phi$. $PPC=1$ if every phase in $\Phi$ appears at least once in the \lang. In practice, an agent may skip some plan phases, e.g., directly jumping into patching after navigation without reproduction test generation. Therefore, $PPC \in [0,1]$. In Figure~\ref{fig:illustrative-examples}c, the agent skips reproduction and validation, resulting in $PPC=0.5$.
The executions in Figure~\ref{fig:illustrative-examples}a and~\ref{fig:illustrative-examples}b cover all plan phases in $\Phi$ and achieve $PPC=1$. 

Not only is covering all phases important, but so is following the proper order through trajectory execution. $POC$ measures the fraction of phases in $\mathcal{L}^\star(\Phi)$ that appear in the correct relative order: 

\vspace{-8pt}
\begin{equation}
\label{eq:POC}
POC =
\frac{\mathrm{LIS}(i_1, \ldots, i_m)}{m}
\end{equation}

\noindent where $\mathrm{LIS}(\cdot)$ denotes the length of the longest increasing subsequence and $i_k$ denotes the first occurrence index of phase $p_k$ in $\mathcal{L}(T,\Phi)$ (if present). Phase revisits are allowed; $POC$ evaluates the order of first occurrences. In Figure~\ref{fig:illustrative-examples}b, the first occurrence indices of $\colorbox[HTML]{CCC7E6}{N}\colorbox[HTML]{EEC7D4}{R}\colorbox[HTML]{FFED99}{P}\colorbox[HTML]{D9EDCC}{V}$ are $[5,1,8,10]$. The longest increasing subsequence is $[1,8,10]$ with length $3$, yielding $POC=\frac{3}{4}$. Failing to follow the expected order can cause inefficient trajectories or task failure. In Figure~\ref{fig:illustrative-examples}b, the agent begins with reproduction before properly navigating the codebase, leading to repeated modifications to the reproduction script (steps~3--4), and a failed edit at step~8. 

Agents operate through iterative reasoning–action–observation cycles~\cite{yao2022react}, in which decisions are \emph{locally} conditioned on the current context rather than the initial instructed plan. Moreover, training strategies can overfit the LLMs to certain actions outside of the instructed plans for specific tasks. Hence, some actions may not map to plan phases in the \lang. For example, an agent may decide to open a pull request after patch validation, which is not part of the instructed plan in existing programming agents~\cite{swe-agent-plan,openhands-plan,trae-agent-plan}. In such a case, \lang may contain \emph{unknown} letters that are considered gibberish with respect to the specified plan phases. Including additional actions beyond those in the recommended plan is not necessarily negative, but can be distracting. Therefore, $PPF$ penalizes the appearance of phases outside the plan alphabet:

\vspace{-8pt}
\begin{equation}
\label{eq:PPF}
PPF = \frac{|\Phi|}
{\left| \Phi \cup \{\mathcal{L}(T,\Phi)_t \mid 1 \le t \le n\} \right|}
\end{equation}

\noindent \sloppy $PPF \in (0,1]$, and $PPF=1$ if every phase appearing in the \lang belongs to $\Phi$. The overall compliance score is the geometric mean of its three component metrics:

\vspace{-8pt}
\begin{equation}
\label{eq:PC}
PC = (PPC\;.\;POC\;.\;PPF)^{1/3}
\end{equation}

\noindent $PC \in [0,1]$, where score $PC=1$ indicates perfect plan compliance. Geometric mean aggregates sub-metrics multiplicatively, ensuring equal weighting and preventing compensation across dimensions. Low compliance in any dimension proportionally reduces the overall score. Lower $PC$ scores reflect deviations in missing phases, spurious phases, or violations of the logical phase ordering.

\section{Empirical Setup}
\label{sec:setup}

\noindent \textbf{Models and Scaffold.} 
To capture a multi-dimensional analysis of plan compliance, we evaluate the \SA scaffold~\cite{SWE-agent} across four 
diverse LLMs: \emph{\gptf}~\cite{gpt5_mini} (closed-source frontier reasoning model), \emph{\Rone}~\cite{deepseek_r1_0528} (open-source reasoning model), \emph{\Vthree}~\cite{deepseek_v3} (open-source general-purpose model), and \emph{\devstral~[24B]}~\cite{devstral_2512} (distilled model specialized in coding). We use the default settings of the models and agent~(see our artifact for details~\cite{website}). 
\SA provides a standardized execution environment, supports multiple LLMs, and embeds a default plan in its system prompt. These properties make it a natural testbed for studying the role of planning in programming agents.

\begin{figure*}[t]
\centering
\vspace{-10pt}
\includegraphics[width=0.98\linewidth]{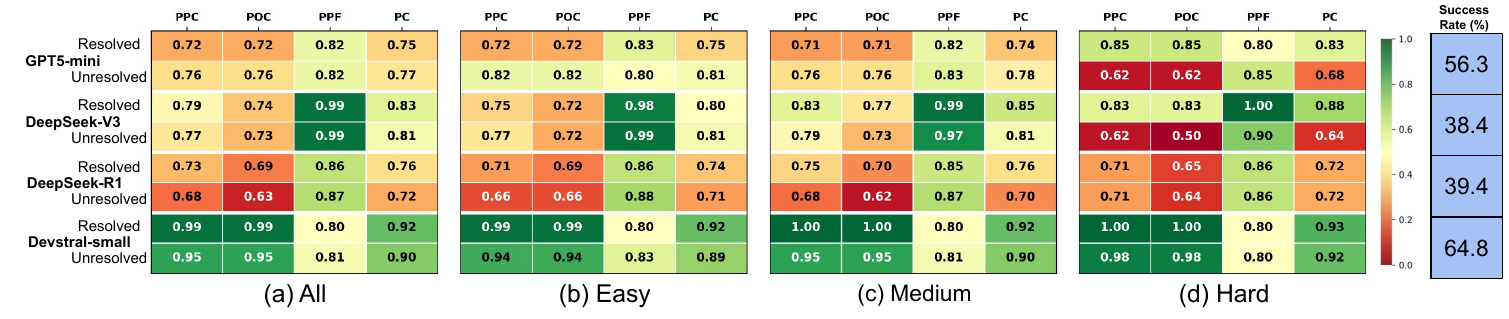}
\vspace{-15pt}
\caption{Standard plan compliance metrics and success rate for studied trajectories across all models.}
\vspace{-10pt}
\label{fig:compliance-heatmap}
\end{figure*}

\vspace{3pt}
\noindent \textbf{Plan Settings.}
We evaluate performance of agents on a given dataset problem under \emph{eight} plan settings: (1) default plan (RQ1), (2) no-plan, i.e., removing the entire plan from the system prompt~(RQ2), (3) removing the reproduction phase (RQ3), (4) removing the validation phase (RQ3), (5) adding a regression test execution phase before navigation, $R_G$, and after validation, $V_G$~(RQ4), (6) adding a change summarization phase $S$ before submission (RQ4), (7) step reordering, i.e., reproduction test generation after patching~(RQ4), and (8) plan reminder, i.e., periodically re-injecting the default plan into the agent's prompt (RQ5). Table~\ref{tab:plan_conditions_full} lists plan settings and their corresponding plan-compliant phase sequence, which our pipeline checks trajectories against. We will explain the rationale for these plan mutations in the corresponding RQs.

\vspace{3pt}
\noindent \textbf{Dataset.}
We evaluate the mentioned LLMs and plan settings for resolving real-world GitHub issues from \swebv~\cite{jimenez_et_al_2024,chowdhury_et_al_2024} and \swebp~\cite{swebenchpro}. Specifically, our primary evaluation (RQ1--RQ6) uses 497 instances\footnote{We exclude the 3 Very Hard instances, none of which is resolved by any studied model under any plan setting and thus provides limited evidence for comparative analysis.} of \swebv, covering three difficulty levels (Easy, Medium, and Hard), providing a realistic setting for agent behavior.
To study the generalizability of findings, we repeat RQ1--RQ5 for Python instances of \swebp.

\vspace{3pt}
\noindent \textbf{Analysis and Metrics.} Along with the plan compliance metrics (Equations~\ref{eq:PPC}--\ref{eq:PC}), we report the success rate~\cite{jimenez_et_al_2024} and \graph metrics (the number of nodes $NC$, temporal edges $TEC$, and loops $LC$ in the \graph)~\cite{graphectory}. Success rate determines the overall impact of plans on the agent's ability to resolve the issue, and \graph metrics provide insights into how plans affect the overall trajectory toward resolution. In addition to metrics, we leverage process-centric \emph{Phase Flow Analysis} by \citet{graphectory} to provide an in-depth analysis of plan-phase changes in trajectories exhibiting plan violations. Phase Flow Analysis can reveal consistent trends in agent trajectories across different problems. The outcome of this analysis is a Sankey diagram illustrating the evolution of trajectories from one plan phase to the next. 

\noindent \textbf{Mapping Validation.} To assess the reliability of our automated action-to-phase mapping, two authors independently annotated a stratified sample from the Default Plan trajectories. For each of the four studied models, we randomly sampled 20 trajectories and selected the first occurrence of each default-plan phase, yielding a total of 320 actions. The annotators labeled these actions without access to the automatically assigned labels. Agreement among the two manual annotations and the automated labels achieved a Fleiss' $\kappa$~\cite{fleiss1971mns} of 0.99, indicating near-perfect agreement and supporting the reliability of the automated mapping.

\section{Standard Plan Compliance and Violation}
\label{sec:plan-compliance-violation}

\begin{figure*}[t]
    \centering
    \footnotesize
    \setlength{\tabcolsep}{5pt} 
    \renewcommand{\arraystretch}{1.35}
    
    \begin{tabular}{@{}l@{\quad}cccc@{}}
         \footnotesize
        \vspace{-0.2cm}
         \rotatebox{90}{
  \begin{minipage}{1.5cm} 
    \centering All
  \end{minipage}
} & 
        \includegraphics[width=0.228\textwidth]{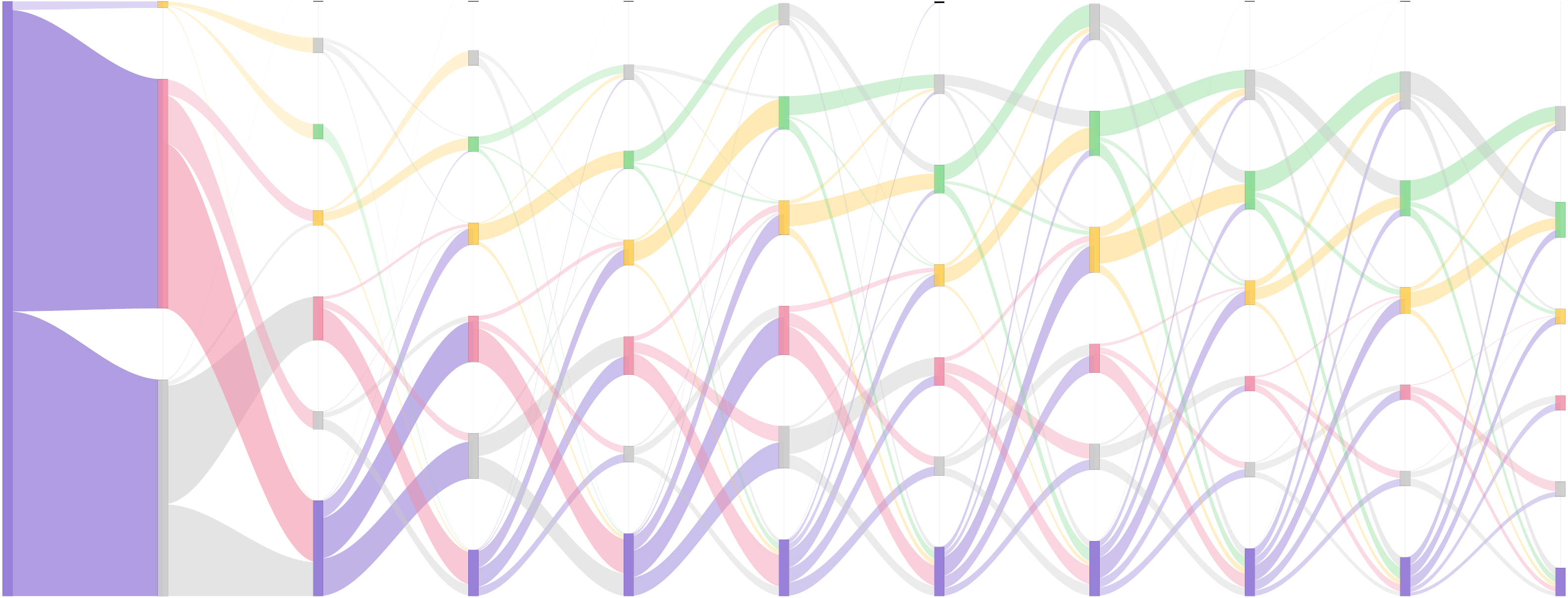} & 
        \includegraphics[width=0.228\textwidth]{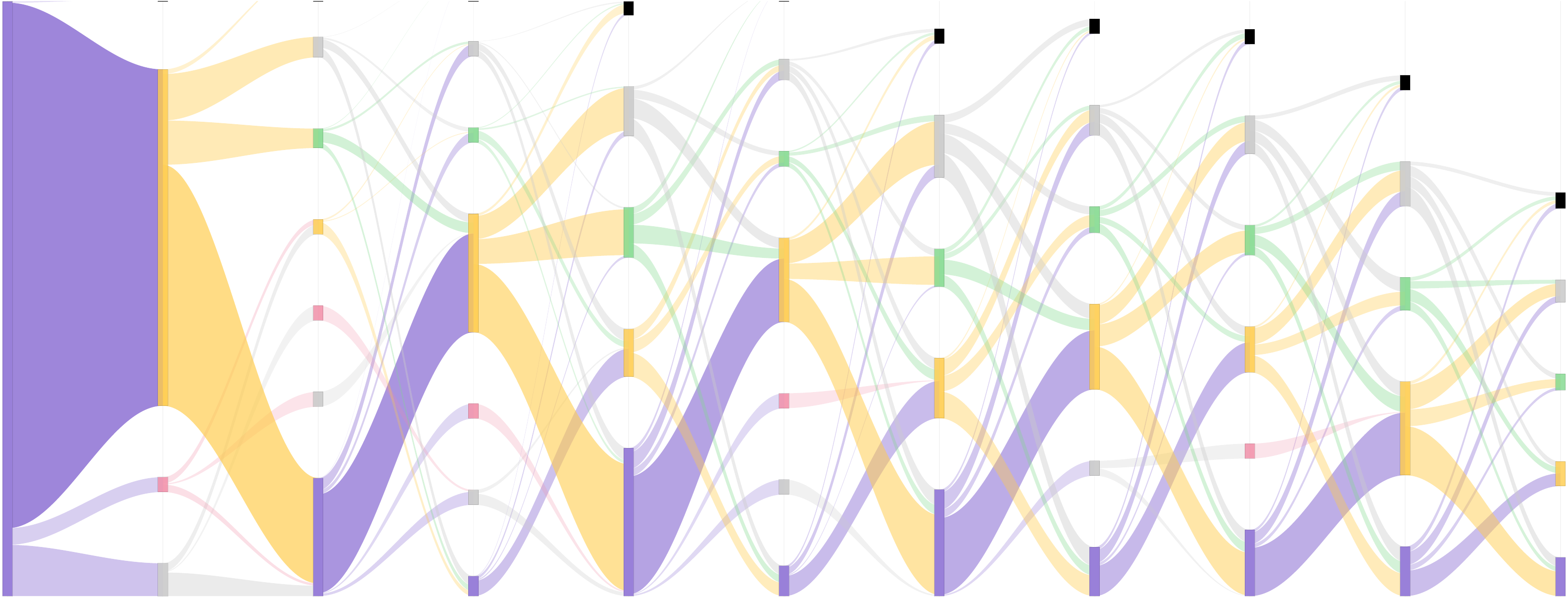} &
        \includegraphics[width=0.228\textwidth]{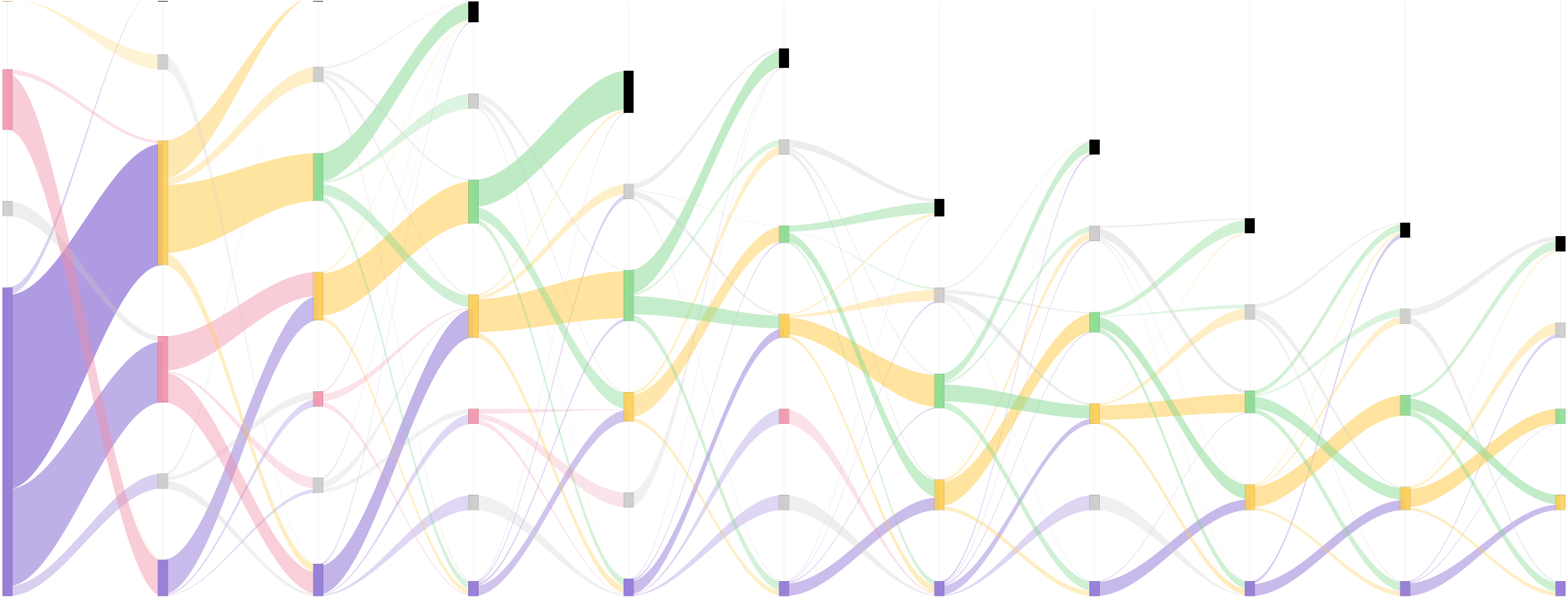} &
        \includegraphics[width=0.228\textwidth]{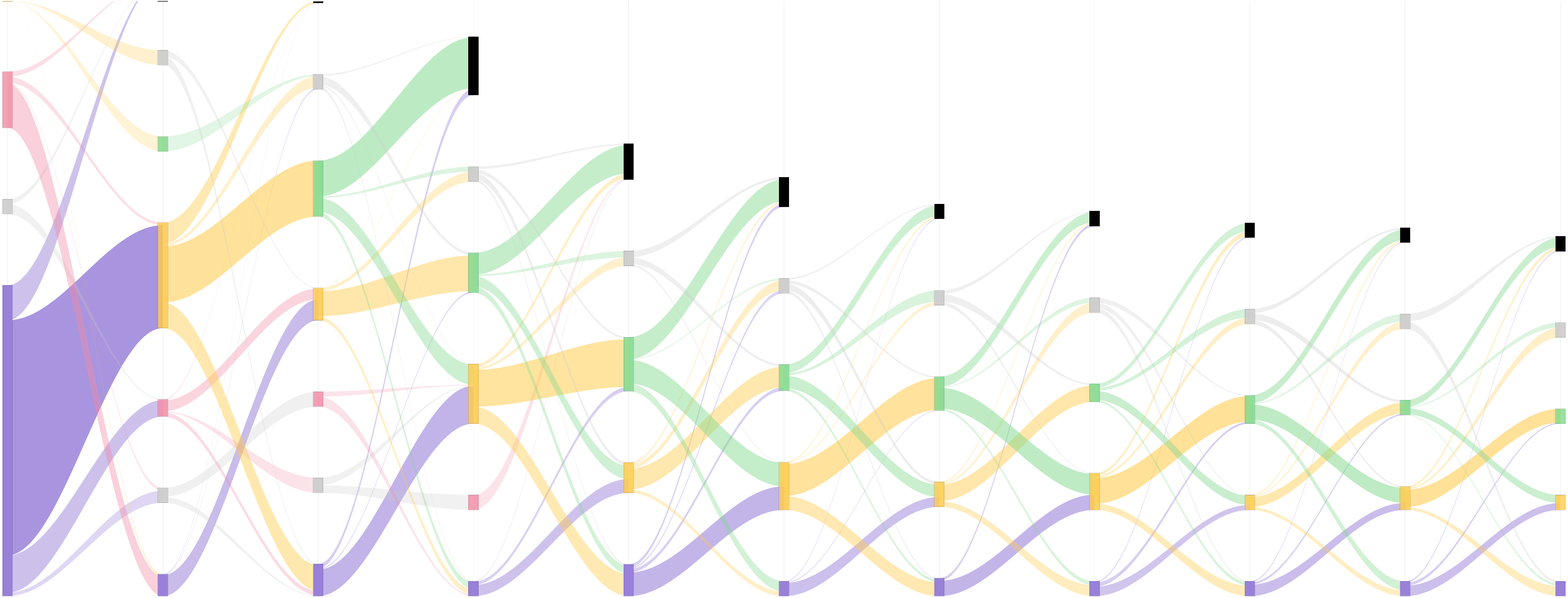} \\[6pt]
        \vspace{-0.2cm}
        \rotatebox{90}{
  \begin{minipage}{1.5cm} 
    \centering Easy
  \end{minipage}
} & 
        \includegraphics[width=0.228\textwidth]{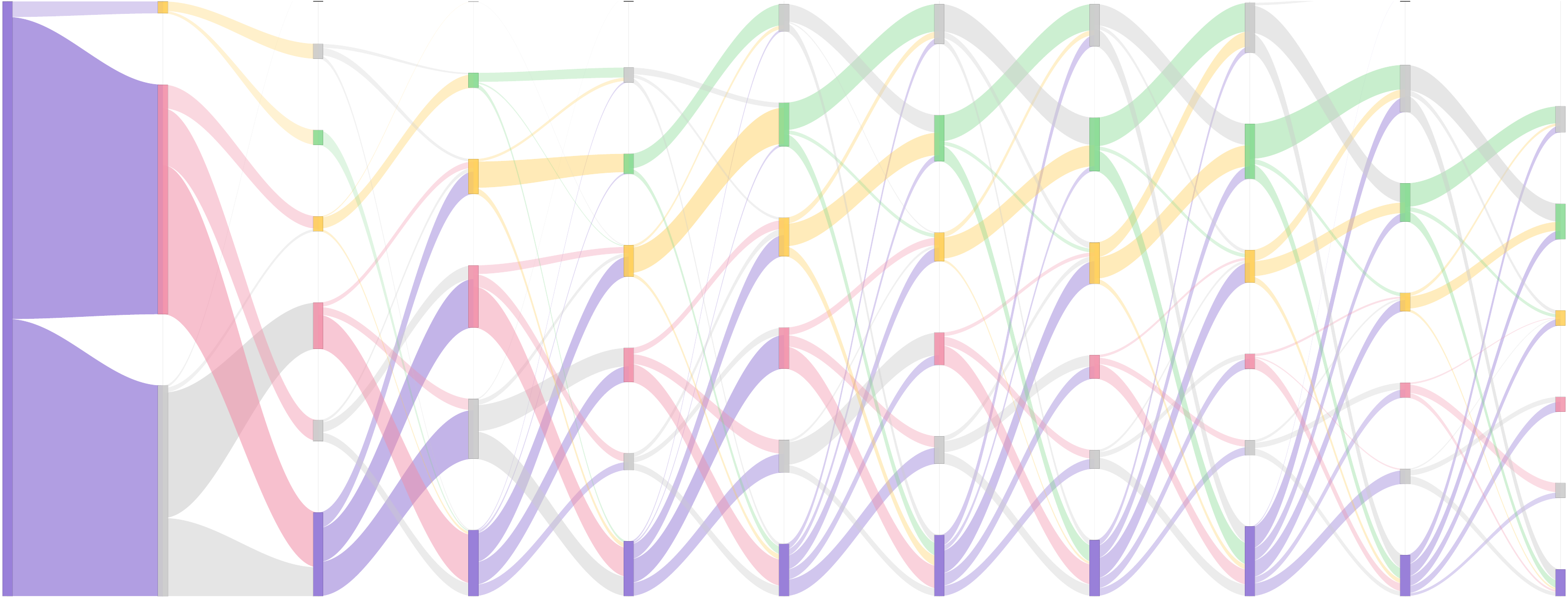} & 
        \includegraphics[width=0.228\textwidth]{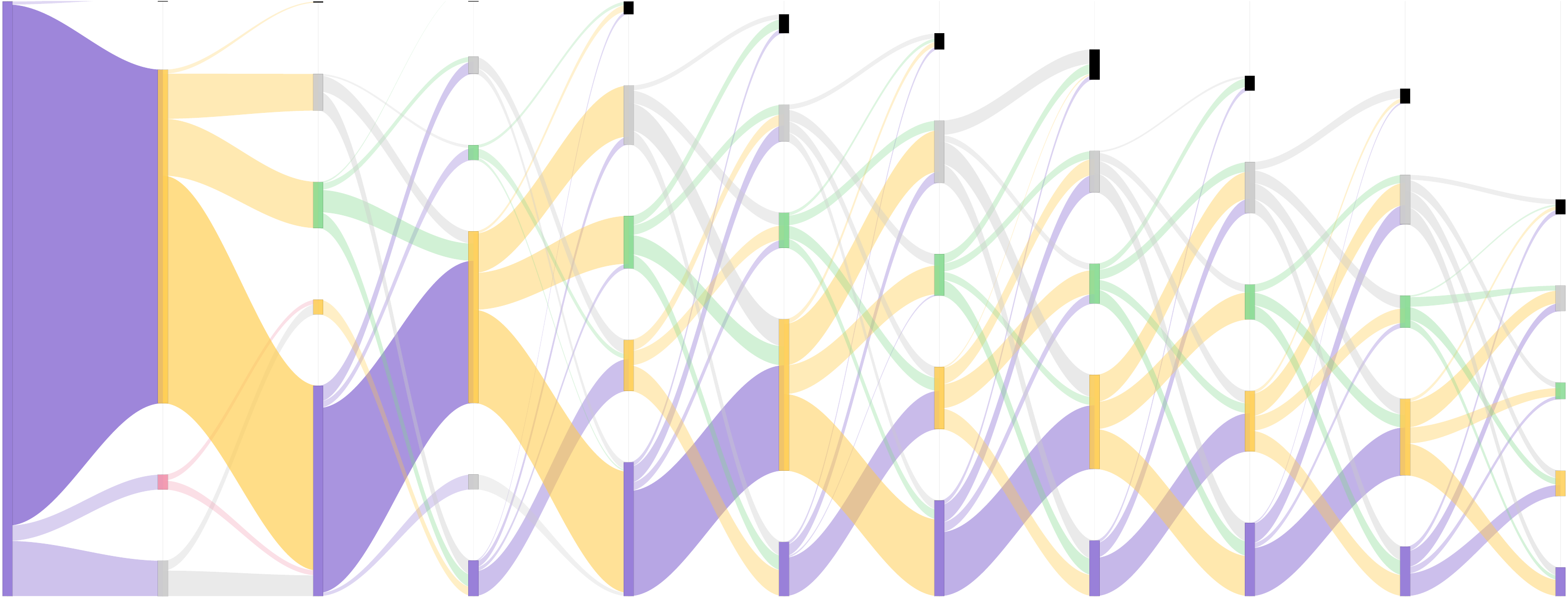} & 
        \includegraphics[width=0.228\textwidth]{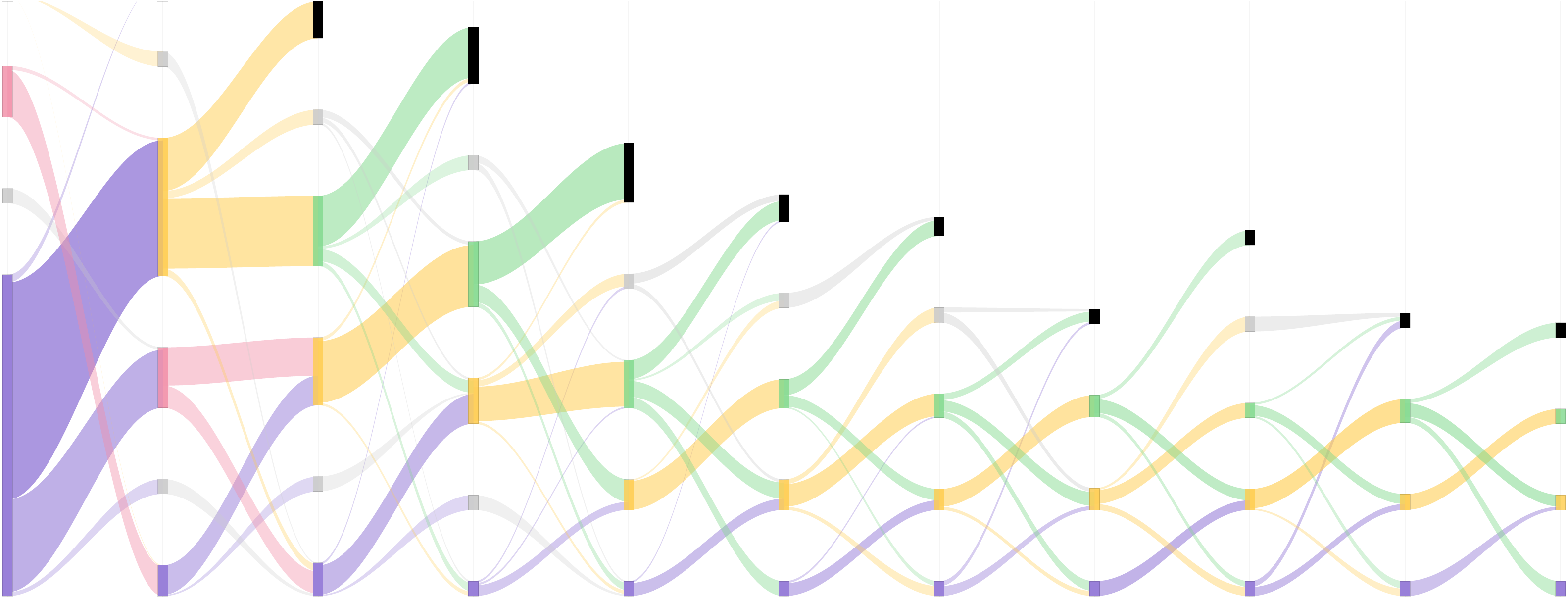} & 
        \includegraphics[width=0.228\textwidth]{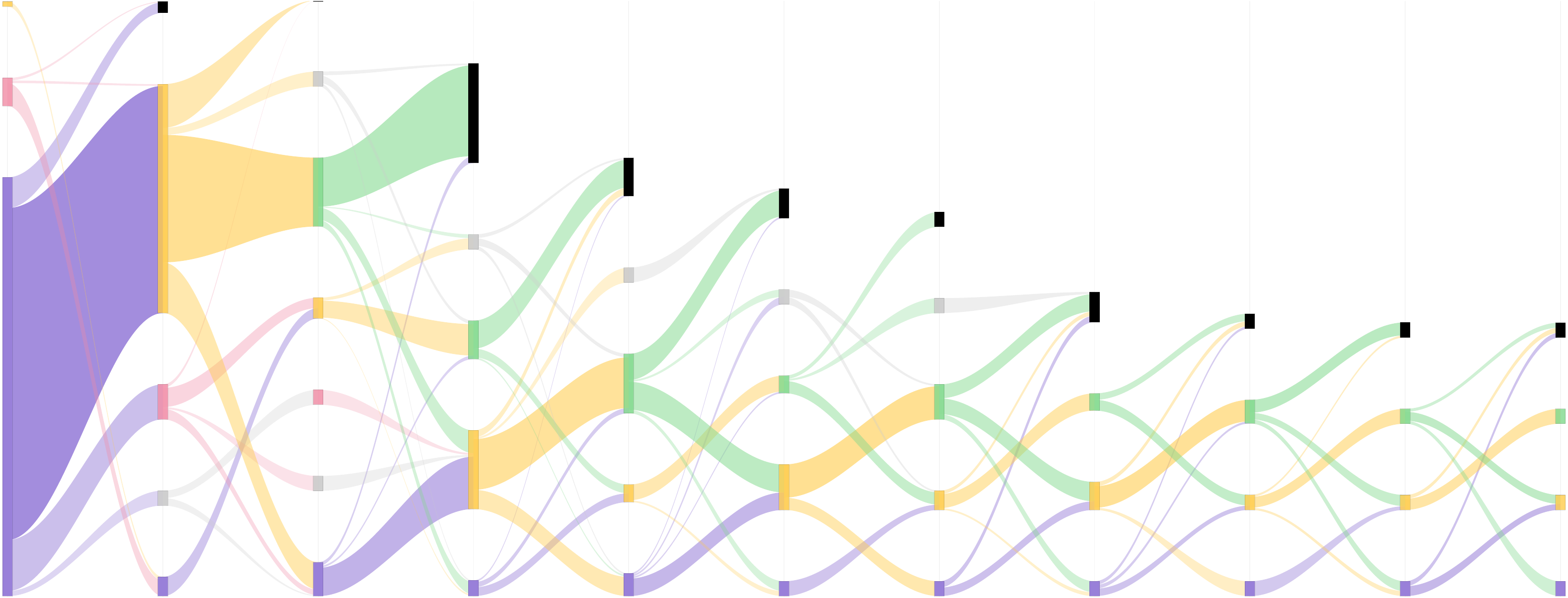}\\[6pt]
        \vspace{-0.2cm}
        \rotatebox{90}{
  \begin{minipage}{1.5cm} 
    \centering Medium
  \end{minipage}
} & 
        \includegraphics[width=0.228\textwidth]{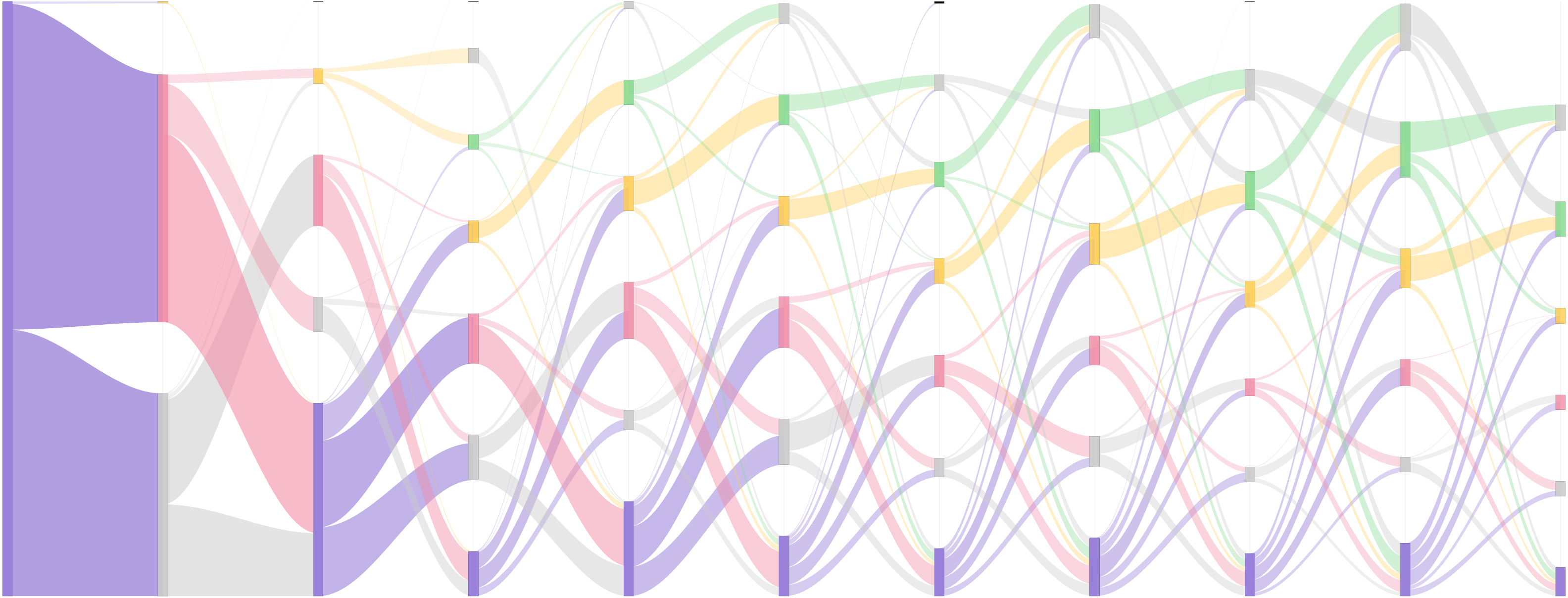} & 
        \includegraphics[width=0.228\textwidth]{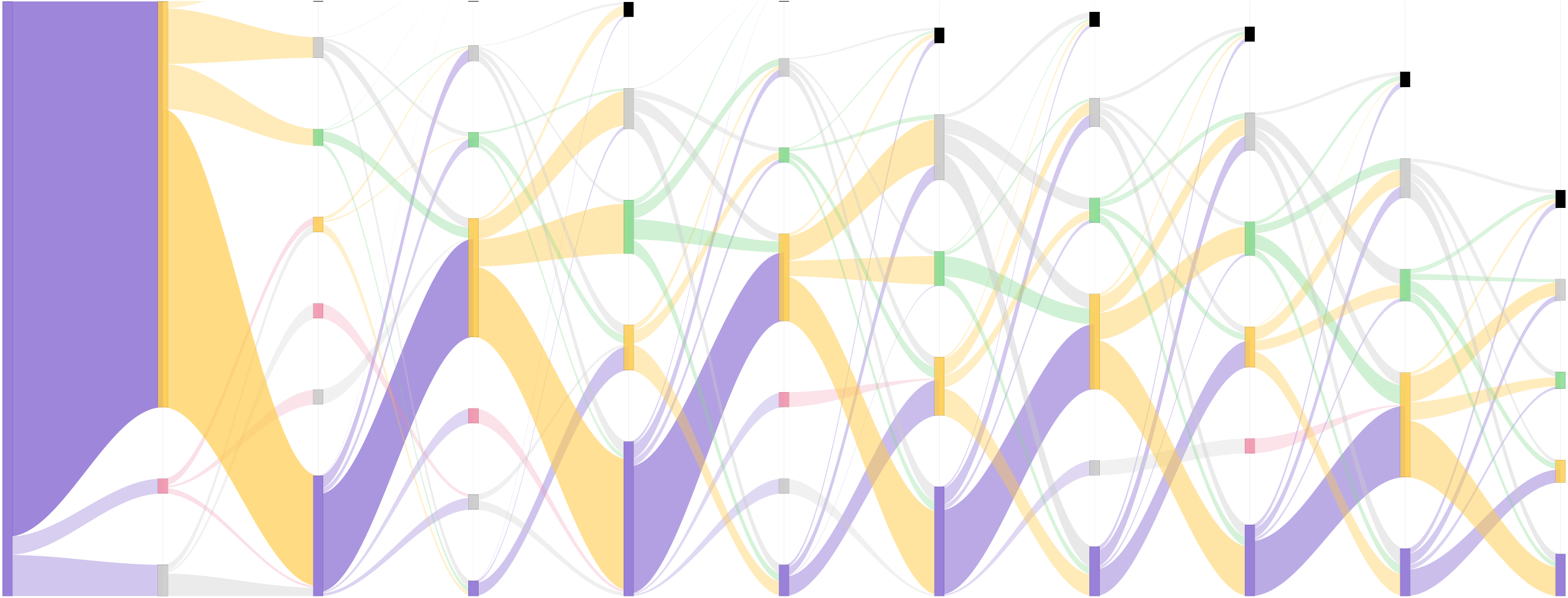} &
        \includegraphics[width=0.228\textwidth]{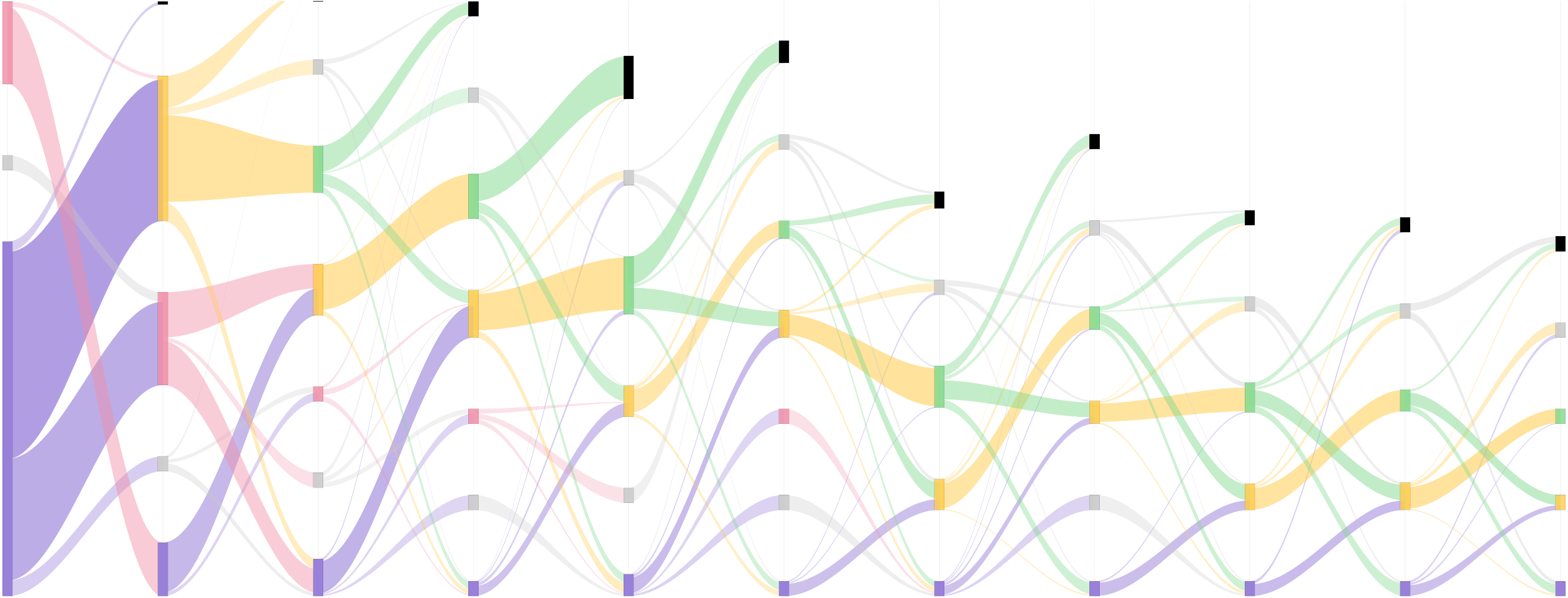} &
        \includegraphics[width=0.228\textwidth]{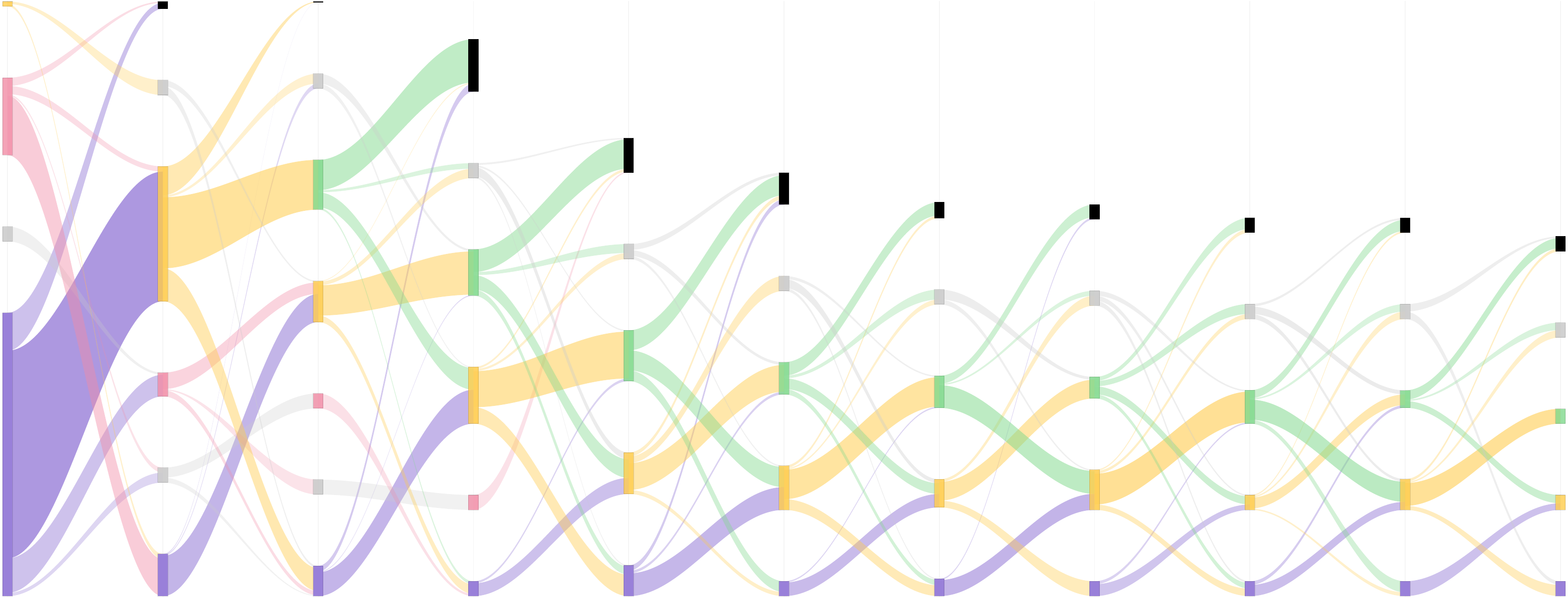} \\[6pt]
        \vspace{-0.2cm}
        \rotatebox{90}{
  \begin{minipage}{1.5cm} 
    \centering Hard
  \end{minipage}
} & 
        \includegraphics[width=0.228\textwidth]{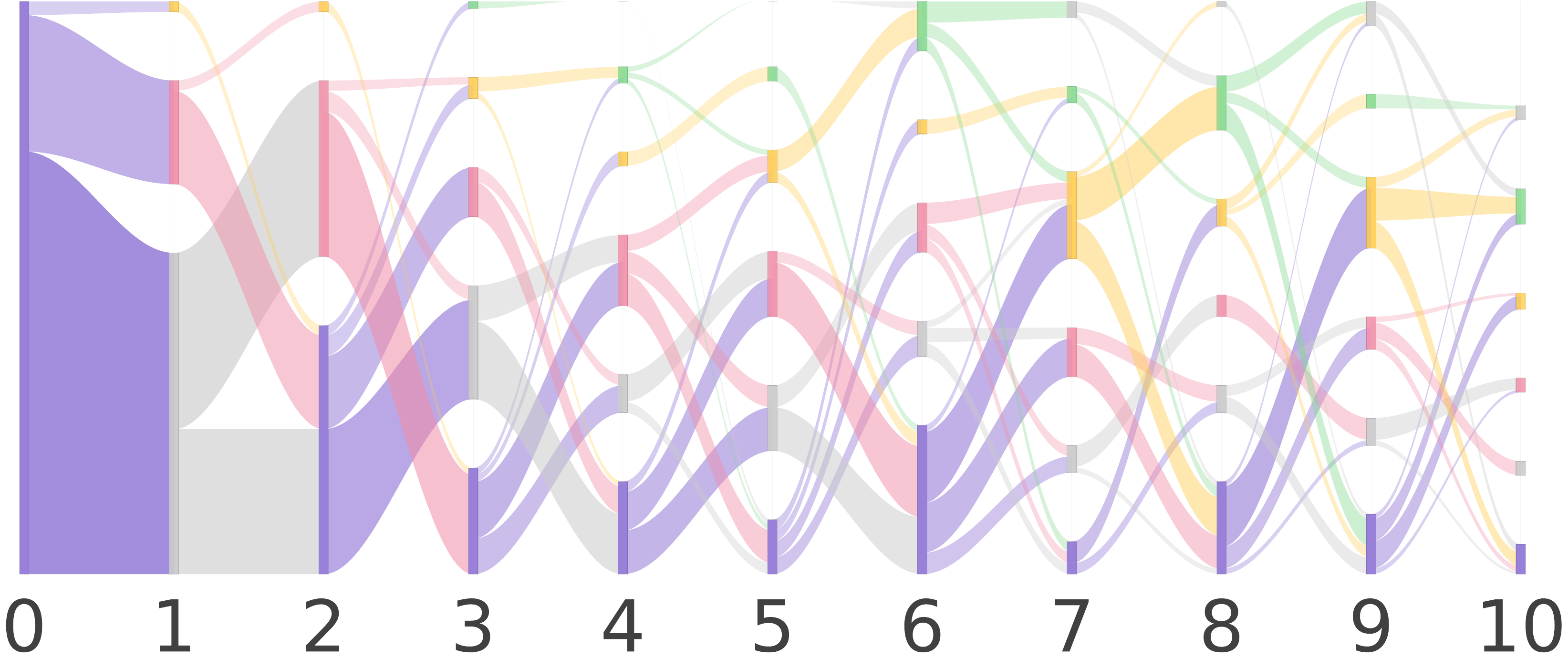} & 
        \includegraphics[width=0.228\textwidth]{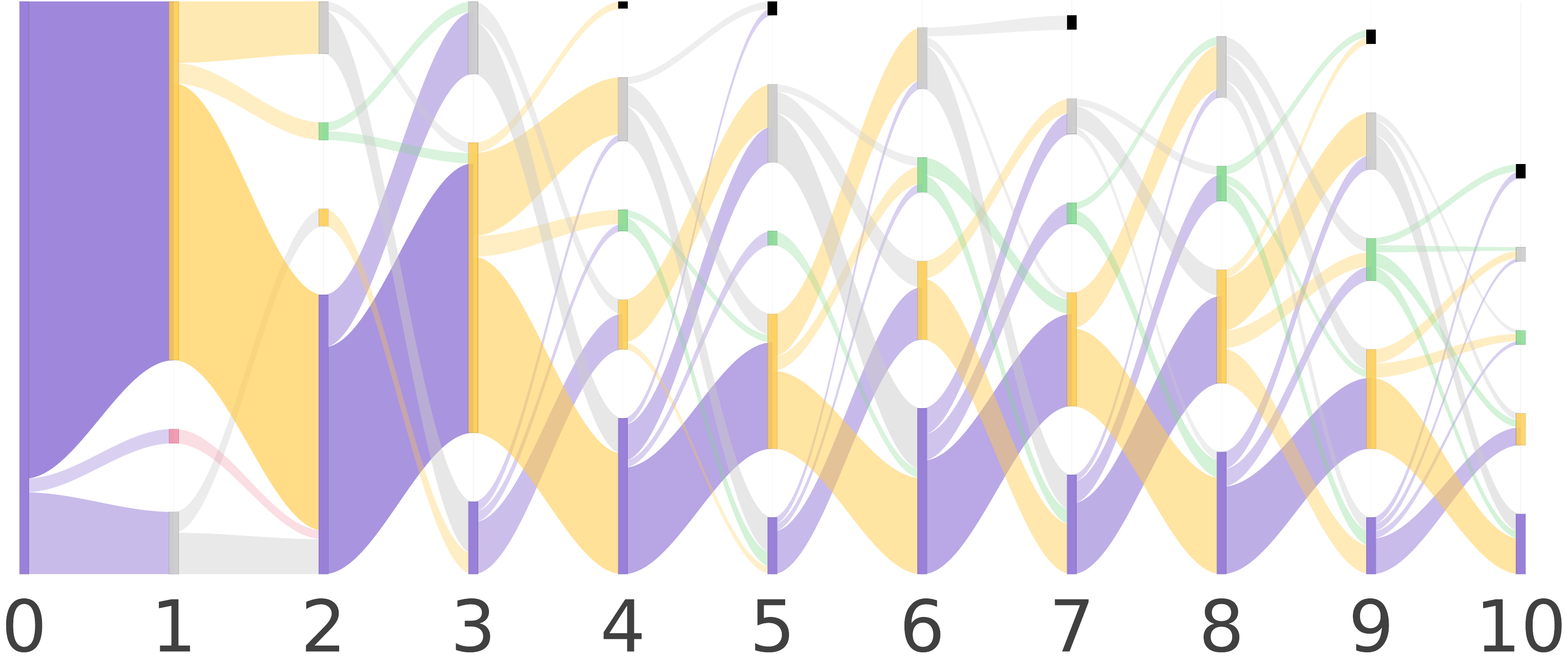} &
        \includegraphics[width=0.228\textwidth]{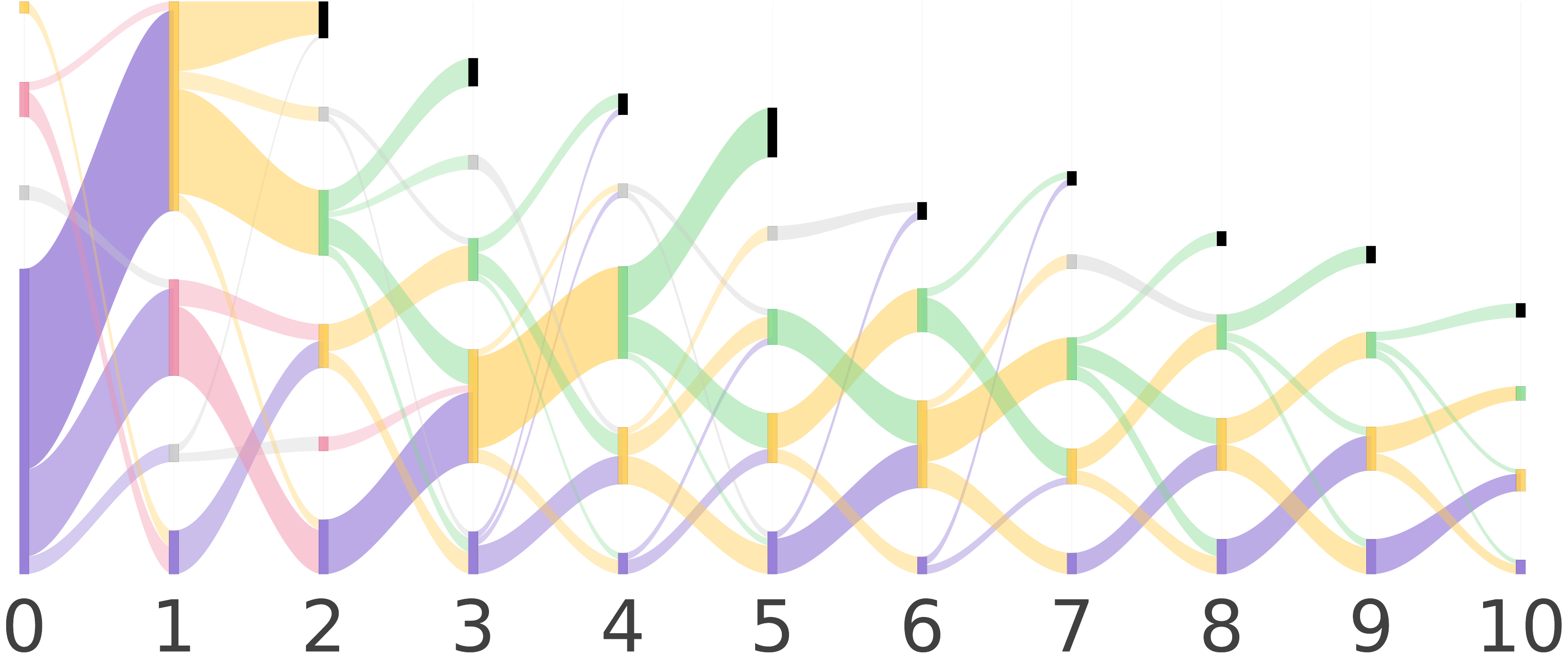} &
        \includegraphics[width=0.228\textwidth]{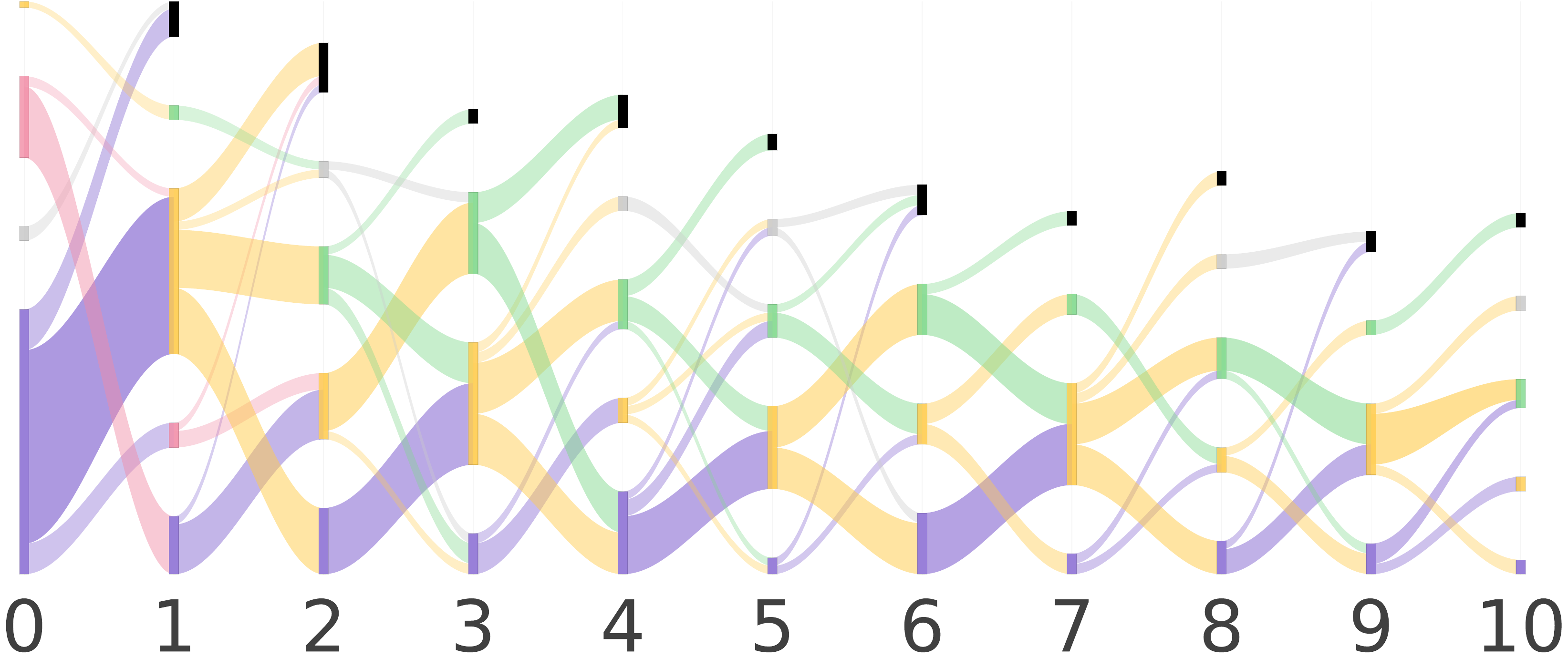} \\[6pt]

        & Devstral-small & GPT5-mini & DeepSeek-V3 & DeepSeek-R1 \\
        
    \end{tabular}
    \vspace{-10pt}
    \caption{Phase flow analysis under Standard plan ($\mathcal{L}^\star(\Phi)=\colorbox[HTML]{CCC7E6}{N}\colorbox[HTML]{EEC7D4}{R}\colorbox[HTML]{FFED99}{P}\colorbox[HTML]{D9EDCC}{V}$). Flow thickness notes the proportion of trajectories going from one phase to another. The black bar indicates trajectory termination, and gray flows represent out-of-plan phases.}
    \vspace{-10pt}
    \label{fig:compliance-sankey}
\end{figure*}

This section first investigates to what extent \SA with different choices of LLMs follows the standard program repair workflow, i.e., \colorbox[HTML]{CCC7E6}{Navigation}, \colorbox[HTML]{EEC7D4}{Reproduction}, \colorbox[HTML]{FFED99}{Patch}, and \colorbox[HTML]{D9EDCC}{Validation}~(\S\ref{sec:rq1-plan}). As an extreme alternative, we remove the entire plan and evaluate how the trajectories change~(\S\ref{sec:rq2-noplan}). 

\vspace{-10pt}
\subsection{RQ1. Standard (Default) Plan Setting}
\label{sec:rq1-plan}
The standard program repair plan that has been used for years by software developers is localizing the bug (\emph{navigating} through files, classes, methods, and lines to pinpoint the bug location) and attempting to \emph{reproduce} it, \emph{patch} it, and \emph{validate} the patch through test execution. Existing scaffolds instruct agents to follow a similar plan with the given order in their default system prompt ($\Phi=\{N,R,P,V\}$ and $\mathcal{L}^\star(\Phi)=\colorbox[HTML]{CCC7E6}{N}\colorbox[HTML]{EEC7D4}{R}\colorbox[HTML]{FFED99}{P}\colorbox[HTML]{D9EDCC}{V}$).   
Figure~\ref{fig:compliance-heatmap}a presents the average $PPC$, $POC$, $PPF$, and $PC$ values (Equations~\ref{eq:PPC}--\ref{eq:PC}) calculated for {2,000 trajectories under this plan. Figures~\ref{fig:compliance-heatmap}b--\ref{fig:compliance-heatmap}d show breakdowns by problem difficulty levels (Easy, Medium, and Hard). Beyond quantitative metrics, Figure~\ref{fig:compliance-sankey} shows the plan phase flow of the agents for all the trajectories. 

\noindent \textbf{Finding 1. Standard plan compliance varies across models.} \devstral consistently follows the plan phases in the given order, demonstrated by high $PPC$ and $POC$ values. However, it exhibits out-of-plan phases to a notable degree in its trajectories (gray flows in Figure~\ref{fig:compliance-sankey}), with low overall $PC$. \gptf, on the other hand, may adapt its strategy depending on the difficulty of the problem. Its trajectories show out-of-plan phases (lower $PPF$), and it usually skips \colorbox[HTML]{EEC7D4}{Reproduction} (lower $PPC$ and $POC)$. \Vthree exhibits a near-perfect $PPF=0.99$ but substantially lower $PPC$ and $POC$, i.e., restricts itself to plan phases but frequently omits some or executes them out of order. \Rone consistently demonstrates lower plan compliance than others, both in following the instructed plan phases and in doing so in the correct order. 

\noindent \textbf{Finding 2. Standard plan compliance is overall higher on resolved instances.} Intuitively, following the instructed standard plan that reflects decades of best practices should lead to successful bug repair. The Mann--Whitney U test~\cite{mann1947test} confirms the significance of this observation for \devstral and \Rone, where resolved instances consistently exhibit higher plan compliance ($p=1e-5$ and $p=0.032$, respectively). 

The correlation is positive for \Vthree, but with less statistical significance ($p=0.60$). The exception is \gptf, where unresolved trajectories are usually more compliant with the plan, demonstrating negative correlation but with low statistical significance ($p=0.285$). Phase flow analysis (Figure~\ref{fig:compliance-sankey}) demystifies this observation: \gptf adapts its strategy based on problem difficulty. For easier problems, where the issue description also likely contains all the information to localize the bug, it often skips \colorbox[HTML]{EEC7D4}{Reproduction} and transitions from navigation to patch. For harder problems, it follows the instructed plan more closely, with \emph{thicker} \colorbox[HTML]{CCC7E6}{Navigation}-to-\colorbox[HTML]{EEC7D4}{Reproduction} flows in earlier phase changes. 

\noindent \textbf{Finding 3. Necessity for process-centric plan compliance metrics.} \graph metrics <node count, edge count, loop count> are in general higher for \devstral (<86,179,64>) and \gptf (<38,49,18>) compared to \Vthree (<15,27,4>) and \Rone (<14,21,4>). Pearson correlation~\cite{pearson-correlation} shows a very weak positive correlation ($0<r\le0.2$) between plan compliance $PC$ and \graph metrics. This confirms the need for a new process-centric metric to specifically target plan compliance, as an orthogonal factor to trajectory complexity.  

\noindent \textbf{Finding 4. The standard plan, in its current form, is incomplete.} We observe that \gptf and \devstral, \emph{in addition} to creating and executing new tests as instructed by the plan, frequently run existing tests in the repository (lower $PPF$ compared to other models). The practice is, in fact, useful for better reproduction test generation and patch validation~\cite{chen2025old}. This finding motivates augmenting existing plans with additional, relevant phases, and assessing the impact of this plan on trajectories~(\S\ref{subsec-rq4-addplan}). 

\vspace{-10pt}
\subsubsection{Contributing Factors to Plan Compliance/Violation}
\label{subsubsec:contributing-factors-to-plan-compliance}

\noindent \textbf{Fine-tuning paradigm.} Depending on the LLM, agents may skip specific plan phases, perform them in a different order, or exhibit out-of-plan actions. Except for \devstral, \SA with other LLMs tends to \emph{skip} \colorbox[HTML]{EEC7D4}{Reproduction} (less presence in Figure~\ref{fig:compliance-sankey}). \SA with DeepSeek models often performs \colorbox[HTML]{EEC7D4}{Reproduction} before \colorbox[HTML]{CCC7E6}{Navigation}.  RQ2~(\S\ref{sec:rq2-noplan}) investigates this under a controlled setting. This also motivates modifying the plan by removing some steps~(\S\ref{subsec-rq3-remove-plan}) to further investigate plan compliance across models. 

\noindent \textbf{Context window pressure.} As trajectories grow, the initial plan must compete with an increasingly long history of thoughts, tool calls, file contents, and error messages, which can make the plan less salient later in execution. Deviation is further encouraged by the agent's locally-conditioned decision process, in which each action is chosen primarily based on the current context and recent tool feedback rather than explicit adherence to the original global workflow. We further investigate this speculation in RQ5 by frequent plan reminders~(\S\ref{subsubsec:with-reminder}).

\noindent \textbf{Data contamination and overfitting.} Backbone LLMs may overfit to the workflow defined by the standard plan. There is also a risk of data contamination when a successful trajectory for solving individual problems in \swebv is used to fine-tune them~\cite{article2,article1}. Therefore, plan compliance may not be rooted in their ability to follow plan instructions~\cite{prathifkumar2025does} or the plan positively impacting their reasoning to accomplish the task~\cite{liang2025swe}. We will evaluate the impact of this factor by repeating the experiments on a \emph{less contaminated} \swebp dataset~(\S\ref{subsec-rq6-contamination}).


\subsection{RQ2. No Plan Setting}
\label{sec:rq2-noplan}

\begin{figure*}[t]
    \centering
    \footnotesize
    \setlength{\tabcolsep}{5pt} 
    \renewcommand{\arraystretch}{1.35}
    
    \begin{tabular}{cccc}
         \footnotesize
         
        \includegraphics[width=0.23\textwidth]{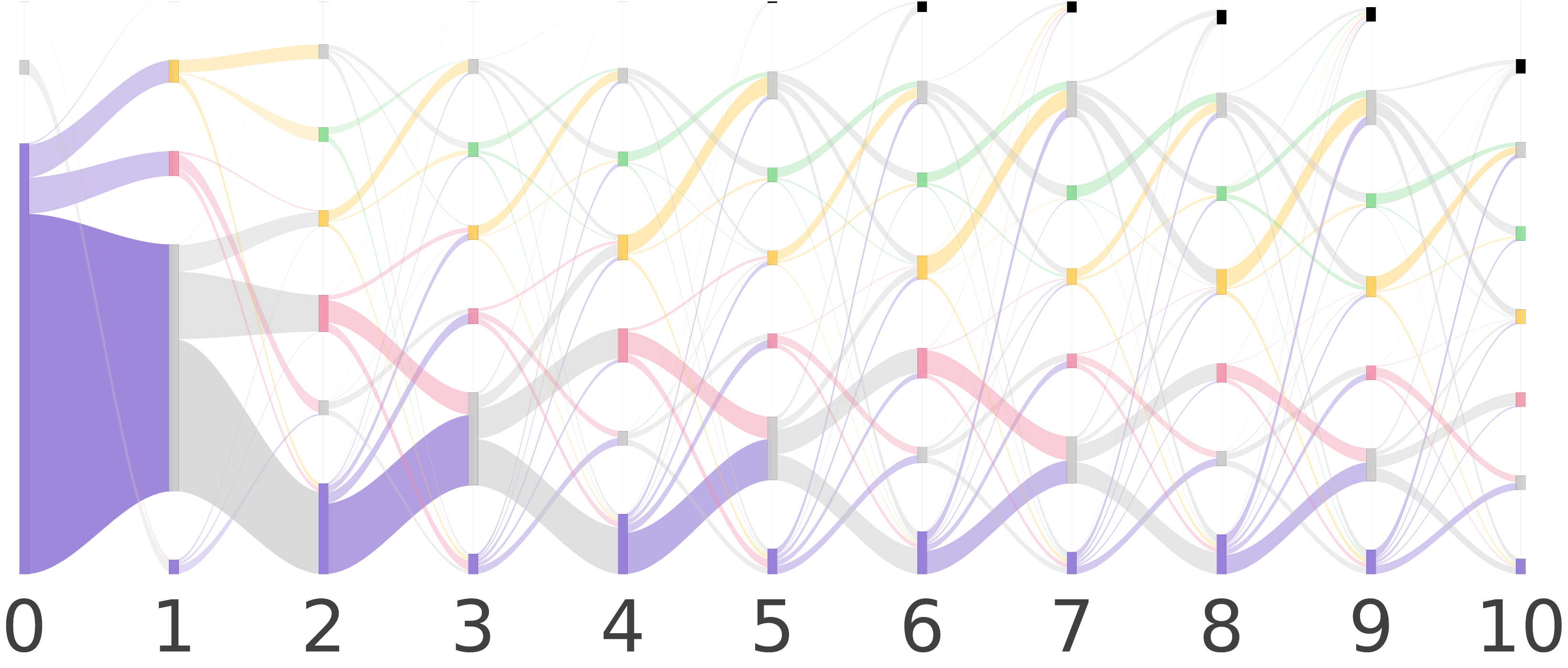} & 
        \includegraphics[width=0.23\textwidth]{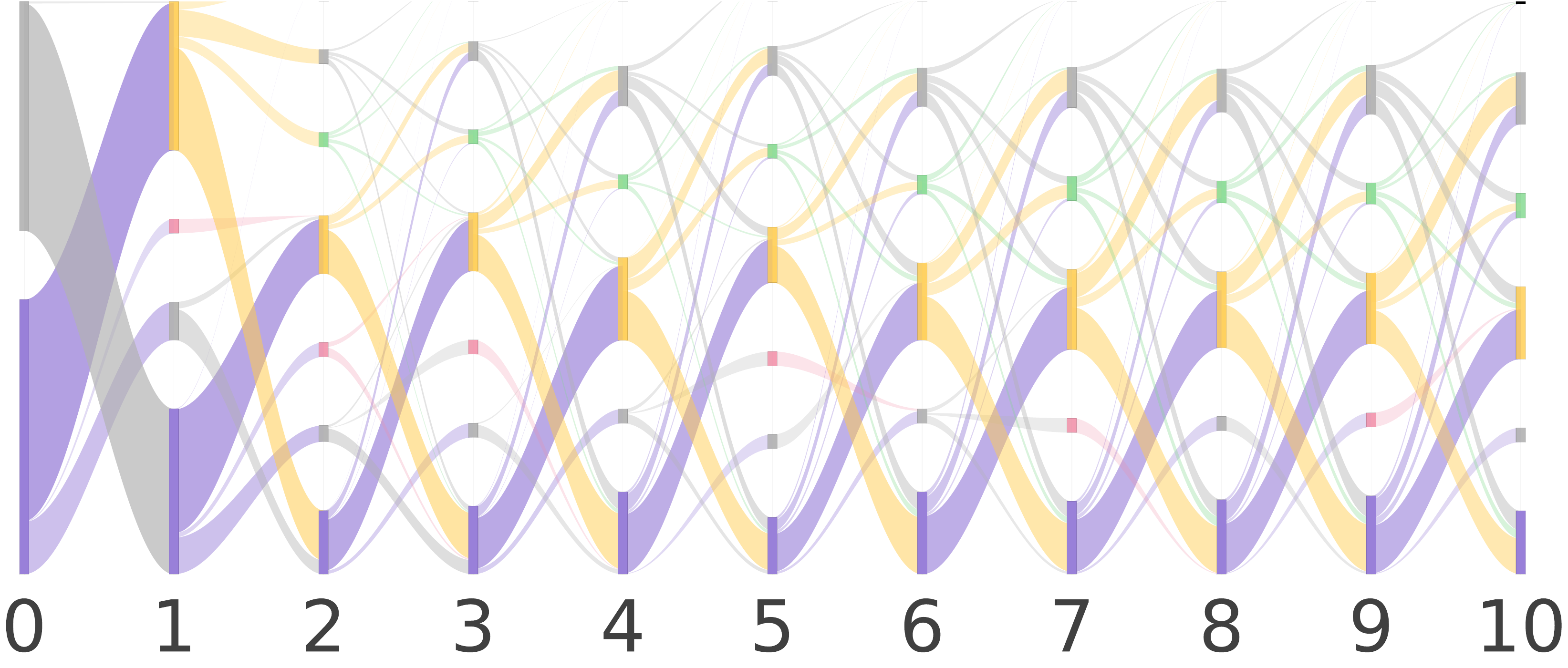} &
        \includegraphics[width=0.23\textwidth]{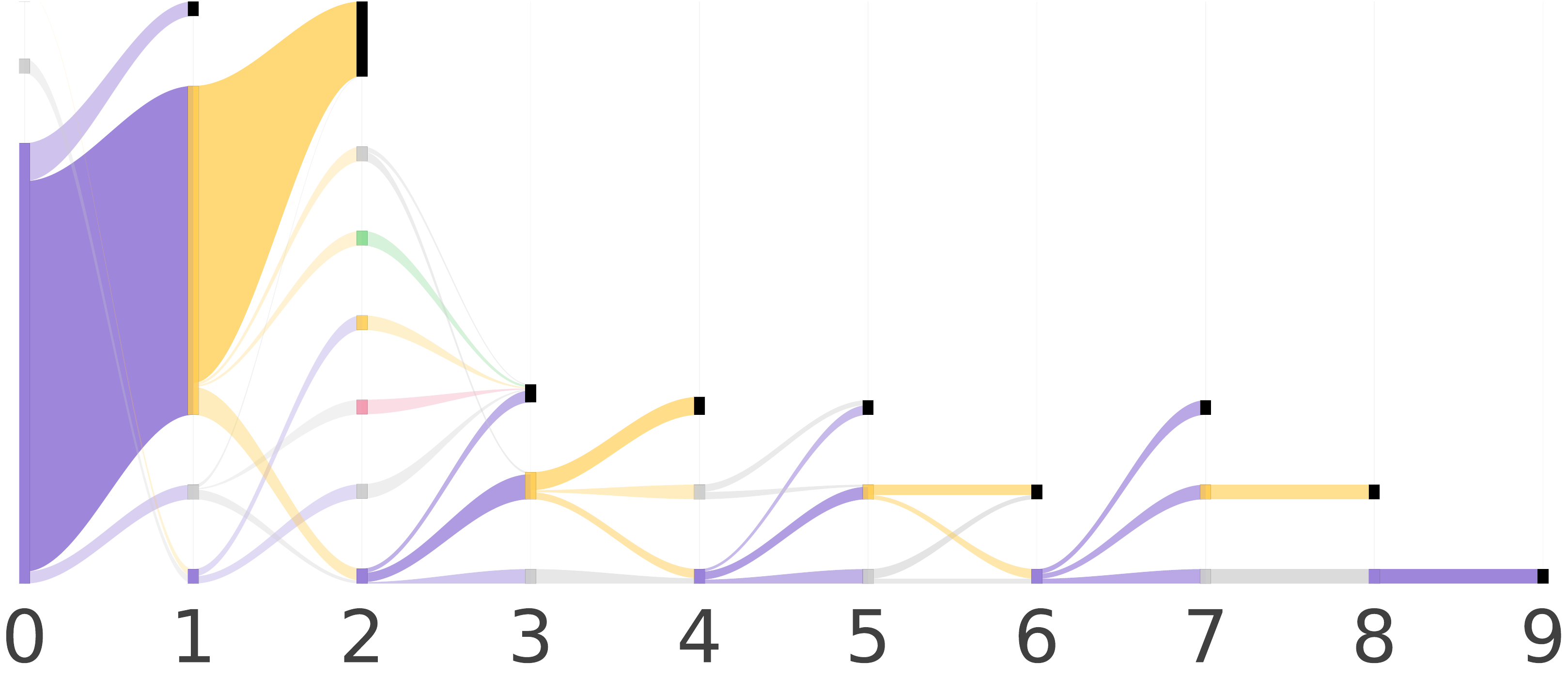} &
        \includegraphics[width=0.23\textwidth]{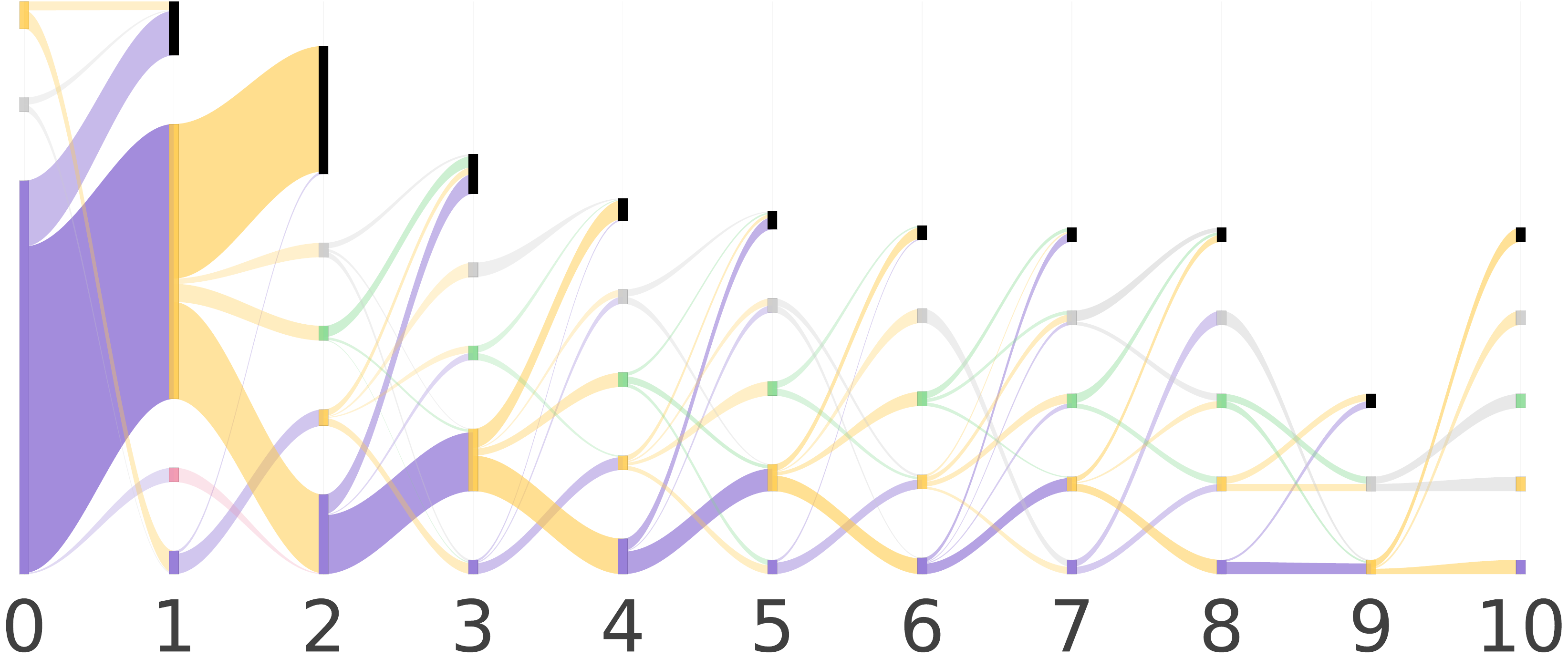} \\[6pt]

        Devstral-small & GPT5-mini & DeepSeek-V3 & DeepSeek-R1 \\
    \end{tabular}
    \vspace{-10pt}
    \caption{Phase flow analysis under No Plan setting. Agents still exhibit traces of Standard plan phases ($\Phi=\{\colorbox[HTML]{CCC7E6}{N},\colorbox[HTML]{EEC7D4}{R},\colorbox[HTML]{FFED99}{P},\colorbox[HTML]{D9EDCC}{V}\}$).}
    \vspace{-10pt}
    \label{fig:sankey-no-plan}
\end{figure*}

\begin{figure}[t]
    \centering
    \includegraphics[width=\linewidth]{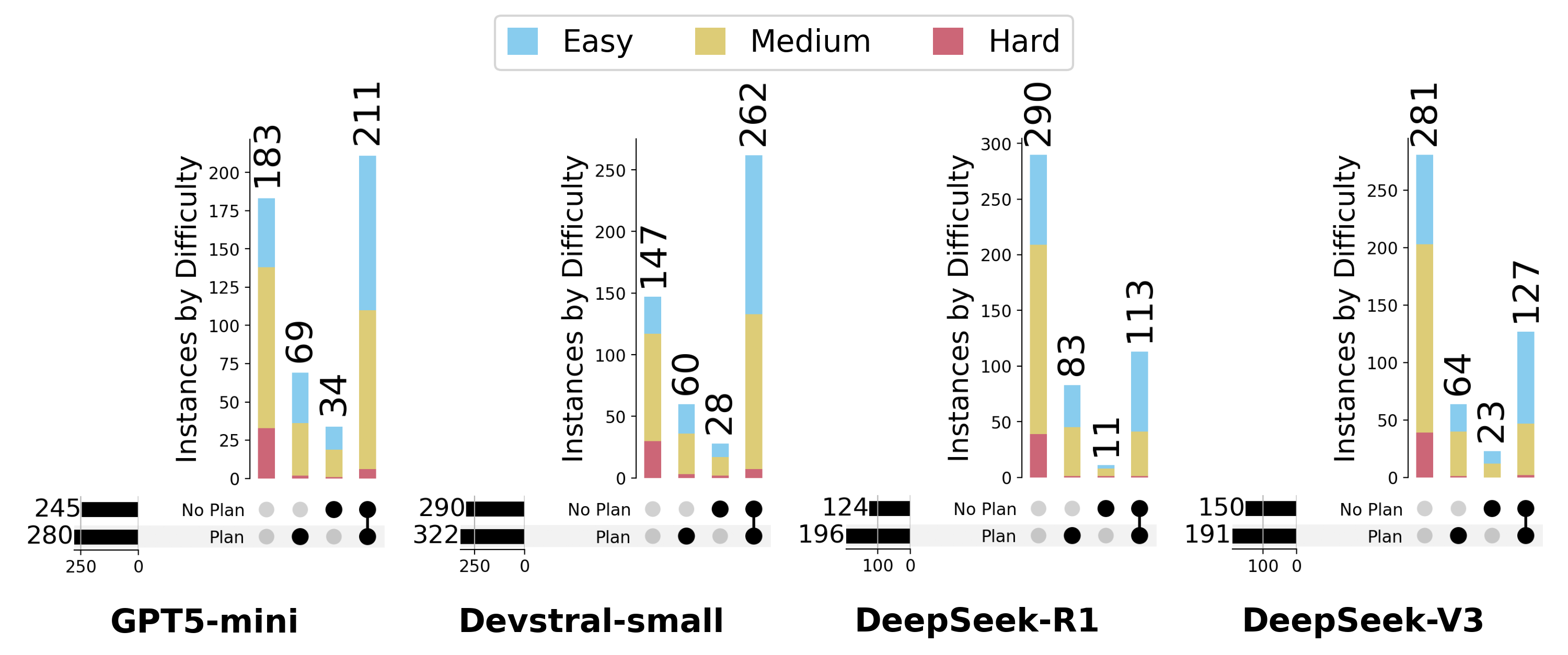}
    \vspace{-20pt}
    \caption{Impact of No Plan setting on the success rate.}
    \vspace{-8pt}
    \label{fig:upset-no-plan}
\end{figure}

The previous research question shows a notable variance in compliance with the Standard plan across agents. Given that the scaffold is identical in all agents, two important factors influencing the observations are (1) the ability of specific LLMs to follow the instructed plan, or (2) a conflict between following plan-prescribed phases and training-prescribed workflows. We investigate the magnitude of the former in \S\ref{sec:plan-variations}. For the latter, we repeated the experiments under a \emph{No Plan} setting to observe how agents perform without any specific plan. We completely remove the default plan from the system prompt of \SA. Thereby, the agent only receives a high-level guideline to fix the issue: given the issue description, make changes to satisfy the issue description requirements\footnote{The system prompt for this experiment is available on the artifact website under artifacts/plan-settings/no\_plan/default.yaml}.  

Figure~\ref{fig:upset-no-plan} compares issue resolution on \swebv with and without the Standard Plan. The left bars show the total number of instances resolved under each setting, while the top bars show the number of instances in each intersection. The dot matrix indicates the corresponding resolution set: black dots mark settings under which the instances are resolved, and gray dots mark settings under which they are unresolved. Thus, from left to right, the top bars represent instances unresolved under both settings, resolved only under the Standard Plan, resolved only under No Plan, and resolved under both settings.
Although the agent receives no plan instruction under this setting, we further investigate if it exhibits any trace of \emph{Standard plan} in its trajectory. The rationale here is that the backbone LLM of the agent may already have seen instructions related to this task during training/fine-tuning. Figure~\ref{fig:sankey-no-plan} shows phase flow analysis of trajectories under the No Plan setting.  

\noindent \textbf{Finding 5. Even when not explicitly instructed, agents follow the Standard plan to a notable degree.} The phase flow analysis in Figure~\ref{fig:sankey-no-plan} shows that \devstral starts with \colorbox[HTML]{CCC7E6}{Navigation}, and most trajectories still follow the Standard workflow $\colorbox[HTML]{CCC7E6}{N}\colorbox[HTML]{EEC7D4}{R}\colorbox[HTML]{FFED99}{P}\colorbox[HTML]{D9EDCC}{V}$, with some phases out of the Standard plan in between.
Similarly, \gptf trajectories also follow a subset of Standard plan, often without \colorbox[HTML]{EEC7D4}{Reproduction}. 
In contrast, \Vthree and \Rone largely reduce their trajectories to $\colorbox[HTML]{CCC7E6}{N}\colorbox[HTML]{FFED99}{P}$ patterns, skipping \colorbox[HTML]{EEC7D4}{Reproduction} or \colorbox[HTML]{D9EDCC}{Validation}. This suggests that different models internalize problem-solving processes differently, depending on their training. In the absence of global plans, the encoded strategy takes over the reasoning to solve the problem. 


\noindent \textbf{Finding 6. The success rate drops in the absence of the standard plan, although to different degrees across models and difficulty levels.} 
Figure~\ref{fig:upset-no-plan} shows that removing the plan consistently reduces performance across all models. The majority of the instances that \SA resolved only under the Standard plan setting are of Medium difficulty. \devstral and \gptf, which exhibit problem-solving strategies similar to the Standard plan, 
show only minor drops when the plan is removed. 
In contrast, DeepSeek models, particularly \Rone, experience a substantial performance drop, despite showing lower compliance when the plan is present. This indicates that the plan, even if not properly followed, can positively impact the local reasoning of the agents, guiding them toward their goal. Without it, reasoning becomes less focused, often resulting in premature convergence: these models demonstrate smaller \graph metric values under the No Plan setting, compared to the Standard plan. 

\noindent \textbf{Finding 7. Agents can fix previously unresolved issues under no-plan setting.}
\Vthree, \Rone, \devstral, and \gptf each resolve additional instances that are not solved under the default plan: 23, 11, 28, and 34, respectively. 
As we will show later~(\S\ref{subsec-rq7-nondeterminism}), this is largely affected by the inherent nondeterminism of LLM-based agents, with 4, 7, 16, and 4 instances deterministically only resolved under the no-plan setting. Manual inspection of these instances reveals a consistent trend across all models: The Standard plan instructs the model to reproduce the bug before the patch. However, reproduction test generation is non-trivial~\cite{ahmed_et_al_2024,ahmed_et_al_2026}. In the instances studied, the models generated incorrect reproduction tests, leading to repeated patch-test failure cycles without success. Under the No Plan setting, the same model skipped the reproduction phase and generated the correct patch. This is alarming but interesting: the solution under No Plan can be due to data contamination~\cite{article2}.
\vspace{-5pt}
\section{Plan Variations}
\label{sec:plan-variations}

\begin{figure*}[t]
    \centering
    \footnotesize
    \setlength{\tabcolsep}{5pt} 
    \renewcommand{\arraystretch}{1.35}
    
    \begin{tabular}{@{}l@{\quad}cccc@{}}
         \footnotesize
                 \vspace{-0.2cm}
        \rotatebox{90}{
  \begin{minipage}{1.5cm} 
    \centering No Reproduce
  \end{minipage}
} & 
        \includegraphics[width=0.228\textwidth]{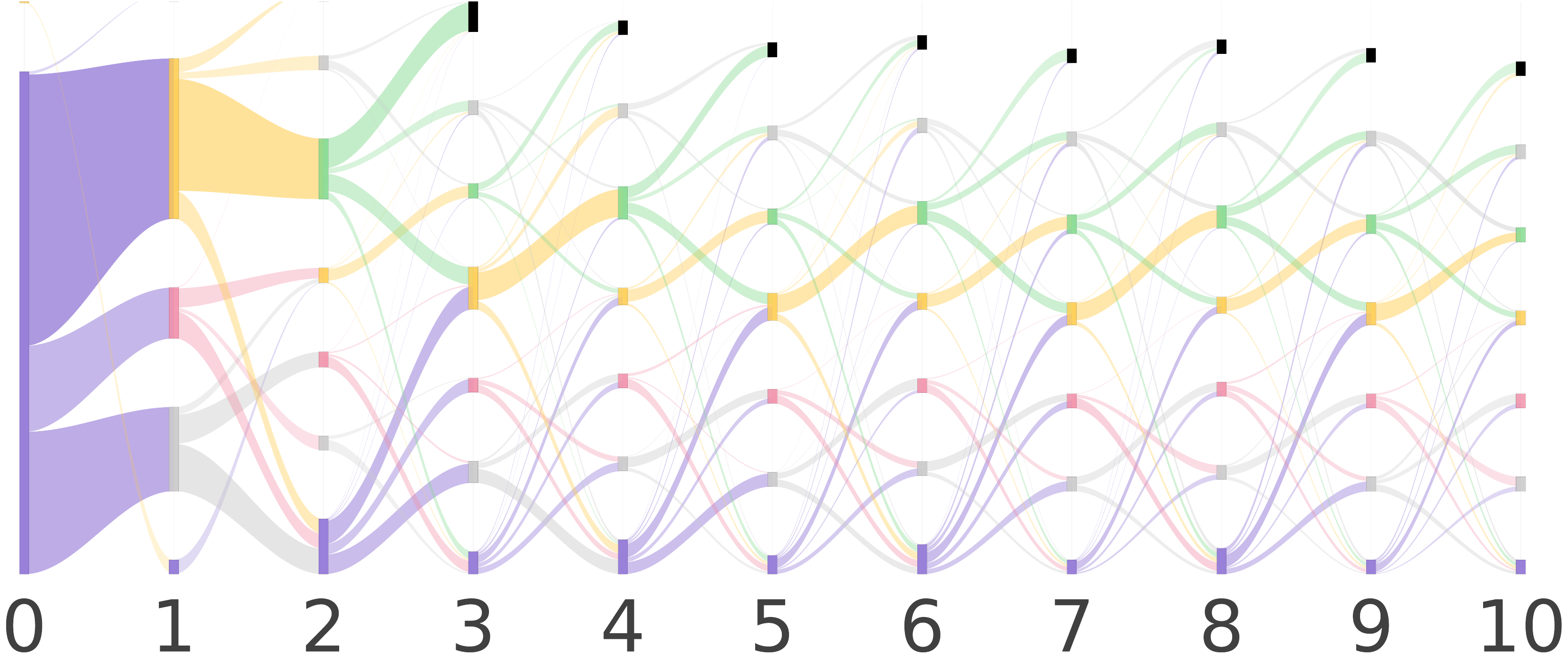} & 
        \includegraphics[width=0.228\textwidth]{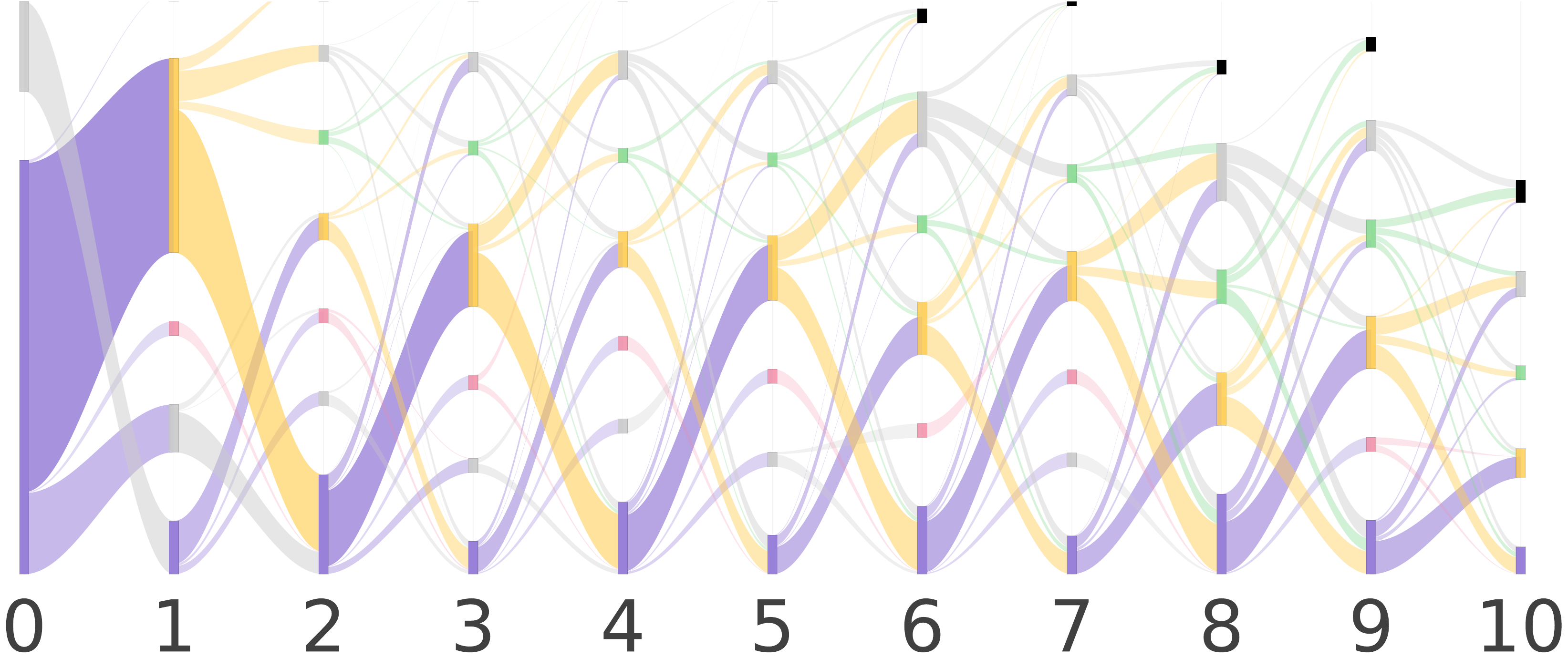} &
        \includegraphics[width=0.228\textwidth]{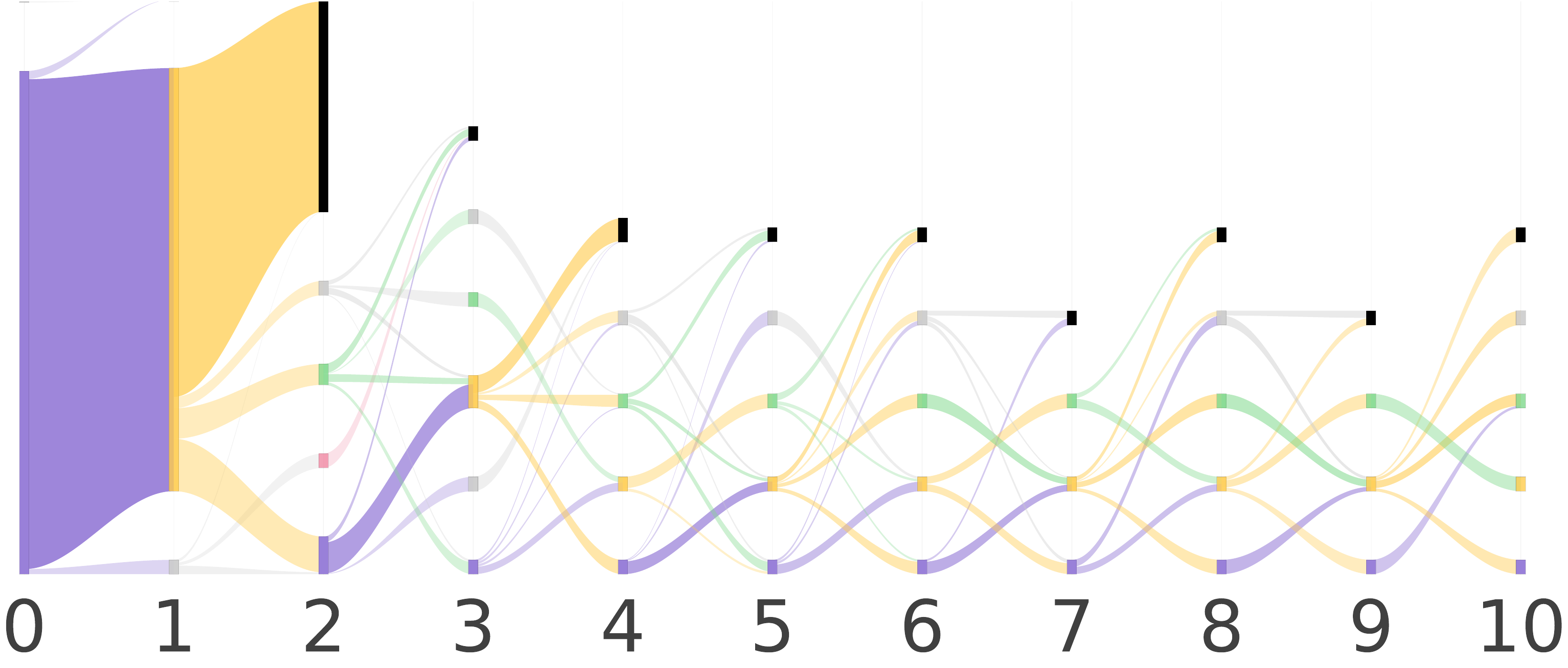} &
        \includegraphics[width=0.228\textwidth]{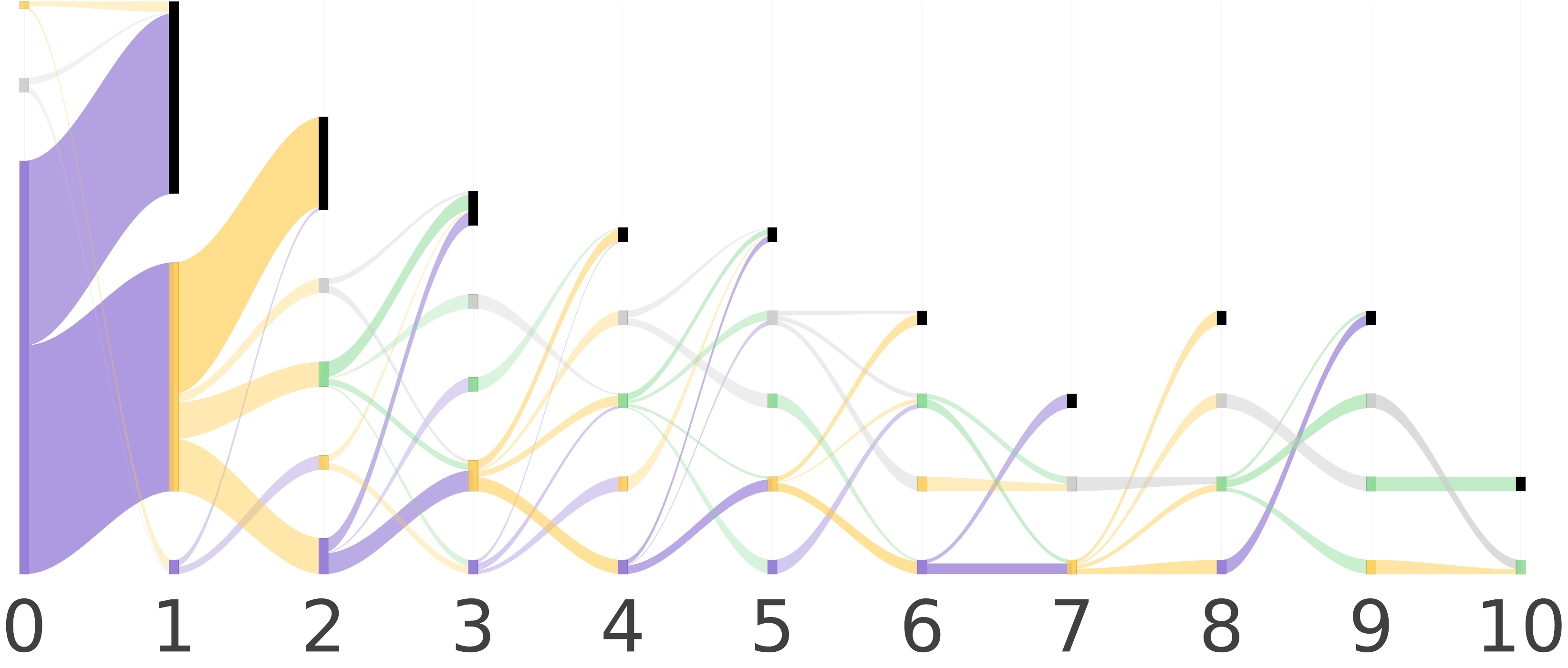} \\[6pt]
        \vspace{-0.2cm}
        \rotatebox{90}{
  \begin{minipage}{1.5cm} 
    \centering No Validation
  \end{minipage}
} & 
        \includegraphics[width=0.228\textwidth]{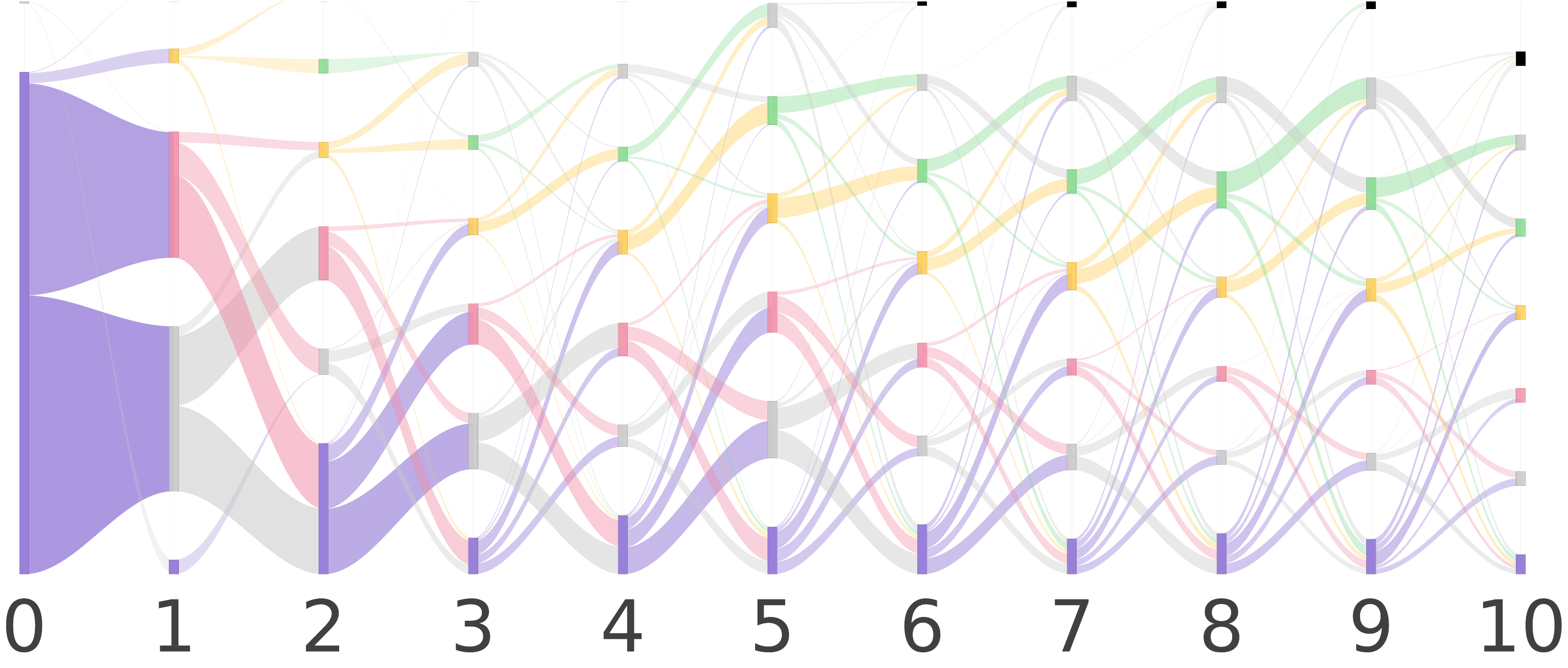} & 
        \includegraphics[width=0.228\textwidth]{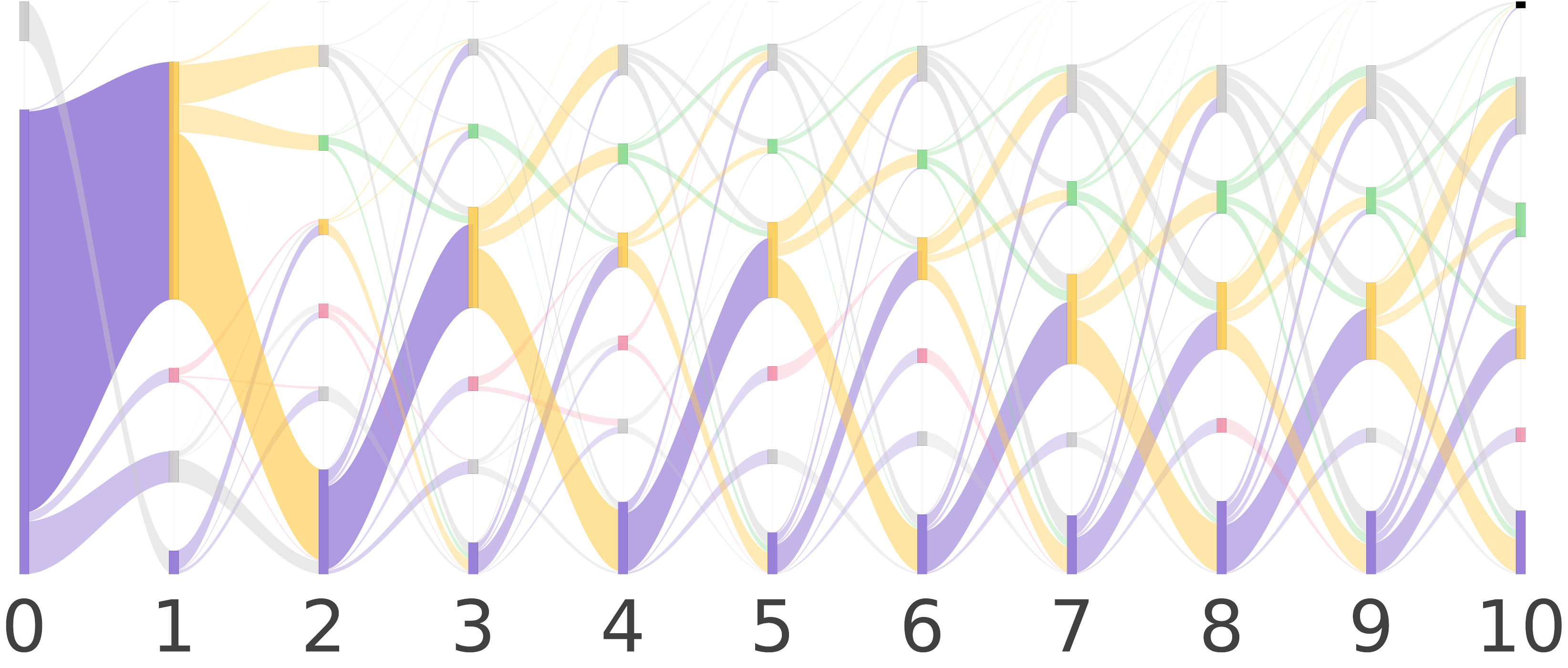} &
        \includegraphics[width=0.228\textwidth]{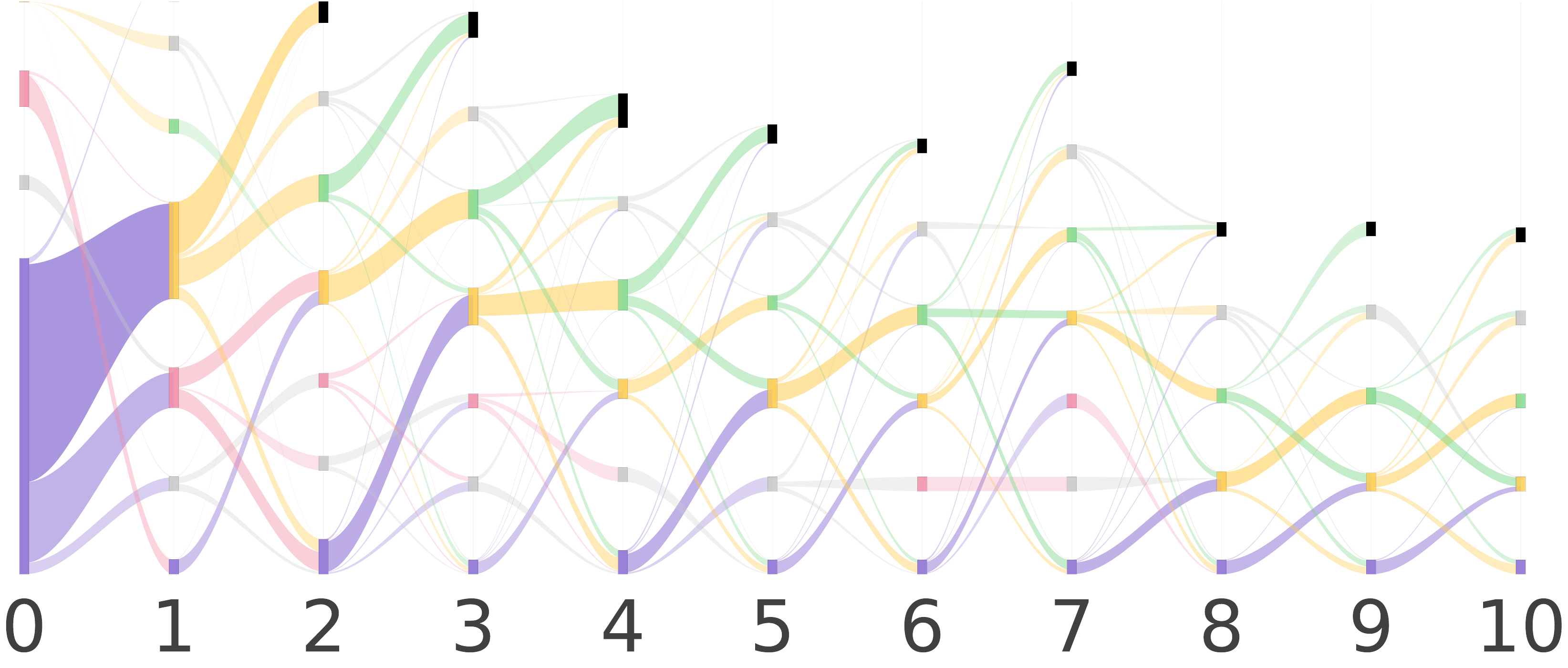} &
        \includegraphics[width=0.228\textwidth]{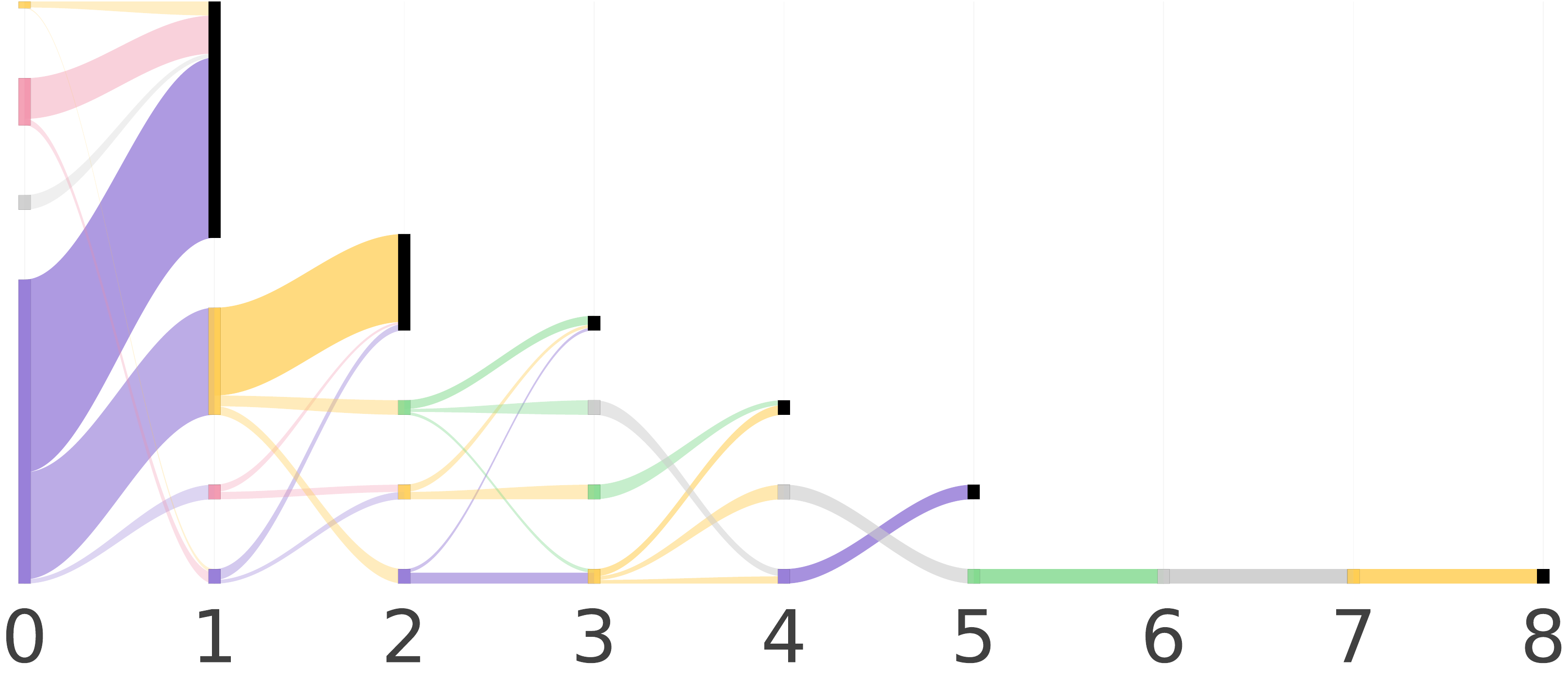} \\[6pt]

        & Devstral-small & GPT5-mini & DeepSeek-V3 & DeepSeek-R1 \\
        
    \end{tabular}

    \vspace{-10pt}
    \caption{Phase flow analysis under No Reproduction ($\mathcal{L}^\star(\Phi)=\colorbox[HTML]{CCC7E6}{N}\colorbox[HTML]{FFED99}{P}\colorbox[HTML]{D9EDCC}{V}$) and No Validation plan setting ($\mathcal{L}^\star(\Phi)=\colorbox[HTML]{CCC7E6}{N}\colorbox[HTML]{EEC7D4}{R}\colorbox[HTML]{FFED99}{P}$). Trajectories still show traces of removed phases from the Standard plan.}
    \vspace{-10pt}
    \label{fig:step-removed-compliance-sankey}
\end{figure*}

\begin{figure*}[t]
    \centering
    \includegraphics[width=\linewidth]{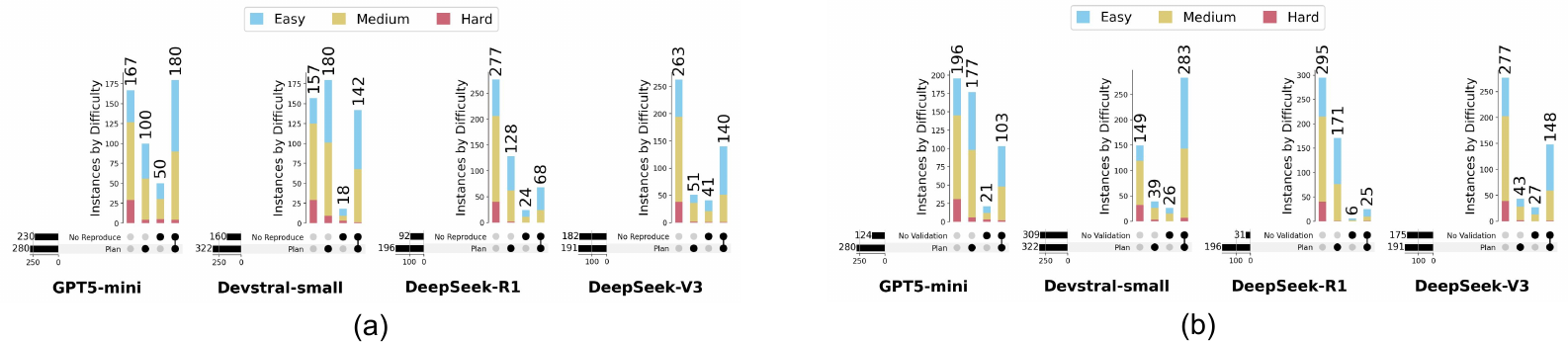}
    \vspace{-20pt}
    \caption{Impact of No Reproduction plan (a) and No Validation plan (b) on the success rate.}
    \vspace{-8pt}
    \label{fig:upset-no-reproduction-no-validation}
\end{figure*}

\begin{figure}[t]
    \centering
    \vspace{-5pt}
    \includegraphics[width=0.7\linewidth]{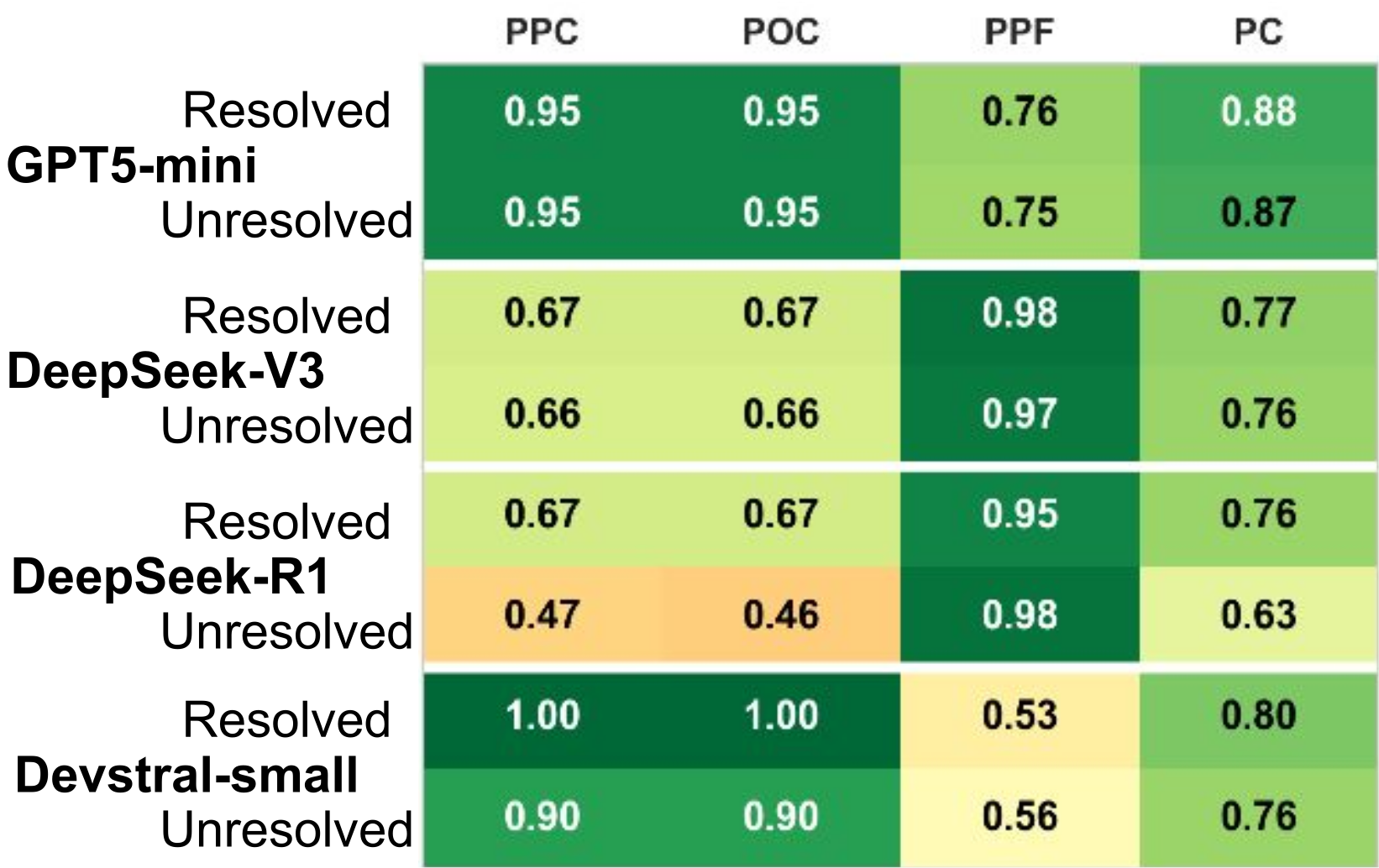}
    \vspace{-8pt}
    \caption{No Reproduction plan compliance metrics.}
    \label{fig:heatmap-no-reproduction}
\end{figure}

RQ2 shows that agents, even if not explicitly instructed to follow the standard plan, still incorporate it in their problem-solving strategy. As discussed, this is likely due to the training objectives of backbone LLMs. For a more controlled plan compliance analysis, we create variations of the Standard plan with small changes (removing one plan phase in \S \ref{subsec-rq3-remove-plan} or adding one phase outside of the Standard plan in \S \ref{subsec-rq4-addplan}). We then investigate compliance with the mutated plan and the impact of isolated plan changes on success rate. 


\vspace{-5pt}
\subsection{RQ3. Reduced Plan Settings}
\label{subsec-rq3-remove-plan}

We study two reduced plan variations by removing either the \colorbox[HTML]{EEC7D4}{Reproduction} or the \colorbox[HTML]{D9EDCC}{Validation} phase. Removing Navigation and Patching is unlikely to reveal notable observations, as these two phases are essential and consistently appear in agent trajectories from observations in the No Plan setting.

\vspace{-5pt}
\subsubsection{RQ3.1. No Reproduction Setting}
\label{subsubsec:no-reproduction}

Figures~\ref{fig:step-removed-compliance-sankey}--\ref{fig:heatmap-no-reproduction} illustrate the results under $\mathcal{L}^\star(\Phi)=\colorbox[HTML]{CCC7E6}{N}\colorbox[HTML]{FFED99}{P}\colorbox[HTML]{D9EDCC}{V}$.
\devstral and \gptf achieve lower scores on $PPF$, since \colorbox[HTML]{EEC7D4}{Reproduction} is encoded in their internal problem-solving strategy. They also achieve near-perfect compliance for $PPC$ and $POC$, i.e., follow all other plan phases in proper order. Phase flow analysis (Figure~\ref{fig:step-removed-compliance-sankey}) shows an interesting observation about the impact of the plan on their trajectory: in the absence of plan, these two models include \colorbox[HTML]{EEC7D4}{Reproduction} more consistently in their trajectories compared to when the plan excludes only that phase (compare Figure~\ref{fig:step-removed-compliance-sankey} with Figure~\ref{fig:sankey-no-plan}). This, consequently, impacts the ability of these models to accomplish the task. As shown in Figure ~\ref{fig:upset-no-reproduction-no-validation}a, their success rate decreases notably.

\begin{figure}[t]
    \centering
    \vspace{-16pt}
    \includegraphics[width=0.7\linewidth]{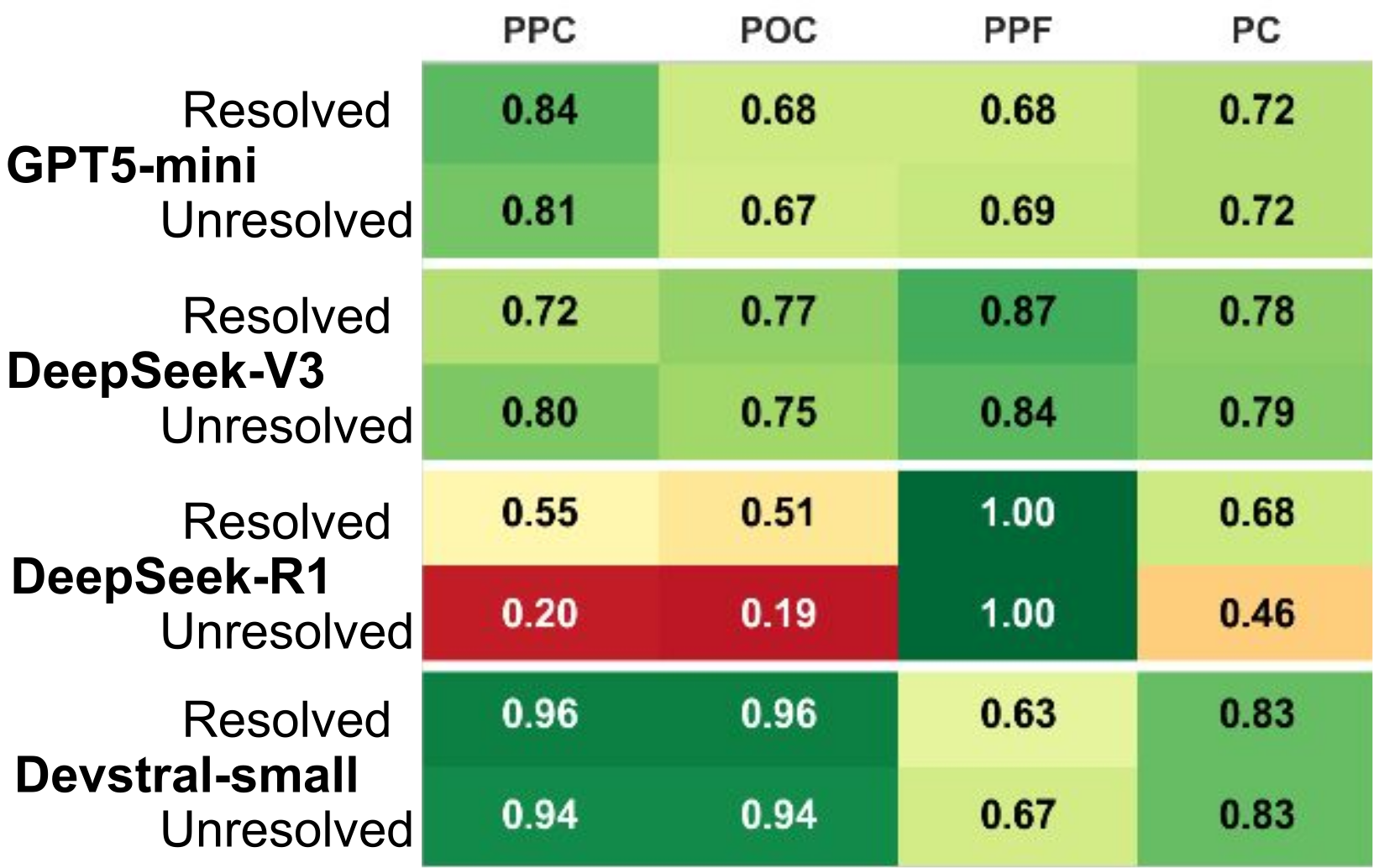}
    \vspace{-8pt}
    \caption{No Validation plan compliance metrics.}
    \label{fig:heatmap-no-validation}
\end{figure}

In contrast, DeepSeek models achieve nearly-perfect $PPF$ values, indicating that their trajectories do not observe \colorbox[HTML]{EEC7D4}{Reproduction}. However, they suffer the most from lower $PPC$ and $POC$, confirmed by phase flow analysis. Similarly, these two models suffer from an incomplete plan, reflected in the drop in success rate. \Rone's performance drop is more substantial. A deeper analysis of its trajectories reveals that, in many cases (349 instances), the model produces malformed tool calls, emitting function calls as plain text rather than in the expected format. This leads to repeated execution errors and eventual termination\footnote{Exit message: ``Exit due to repeated format/blocklist/bash syntax errors''}. The observation suggests that removing \colorbox[HTML]{EEC7D4}{Reproduction} destabilizes \Rone interaction with the scaffold itself, beyond its effect on problem-solving.

\subsubsection{RQ3.2. No Validation Setting}
\label{subsubsec:no-validation}

    
         


\begin{figure*}[t]
    \centering
    \footnotesize
    \setlength{\tabcolsep}{5pt} 
    \renewcommand{\arraystretch}{1.35}
    
    \begin{tabular}{@{}l@{\quad}cccc@{}}
         \footnotesize
                 \vspace{-0.2cm}
        \rotatebox{90}{
  \begin{minipage}{1.5cm} 
    \centering Regression
  \end{minipage}
} & 
        \includegraphics[width=0.228\textwidth]{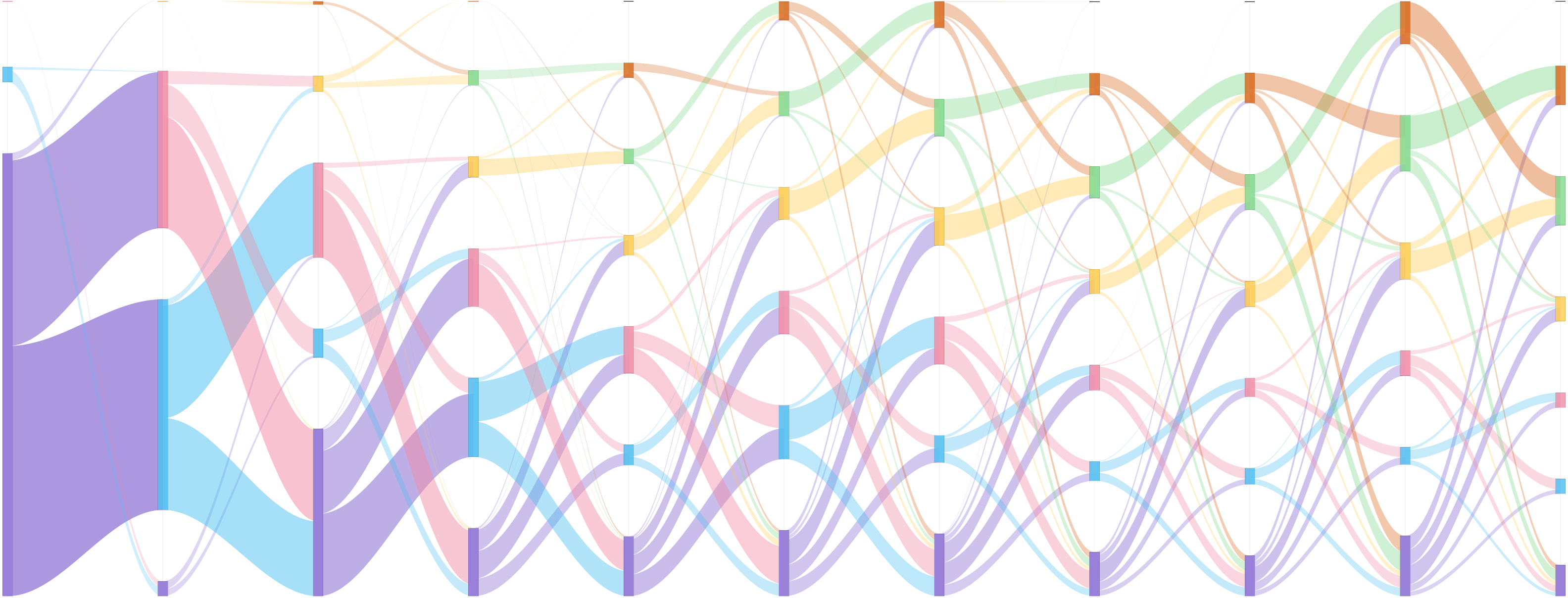} & 
        \includegraphics[width=0.228\textwidth]{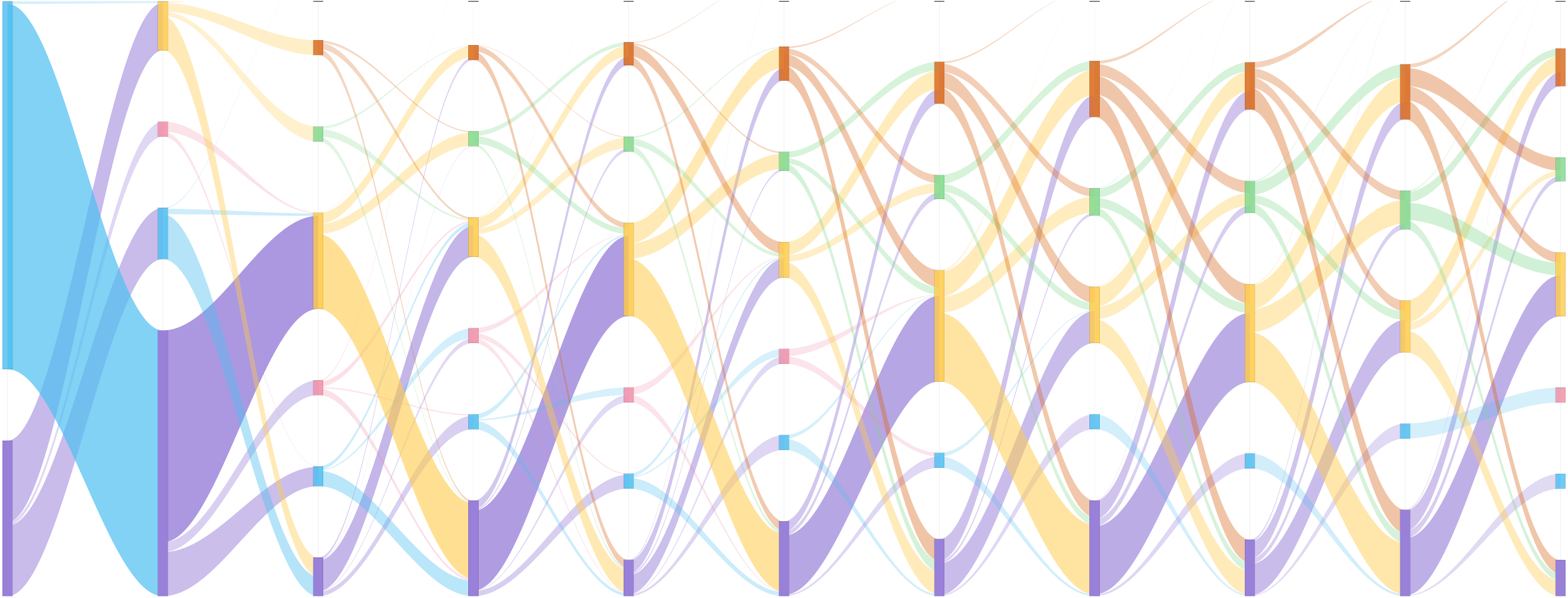} &
        \includegraphics[width=0.228\textwidth]{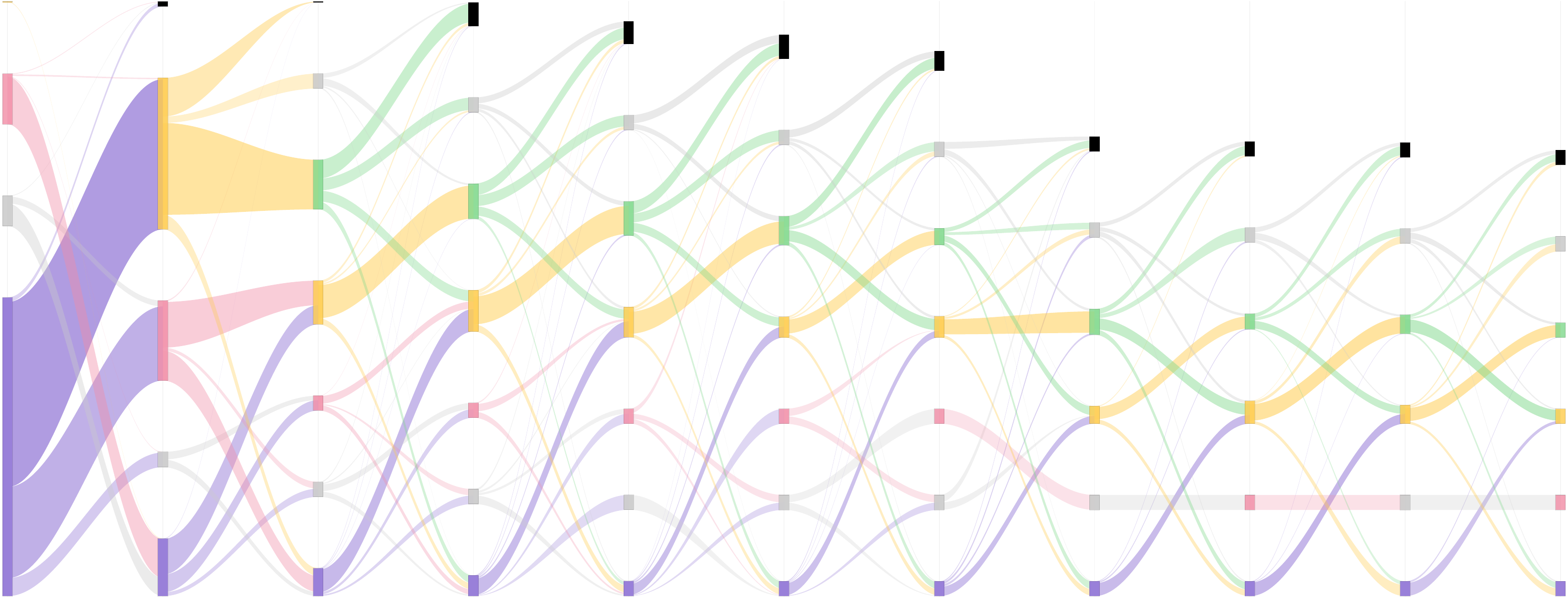} &
        \includegraphics[width=0.228\textwidth]{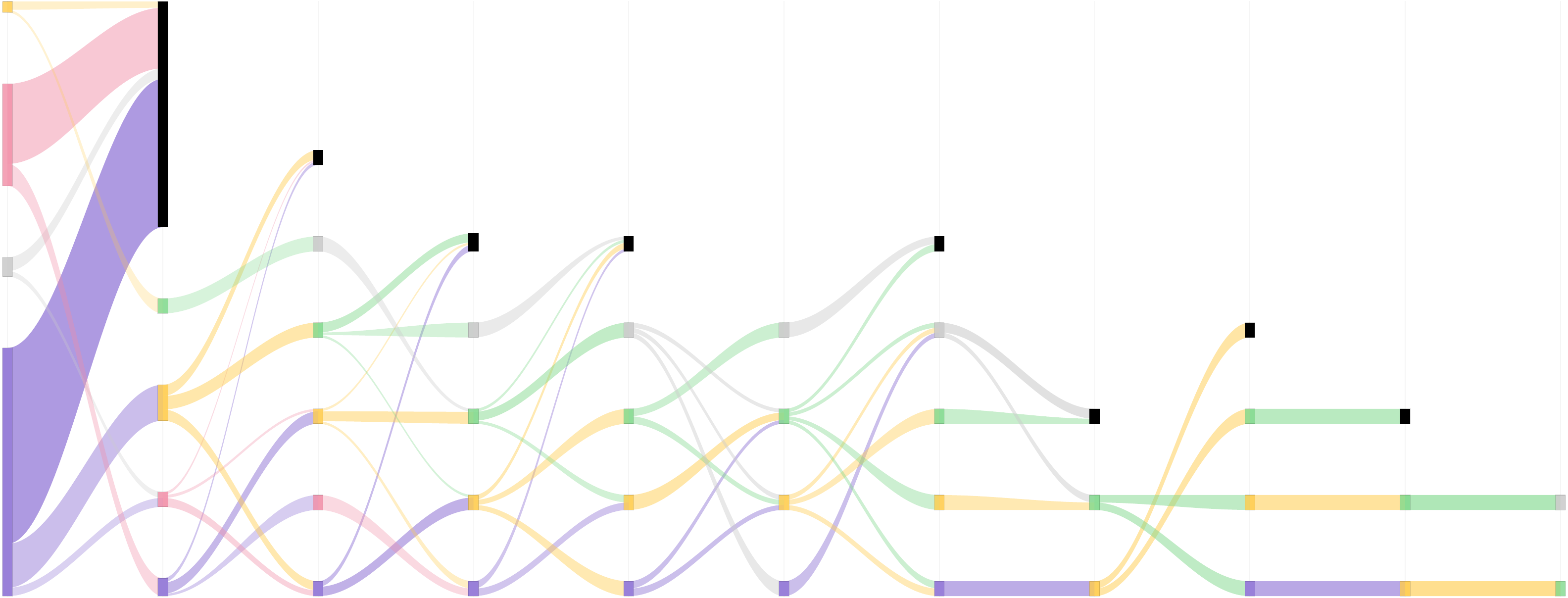} \\[6pt]
        \vspace{-0.2cm}
        \rotatebox{90}{
  \begin{minipage}{1.5cm} 
    \centering Summary
  \end{minipage}
} & 
        \includegraphics[width=0.228\textwidth]{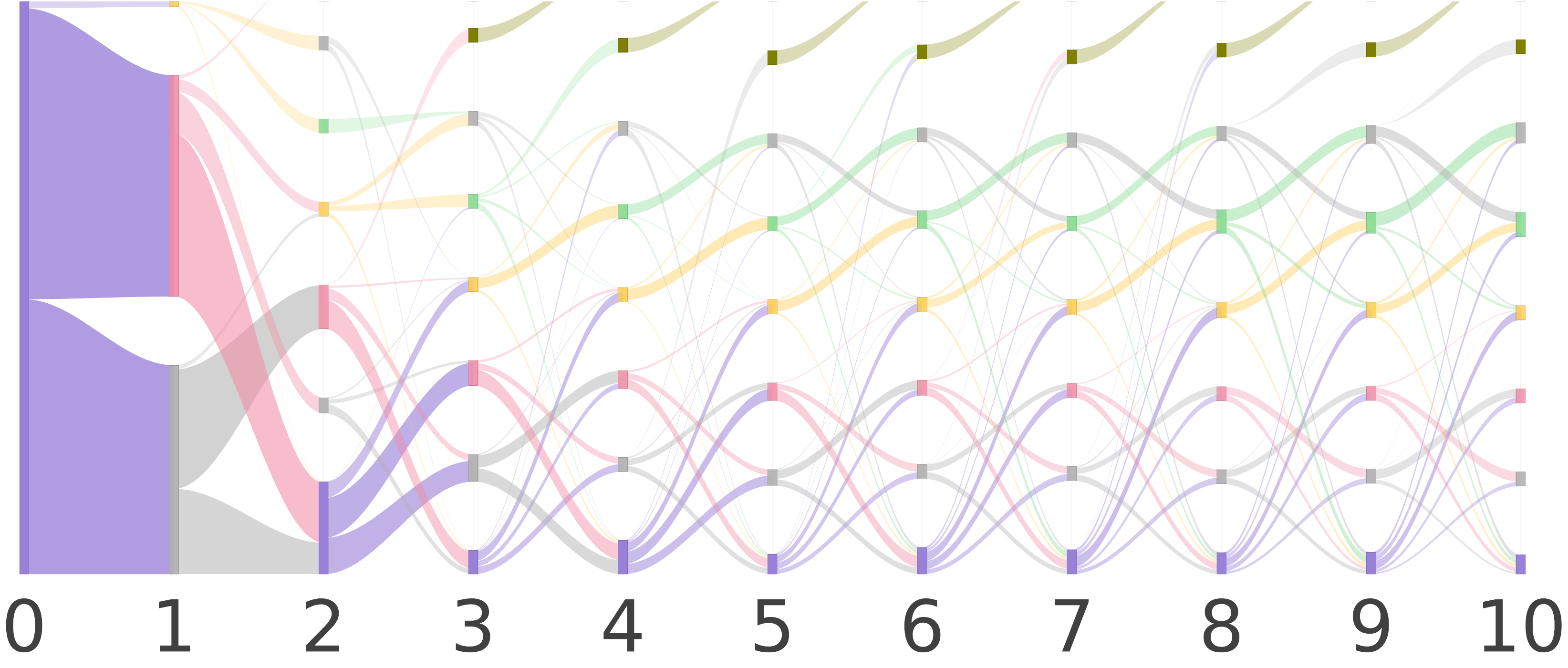} & 
        \includegraphics[width=0.228\textwidth]{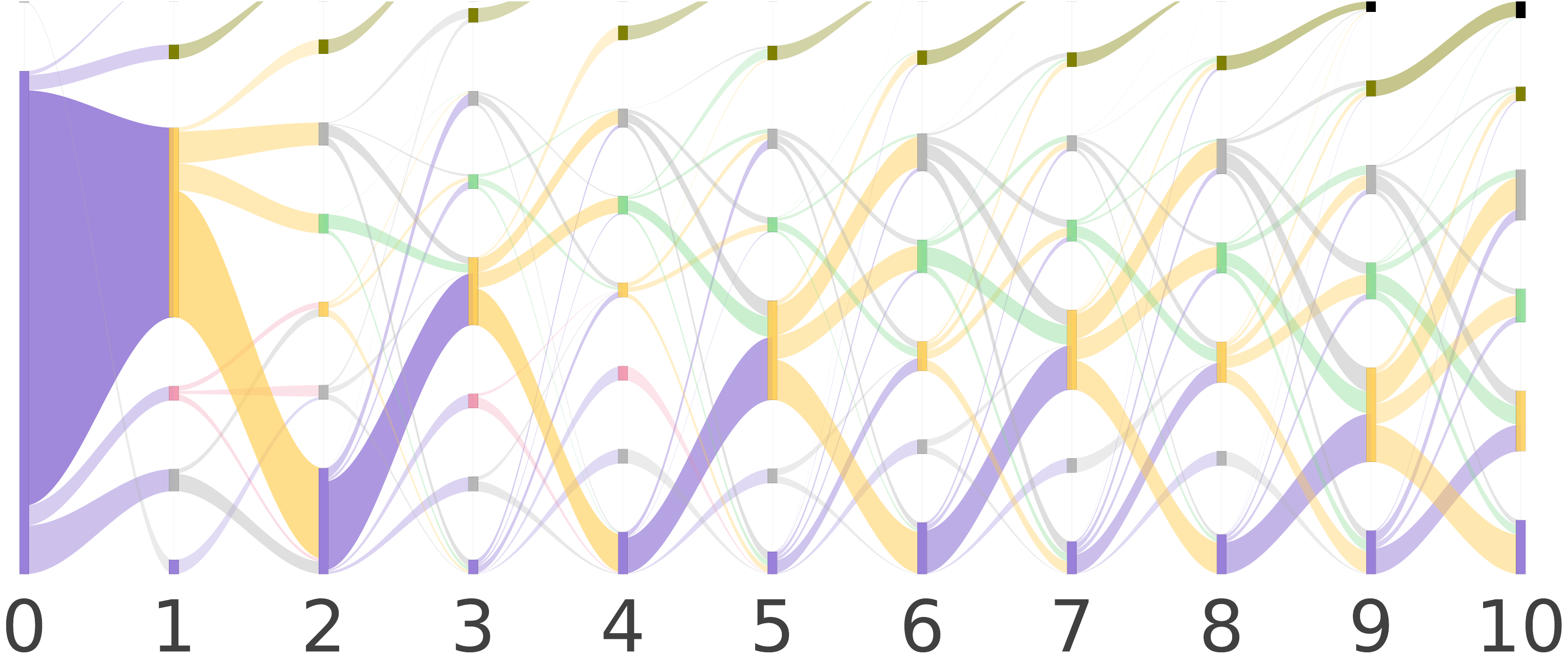} &
        \includegraphics[width=0.228\textwidth]{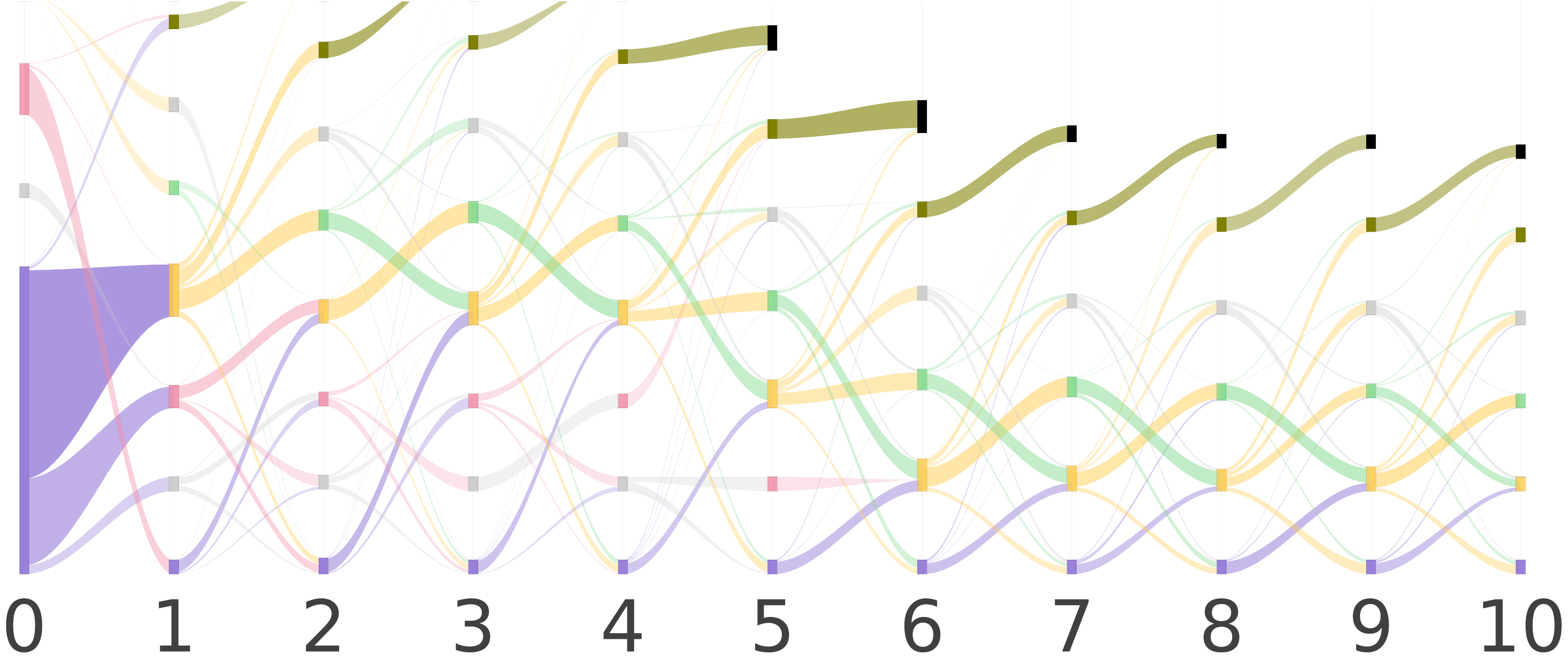} &
        \includegraphics[width=0.228\textwidth]{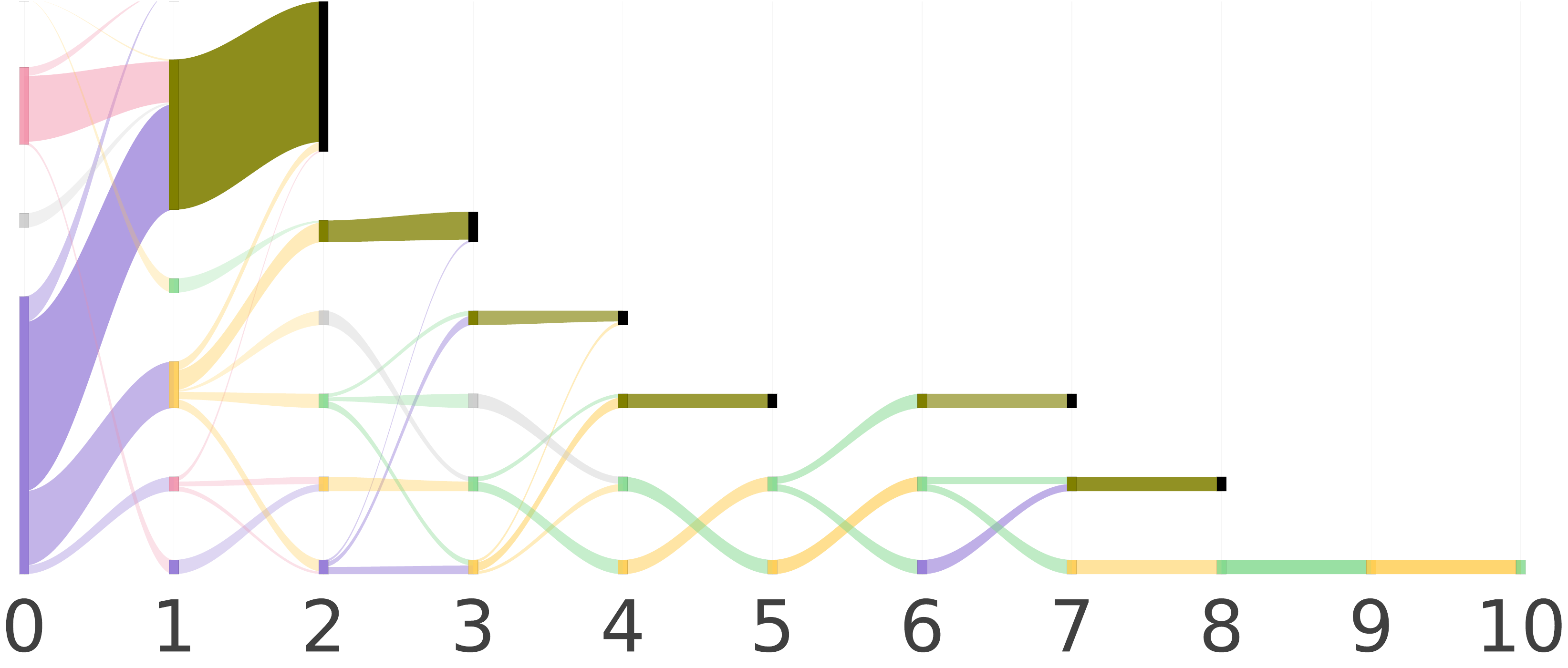} \\[6pt]

        & Devstral-small & GPT5-mini & DeepSeek-V3 & DeepSeek-R1 \\
        
    \end{tabular}

    \vspace{-10pt}
    \caption{Phase flow analysis under Regression Testing ($\mathcal{L}^\star(\Phi)=\colorbox[HTML]{B4DDF7}{\ensuremath{R_G}}\colorbox[HTML]{CCC7E6}{N}\colorbox[HTML]{EEC7D4}{R}\colorbox[HTML]{FFED99}{P}\colorbox[HTML]{D9EDCC}{V}\colorbox[HTML]{E1B28D}{\ensuremath{V_G}}$) and Summary plan setting ($\mathcal{L}^\star(\Phi)=\colorbox[HTML]{CCC7E6}{N}\colorbox[HTML]{EEC7D4}{R}\colorbox[HTML]{FFED99}{P}\colorbox[HTML]{D9EDCC}{V}\colorbox[HTML]{808000}{S}$).}
    \vspace{-10pt}
    \label{fig:step-added-compliance-sankey}
\end{figure*}

\begin{figure*}[t]
    \centering
    \includegraphics[width=\linewidth]{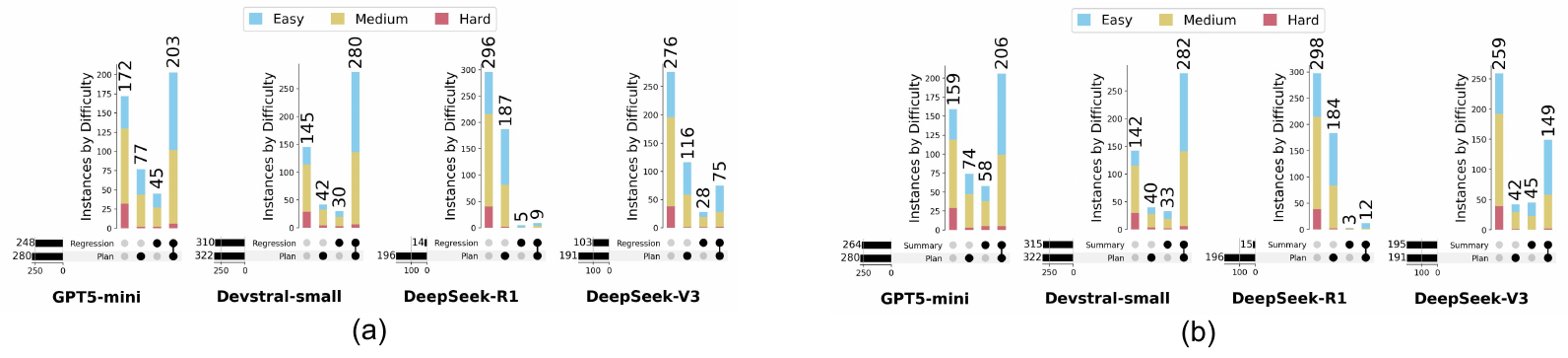}
    \vspace{-20pt}
    \caption{Impact of Regression Testing (a) and Summary plan (b) on the success rate.}
    \vspace{-8pt}
    \label{fig:upset-regression-summary}
\end{figure*}

\begin{figure}[t]
    \centering
    \vspace{-5pt}
    \includegraphics[width=0.7\linewidth]{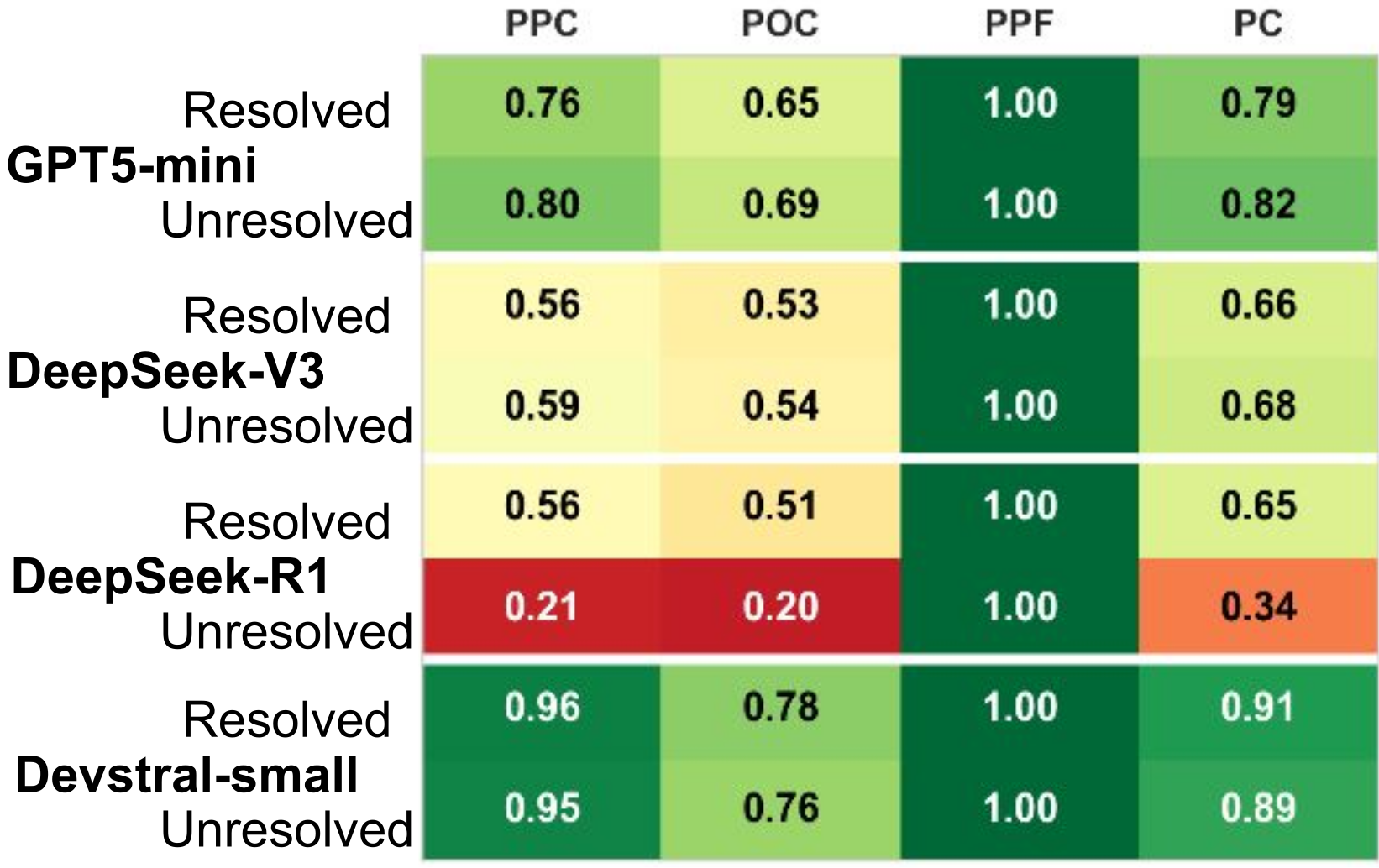}
    \vspace{-8pt}
    \caption{Regression Testing plan compliance metrics.}
    \label{fig:heatmap-regression}
\end{figure}

Figures~\ref{fig:step-removed-compliance-sankey}, \ref{fig:upset-no-reproduction-no-validation}b, and \ref{fig:heatmap-no-validation} illustrate the results under $\mathcal{L}^\star(\Phi)=\colorbox[HTML]{CCC7E6}{N}\colorbox[HTML]{EEC7D4}{R}\colorbox[HTML]{FFED99}{P}$. \gptf suffers greatly without \colorbox[HTML]{D9EDCC}{Validation}, indicating a strong dependence on this phase, as it enables the model to identify incorrect patches and iteratively refine them, rather than prematurely submitting them. 

In contrast, \Vthree shows only slight performance degradation and high $PPF$, consistent with its tendency to skip \colorbox[HTML]{D9EDCC}{Validation} in the No Plan setting. This further reinforces the importance of either alignment of the model’s internalized strategy with the instructed plan or enabling true reasoning and adaptive planning in models. \devstral is less affected than in the No Reproduction setting, suggesting that early-stage grounding through reproduction (e.g., for better bug localization) is more critical to its performance, while validation plays a less central role in shaping its trajectory. We believe this is yet another signal about the data contamination in this model, as it can generate correct patches without validation. \Rone again exhibits the lowest success rate and significantly lower $PPC$ and $POC$, with the majority of trajectories terminating early with repeated tool-calling failures (\S \ref{subsubsec:no-reproduction}).

\noindent \textbf{Finding 8. The negative impact of a bad, incomplete plan is greater on trajectories than the impact of no plan at all.} Overall, removing a specific plan phase can strongly affect models that (1) do not have it in their internal workflow to compensate or (2) are trained to incorporate the plan in their reasoning, as in its absence, their reasoning is incomplete. 

\noindent \textbf{Finding 9. Agents can fix previously unresolved issues under reduced plan settings.} 
Similar to Finding 7, the primary reason for exclusive resolution under the No Reproduction setting is the agents' inability to generate a good reproduction test when instructed by the Standard plan. In contrast, under the No Validation setting, specifically when we eliminate the impact of non-determinism (\S \ref{subsec-rq7-nondeterminism}), agents can rarely resolve previously unsolved issues: \gptf (7), \devstral (19), \Vthree~(5), and \Rone (1).
Analysis of those instances shows that agents still incorporated the validation phase, suggesting remaining impact of nondeterminism. 

\vspace{-4pt}
\subsection{RQ4. Augmented Plan Settings}
\label{subsec-rq4-addplan}

We investigate compliance with plans that include new phases to further demonstrate overfitting to known plans. Arbitrary, task-irrelevant phases may bias the findings; it will be unclear whether an agent struggles with plan compliance or reacts negatively to incoherent instructions. To account for this threat, we only introduce phases relevant to the issue repair task, namely, (1) executing regression tests at the beginning and end (\S \ref{subsubsec:with-regression}) and (2) summarizing changes in a PR-style format before submission (\S \ref{subsubsec:with-summary}). 

\vspace{-4pt}
\subsubsection{RQ4.1. Plan with Regression Test Execution}
\label{subsubsec:with-regression}

\begin{figure}[t]
    \centering
    \vspace{-5pt}
    \includegraphics[width=0.7\linewidth]{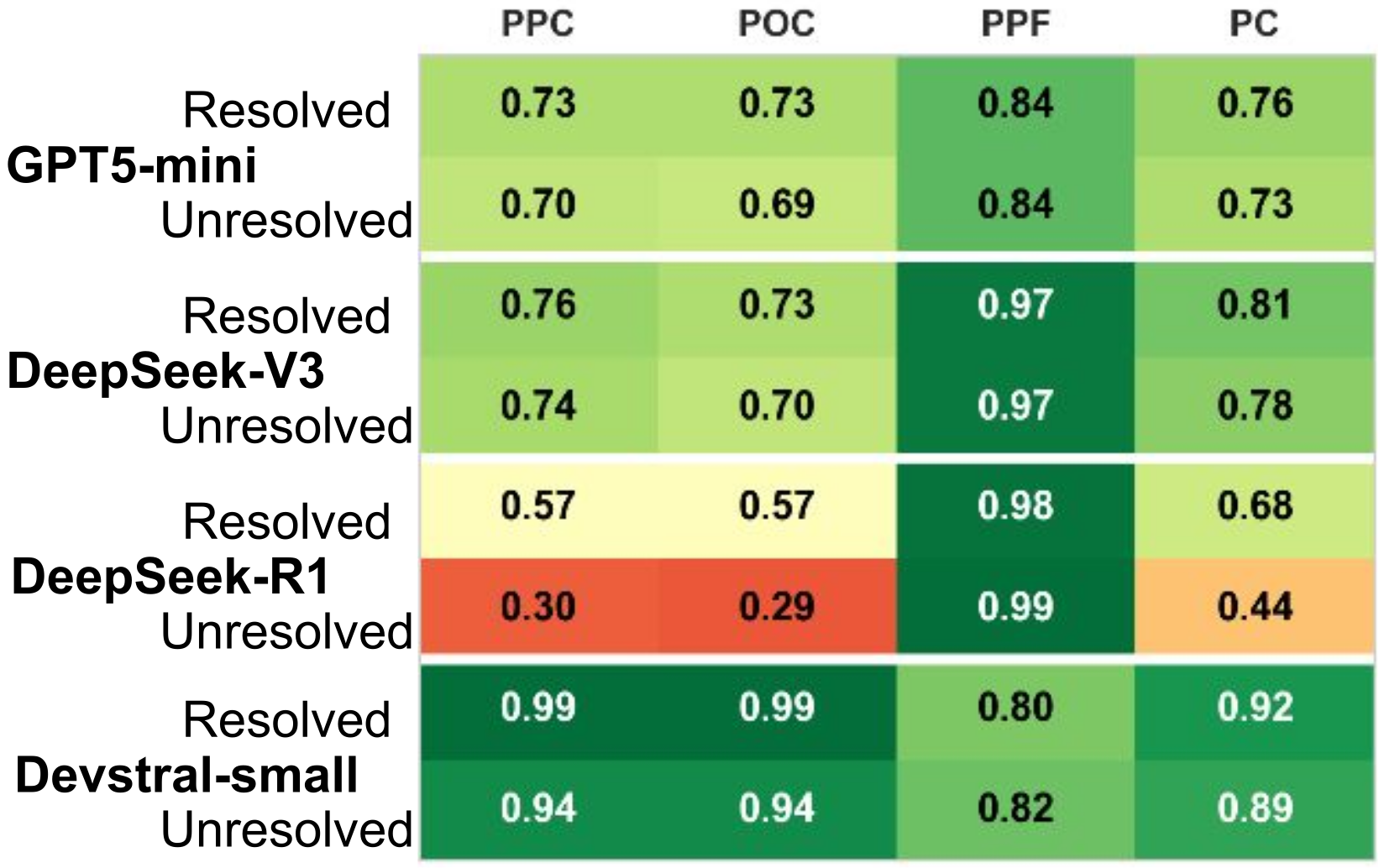}
    \vspace{-8pt}
    \caption{Summary plan compliance metrics.}
    \label{fig:heatmap-summary}
\end{figure}

    
         


\begin{figure*}[t]
    \centering
    \footnotesize
    \setlength{\tabcolsep}{5pt} 
    \renewcommand{\arraystretch}{1.35}
    
    \begin{tabular}{@{}l@{\quad}cccc@{}}
         \footnotesize
                 \vspace{-0.2cm}
        \rotatebox{90}{
  \begin{minipage}{1.5cm} 
    \centering Reordered
  \end{minipage}
} & 
        \includegraphics[width=0.23\textwidth]{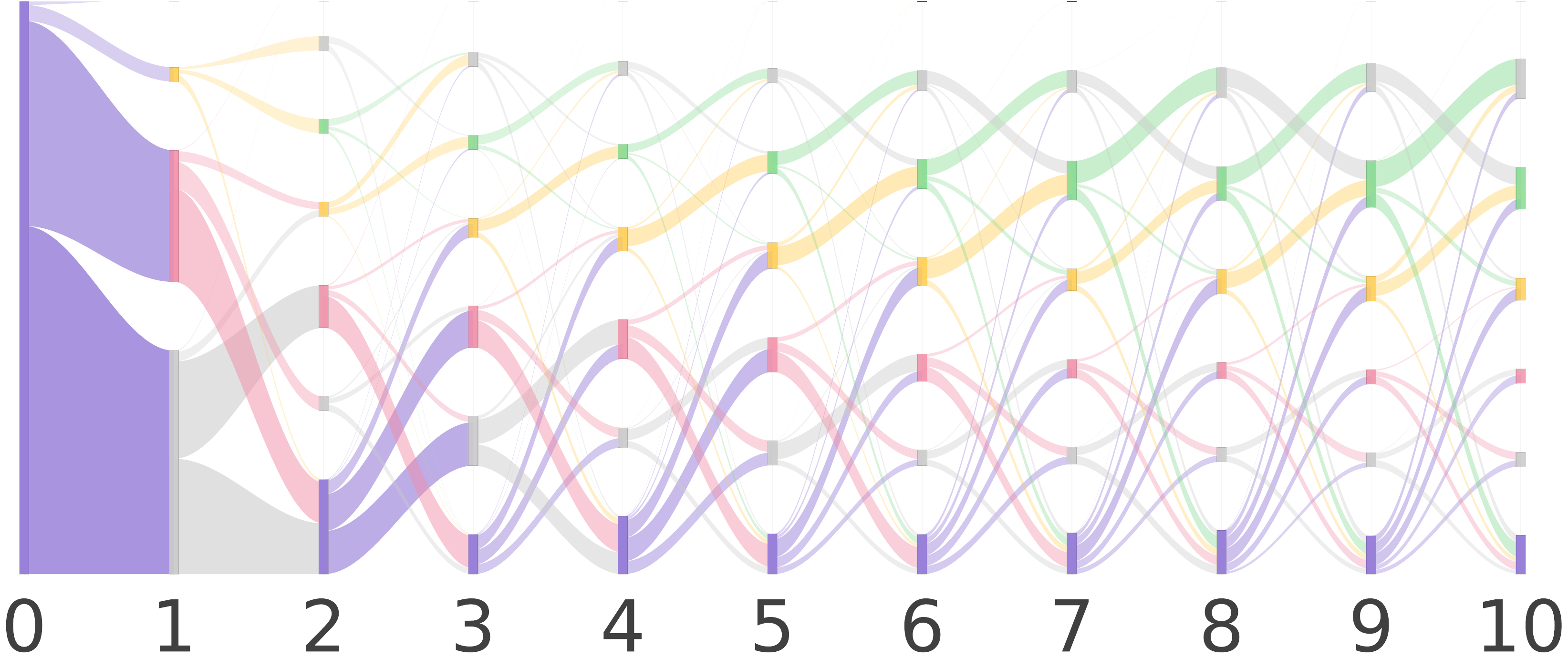} & 
        \includegraphics[width=0.23\textwidth]{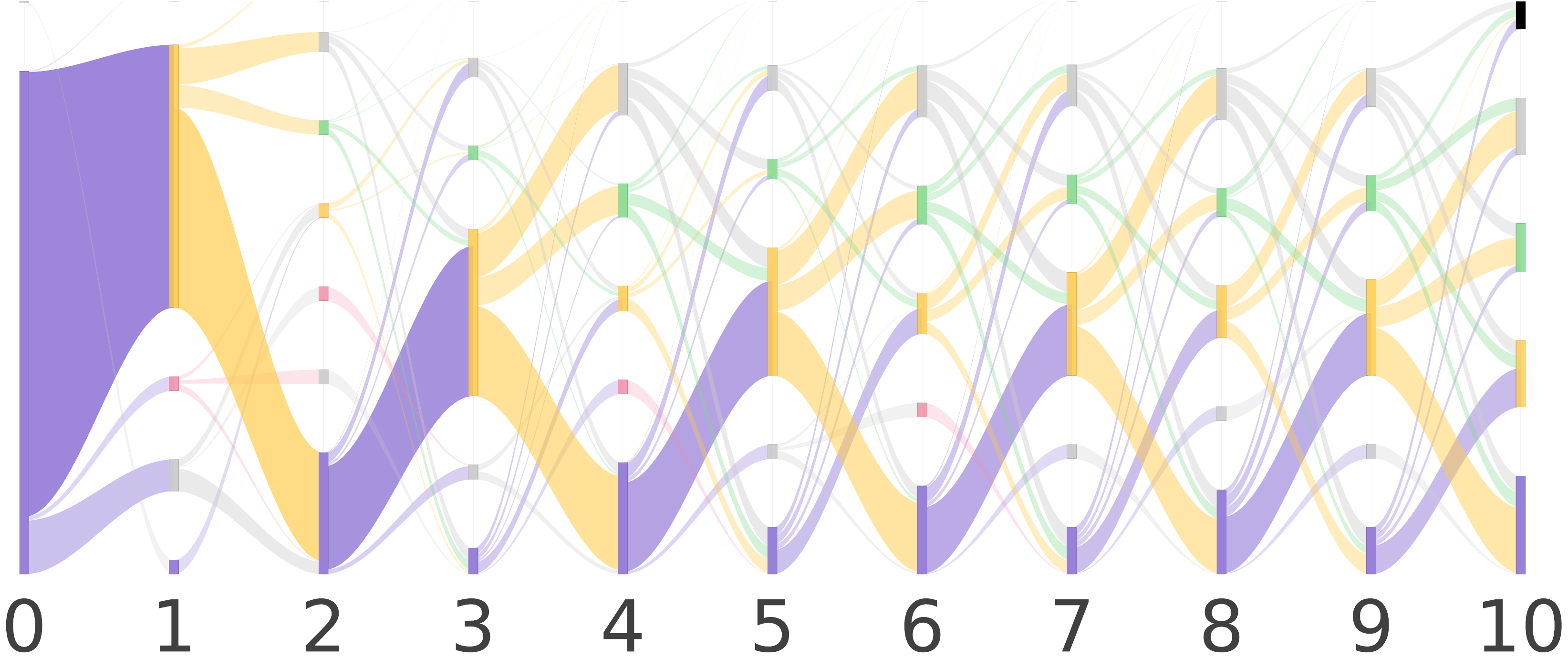} &
        \includegraphics[width=0.23\textwidth]{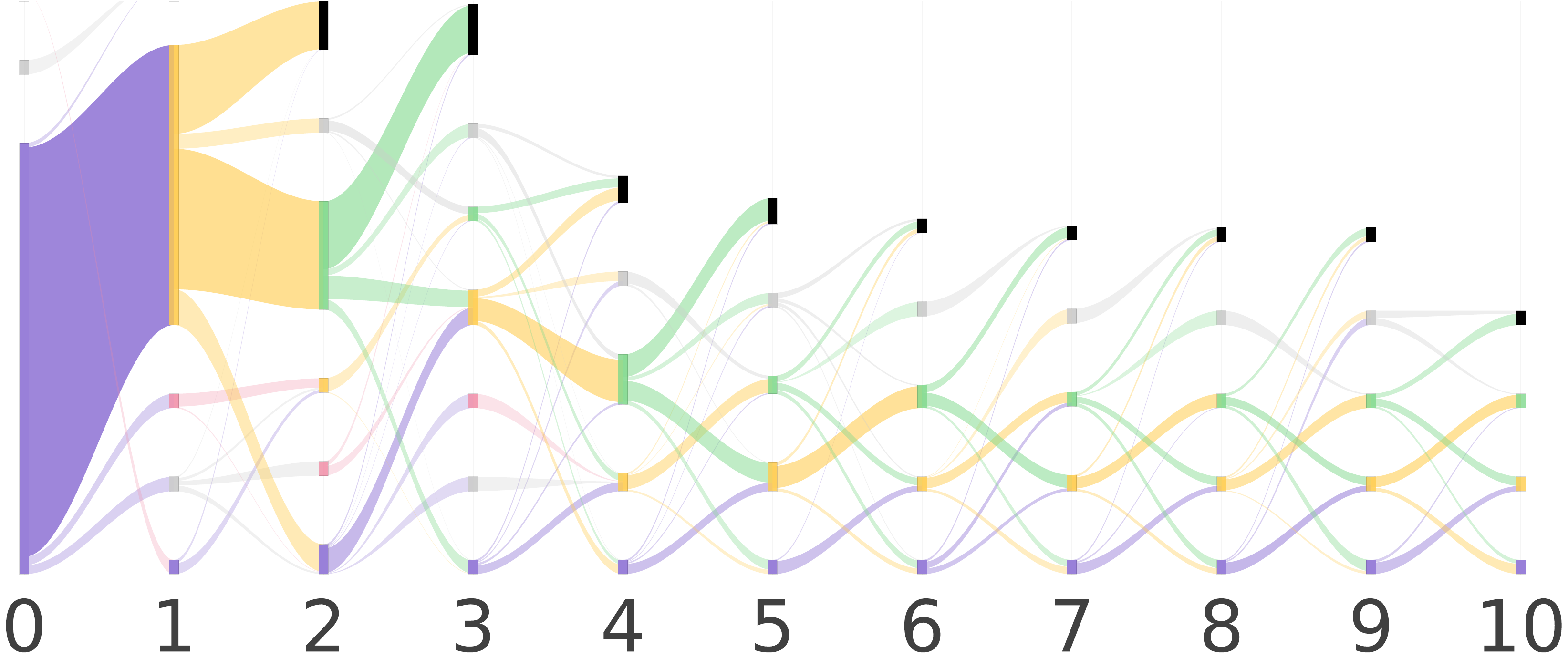} &
        \includegraphics[width=0.23\textwidth]{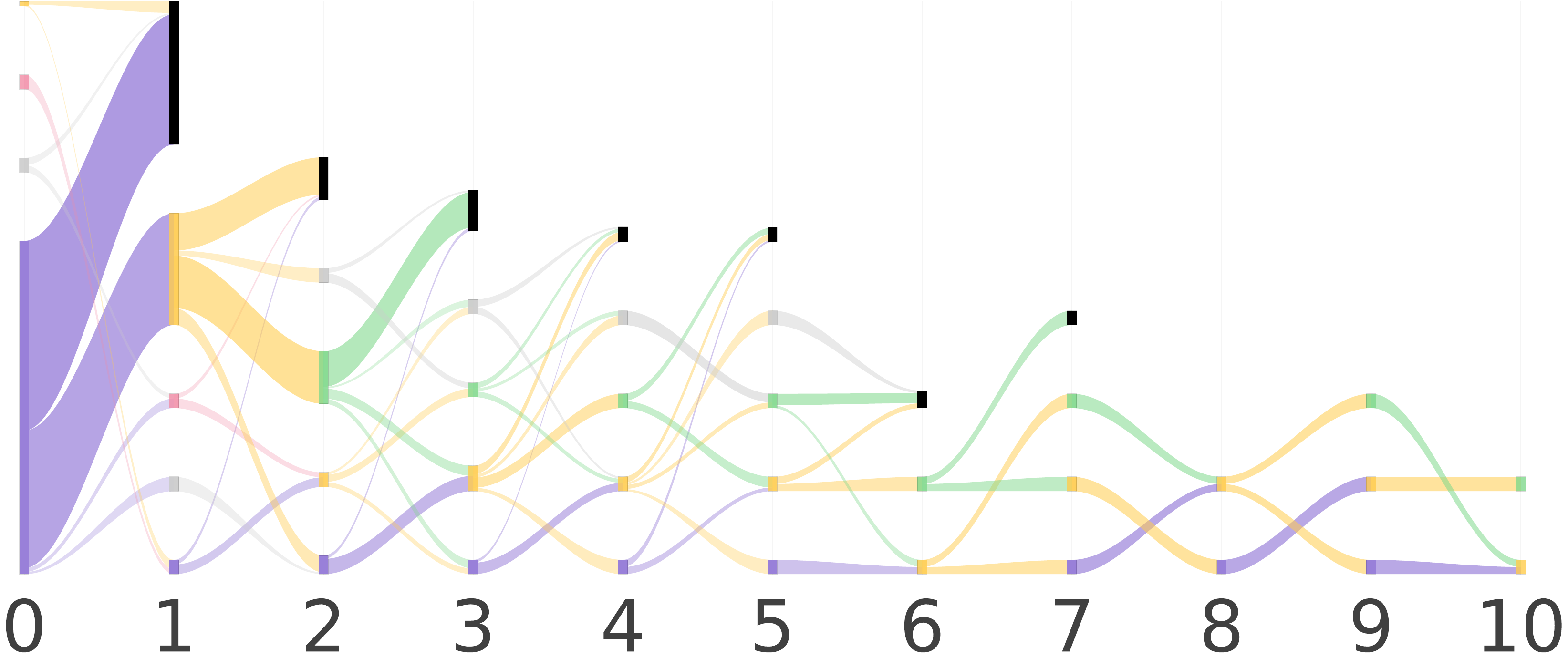} \\[6pt]
        \vspace{-0.2cm}
        \rotatebox{90}{
  \begin{minipage}{1.5cm} 
    \centering Reminded
  \end{minipage}
} & 
        \includegraphics[width=0.23\textwidth]{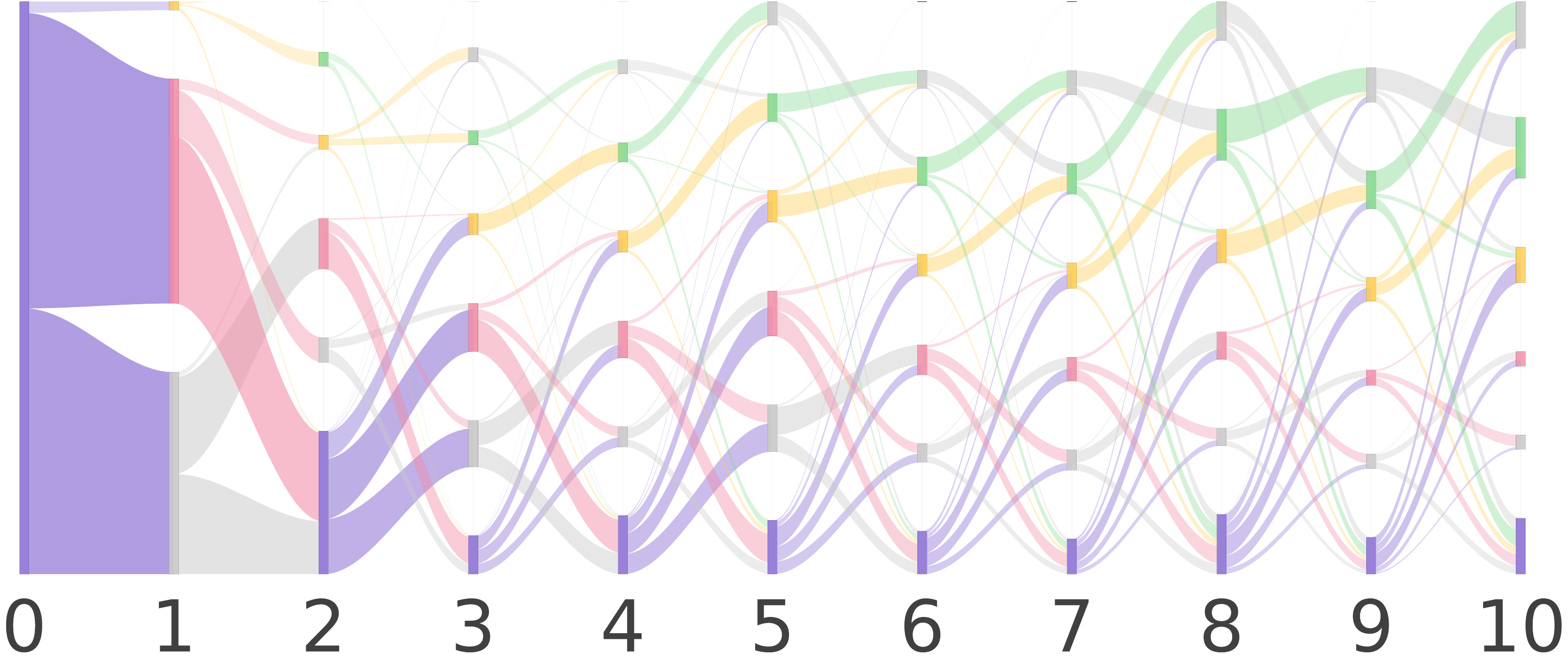} & 
        \includegraphics[width=0.23\textwidth]{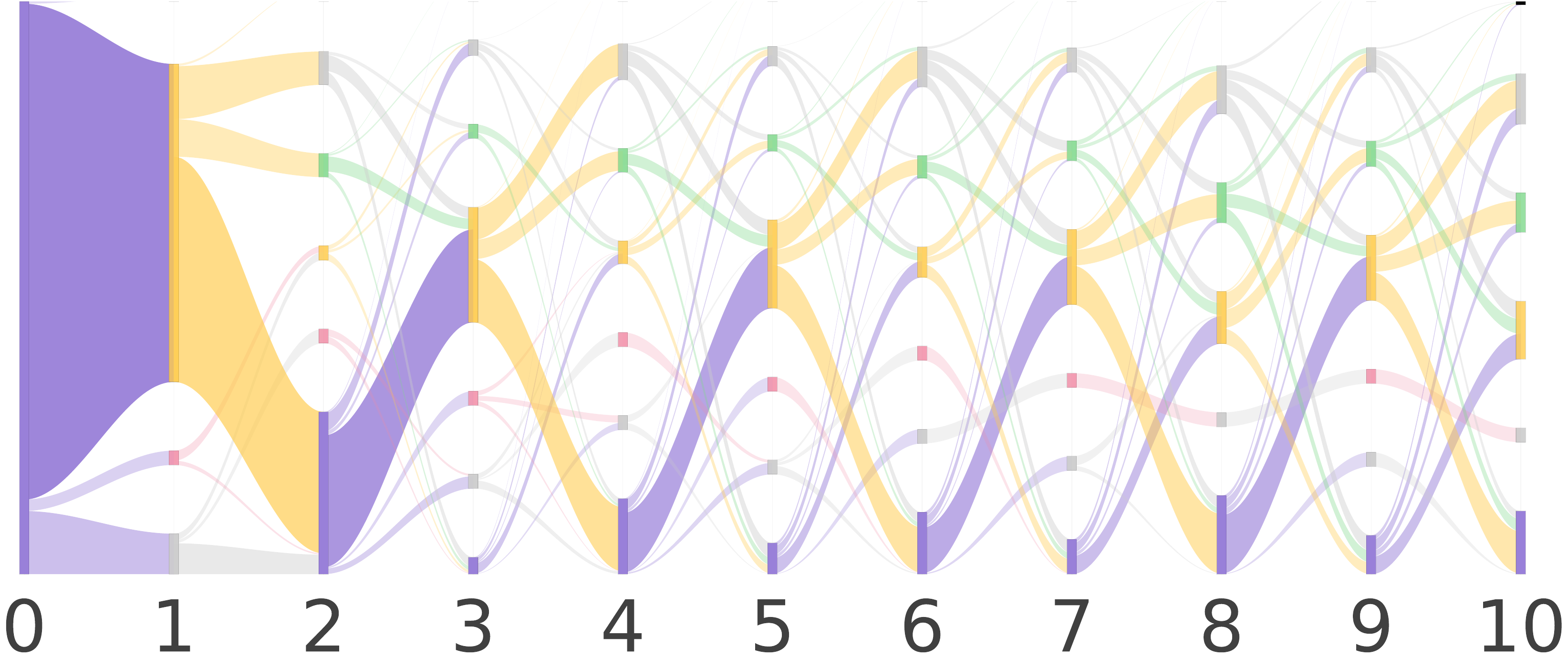} &
        \includegraphics[width=0.23\textwidth]{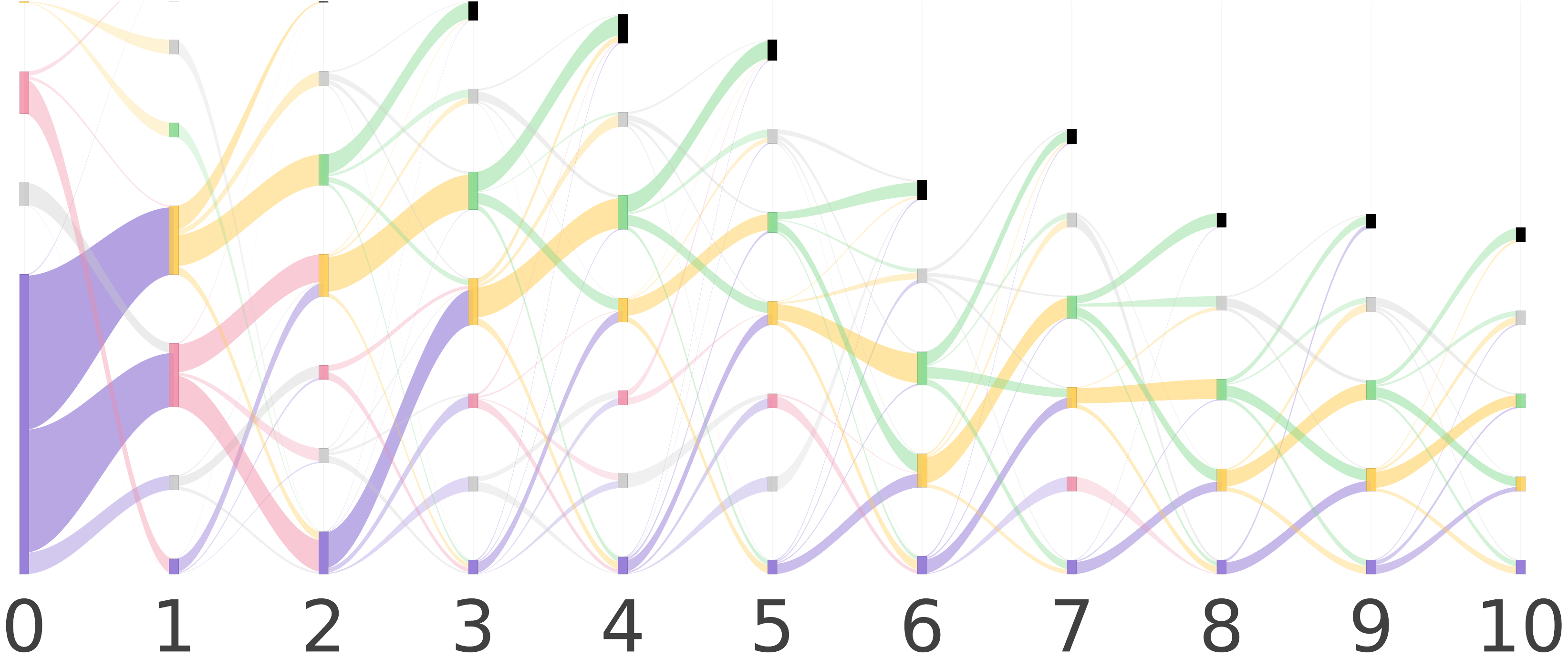} &
        \includegraphics[width=0.23\textwidth]{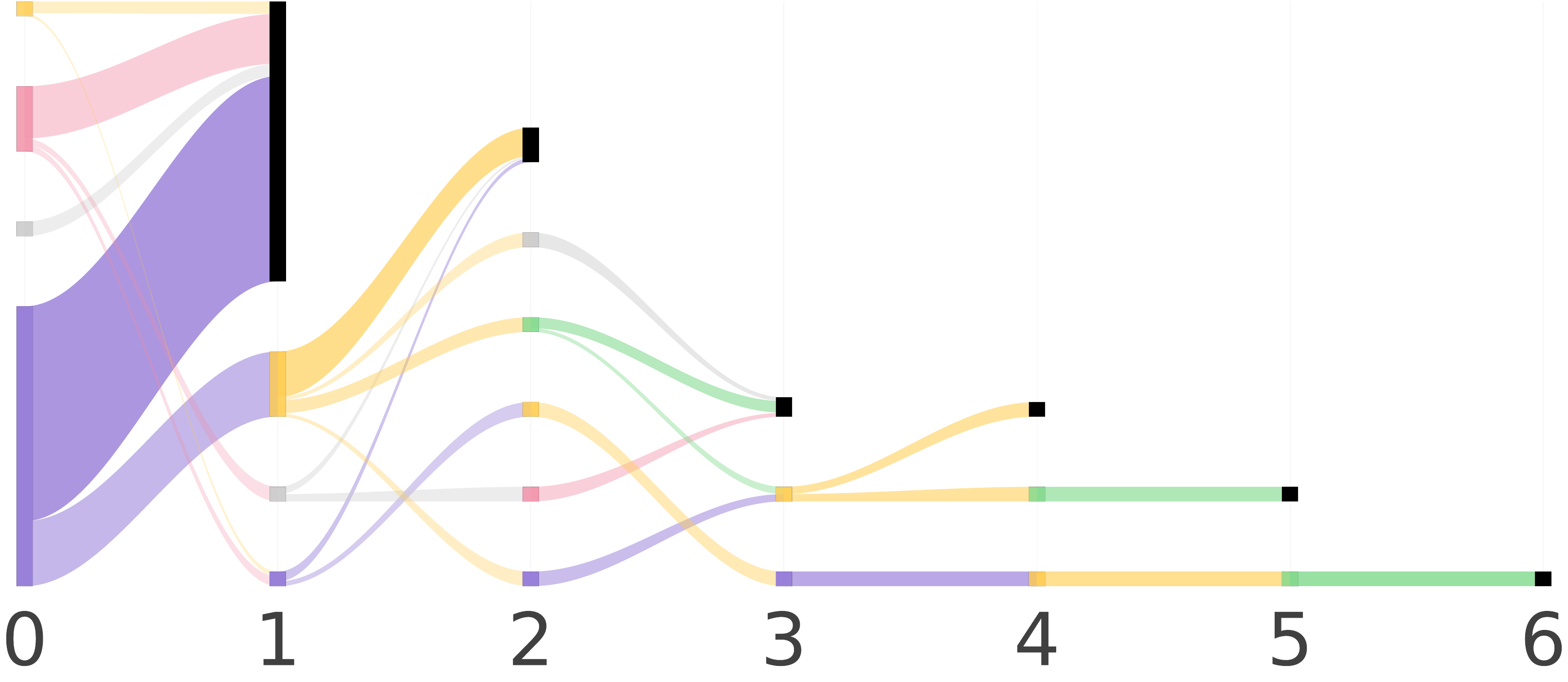} \\[6pt]

        & Devstral-small & GPT5-mini & DeepSeek-V3 & DeepSeek-R1 \\
        
    \end{tabular}

    \vspace{-14pt}
    \caption{Phase flow analysis under Reordered ($\mathcal{L}^\star(\Phi)=\colorbox[HTML]{CCC7E6}{N}\colorbox[HTML]{FFED99}{P}\colorbox[HTML]{EEC7D4}{R}\colorbox[HTML]{D9EDCC}{V}$) and Reminded setting ($\mathcal{L}^\star(\Phi)=\colorbox[HTML]{CCC7E6}{N}\colorbox[HTML]{EEC7D4}{R}\colorbox[HTML]{FFED99}{P}\colorbox[HTML]{D9EDCC}{V}$).}
    \vspace{-12pt}
    \label{fig:reordered-reminded-compliance-sankey}
\end{figure*}

\begin{figure*}[t]
    \centering
    \includegraphics[width=\linewidth]{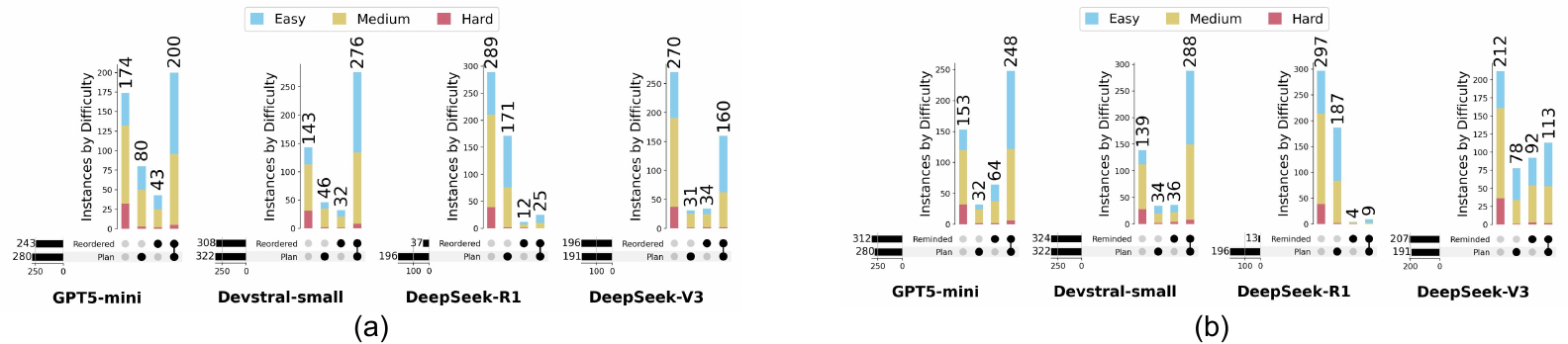}
    \vspace{-20pt}
    \caption{Impact of Reordered plan (a) and Reminded plan (b) on the success rate.}
    \vspace{-8pt}
    \label{fig:upset-reordered-reminded}
\end{figure*}

\begin{figure}[t]
    \centering
    \vspace{-5pt}
    \includegraphics[width=0.7\linewidth]{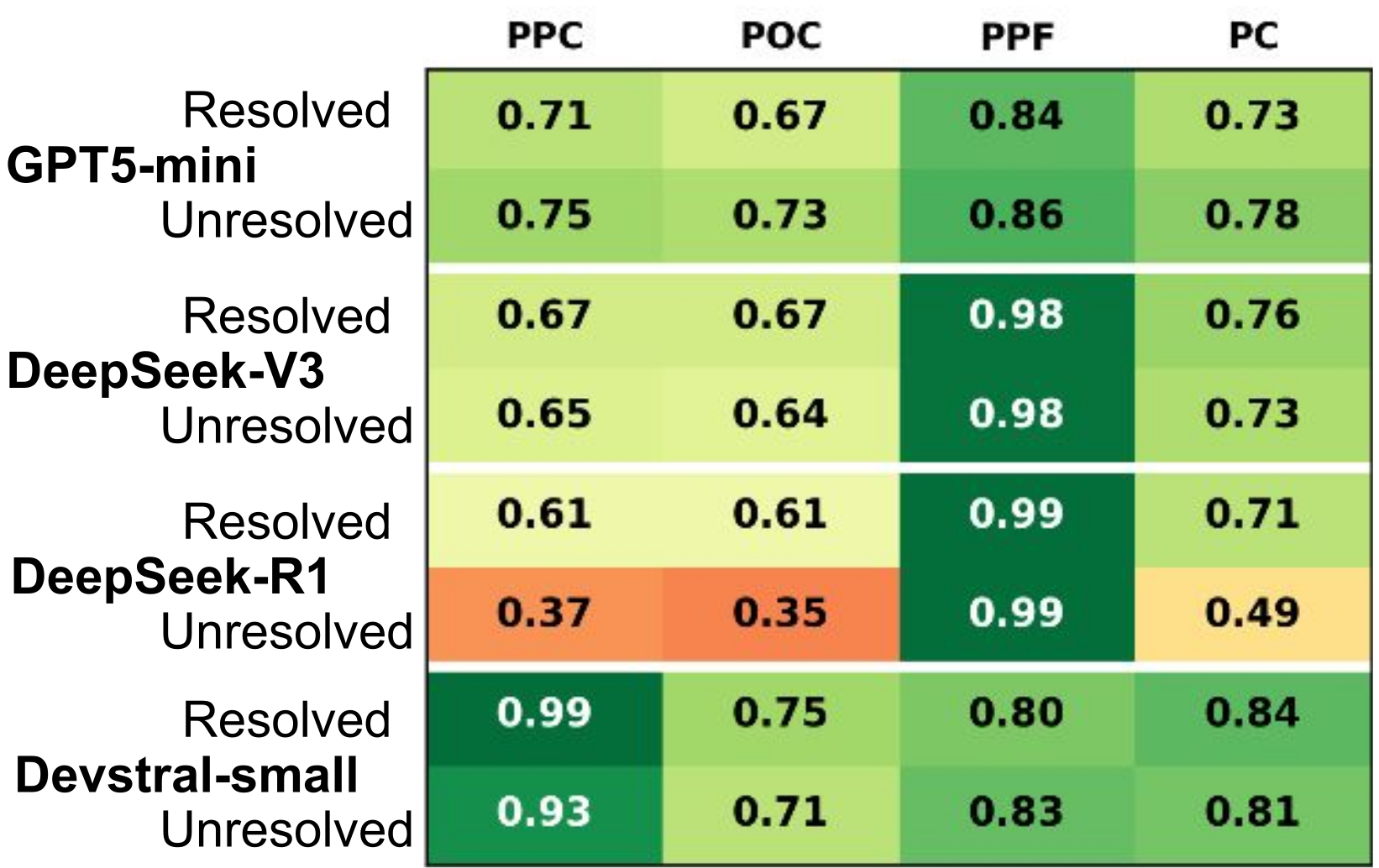}
    \vspace{-8pt}
    \caption{Reordered plan compliance metrics.}
    \vspace{-5pt}
    \label{fig:heatmap-reordered}
\end{figure}

Figures~\ref{fig:step-added-compliance-sankey}--\ref{fig:heatmap-regression} summarize the results. \devstral and \gptf already perform regression testing, even when not explicitly instructed to do so (Finding 4). The small drop in success rate is likely due to nondeterminism (\S \ref{subsec-rq7-nondeterminism}). \devstral has a high $PPC$, including all phases, while \gptf adaptively skips or reorders some phases as discussed (lower $PPC$). Concerning regression testing, \devstral and \gptf show a notable difference: as shown through phase flow analysis (Figure~\ref{fig:step-added-compliance-sankey}), \devstral performs regression testing \emph{after} \colorbox[HTML]{CCC7E6}{Navigation} and after \colorbox[HTML]{D9EDCC}{Validation}, while \gptf consistently runs regression tests as the first step and after \colorbox[HTML]{D9EDCC}{Validation} using new tests. 

\Vthree experiences a higher performance drop, accompanied by low $PPC$, suggesting difficulty in incorporating the regression testing phases. Running existing tests early in the trajectory and their lengthy feedback likely overrides the impact of other plan phases (lower $POC$), shifting focus toward test environment setup and irrelevant execution results rather than effective bug localization and patching. \Rone continues to exhibit severe performance issues in this setting. The persistent tool-calling errors result in a very low success rate, preventing any conclusion regarding plan compliance. We speculate that this behavior is due to optimization for short-term reward~\cite{guo2025deepseek}, which is a known issue in \Rone and specifically in reinforcement learning~\cite{qian2026toolrl,farquhar2025mona}.

\subsubsection{RQ4.2. Plan with Change Summary}
\label{subsubsec:with-summary}

Adding a summary phase, which is independent of the core repair process, yields minimal behavioral changes across most models. The overall plan compliance~($PC$) for \devstral, \gptf, and \Vthree remains nearly unchanged from the Standard setting (Figure~\ref{fig:heatmap-summary}). The success rates are also largely unaffected (Figure~\ref{fig:upset-regression-summary}b), as the summary phase typically occurs at the end of the trajectory and does not impact intermediate reasoning. \Rone, however, exhibits a substantial performance drop, with pervasive tool-calling failures observed in $413$ instances. This suggests that even orthogonal additions to the plan can destabilize this model's reasoning.

\noindent \textbf{Finding 10. Plan augmentation highlights plan overfitting, and is effective only when aligned with a model’s internal strategy.} Introducing additional phases provides limited benefit and can degrade performance if the model does not naturally employ those steps. Adding early phases can introduce unnecessary overhead or distract the model if phases are not well internalized.


\begin{figure}[t]
    \centering
    \vspace{-8pt}
    \includegraphics[width=0.7\linewidth]{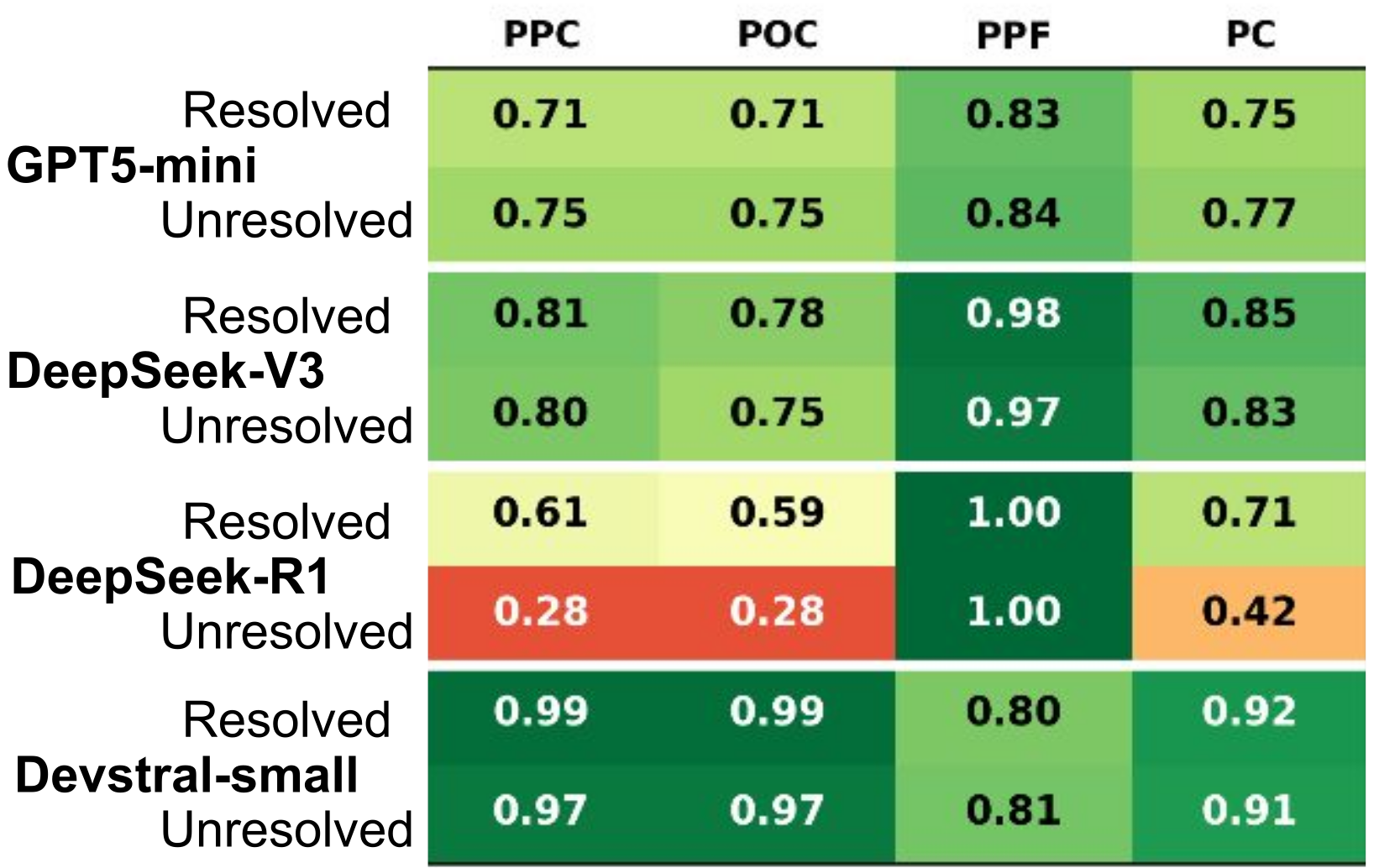}
    \vspace{-5pt}
    \caption{Reminded plan compliance metrics.}
    \vspace{-8pt}
    \label{fig:heatmap-reminded}
\end{figure}

\vspace{-10pt}
\subsection{RQ5. Reordered and Reminded Plan Settings}
\label{subsec-rq5-reorderplan-repeat}

\begin{figure*}[t]
    \centering
    \footnotesize
    \setlength{\tabcolsep}{5pt} 
    \renewcommand{\arraystretch}{1.35}
    
    \begin{tabular}{cccc}
         \footnotesize
         
        \includegraphics[width=0.23\textwidth]{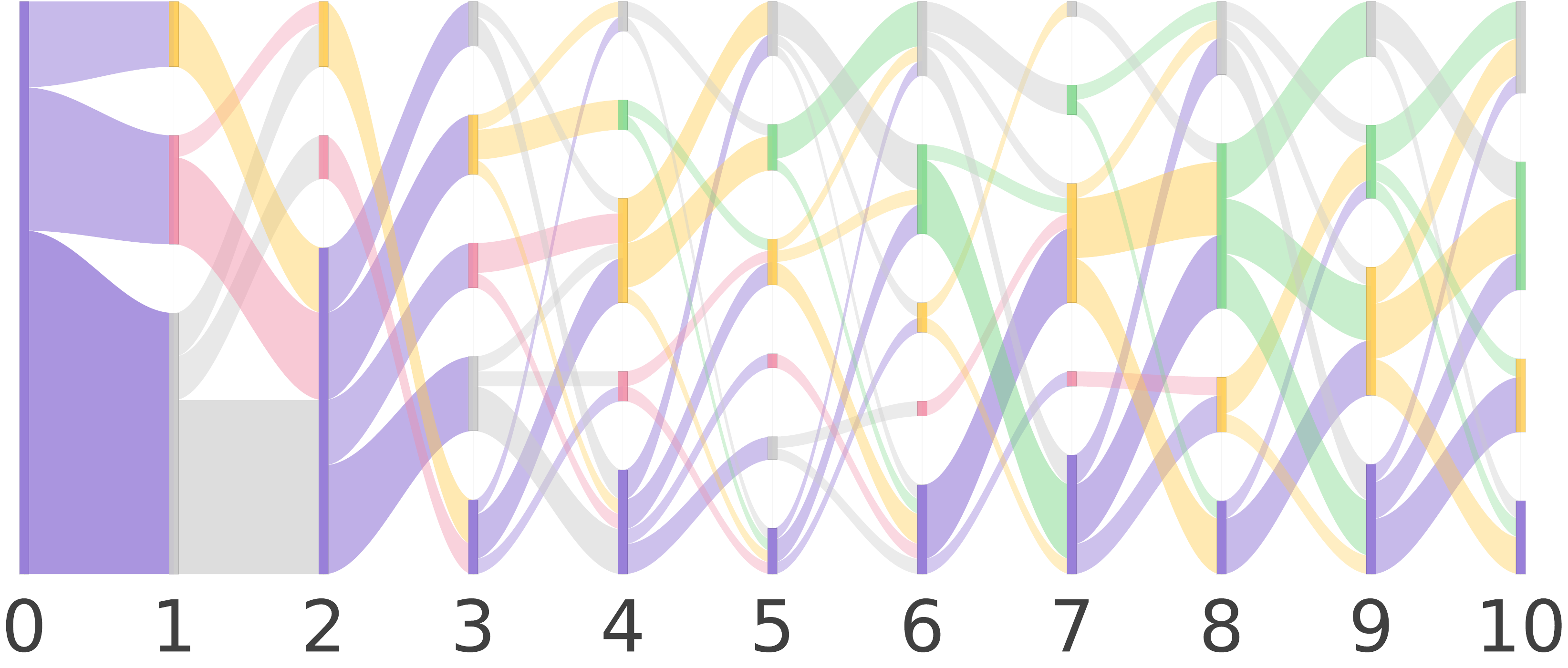} & 
        \includegraphics[width=0.23\textwidth]{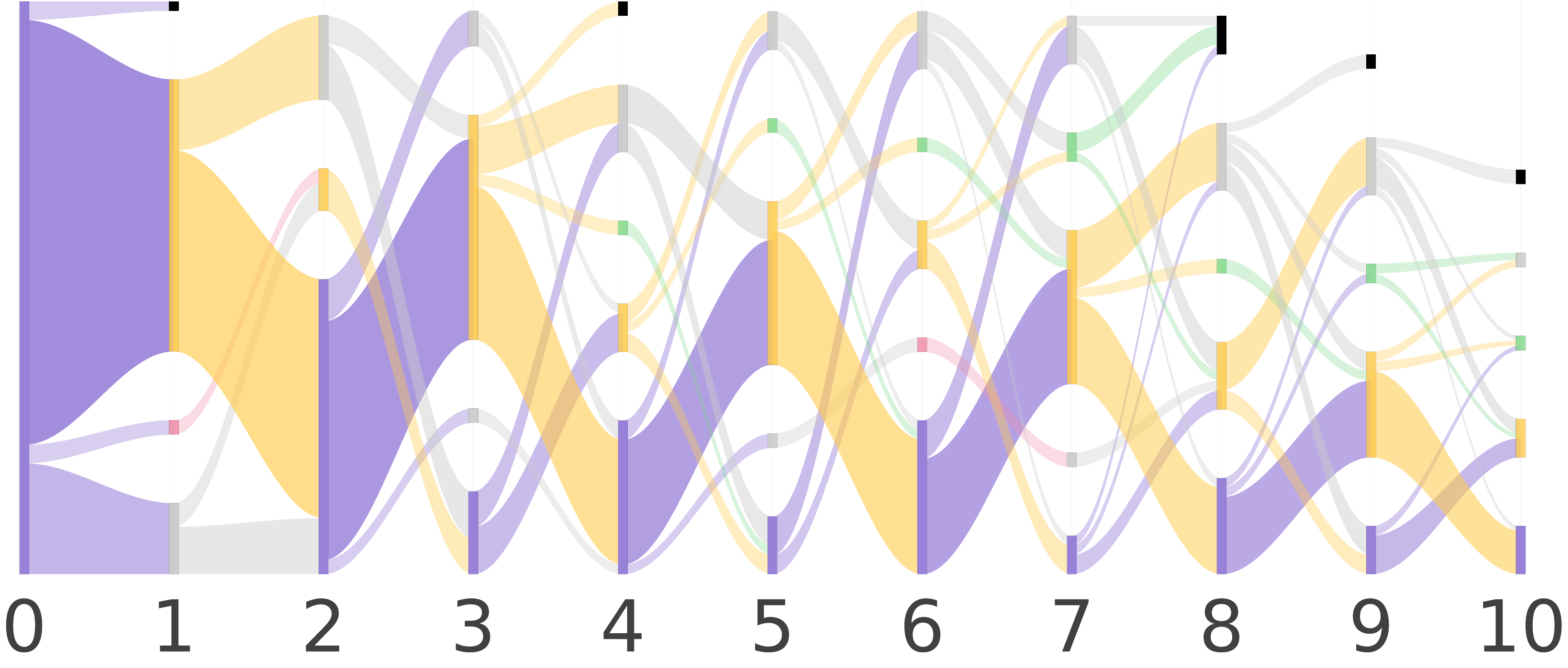} &
        \includegraphics[width=0.23\textwidth]{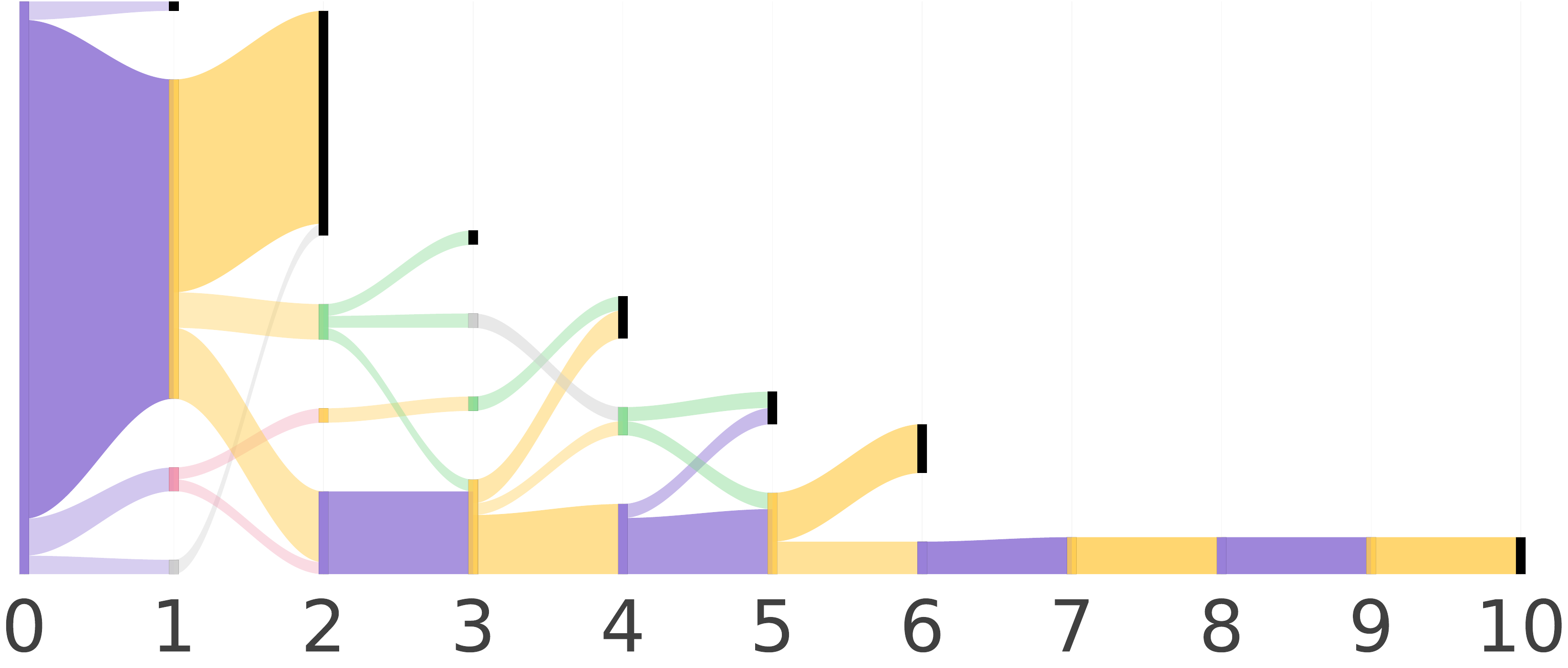} &
        \includegraphics[width=0.23\textwidth]{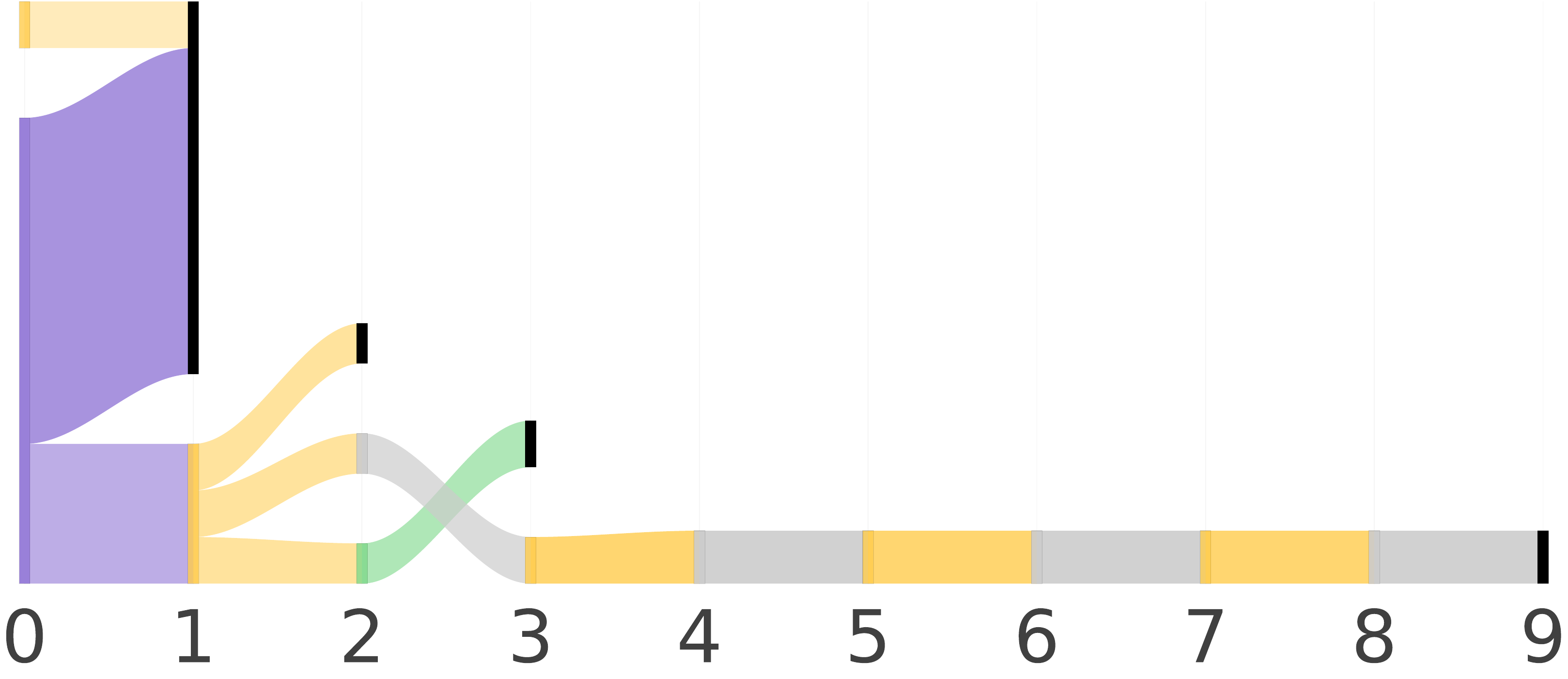} \\[6pt]

        Devstral-small & GPT5-mini & DeepSeek-V3 & DeepSeek-R1 \\
    \end{tabular}
    \vspace{-10pt}
    \caption{Phase flow analysis under Standard plan for \swebp ($\mathcal{L}^\star(\Phi)=\colorbox[HTML]{CCC7E6}{N}\colorbox[HTML]{EEC7D4}{R}\colorbox[HTML]{FFED99}{P}\colorbox[HTML]{D9EDCC}{V})$.}
    \vspace{-10pt}
    \label{fig:sankey-pro}
\end{figure*}

\begin{figure}[t]
    \centering
    \includegraphics[width=0.9\linewidth]{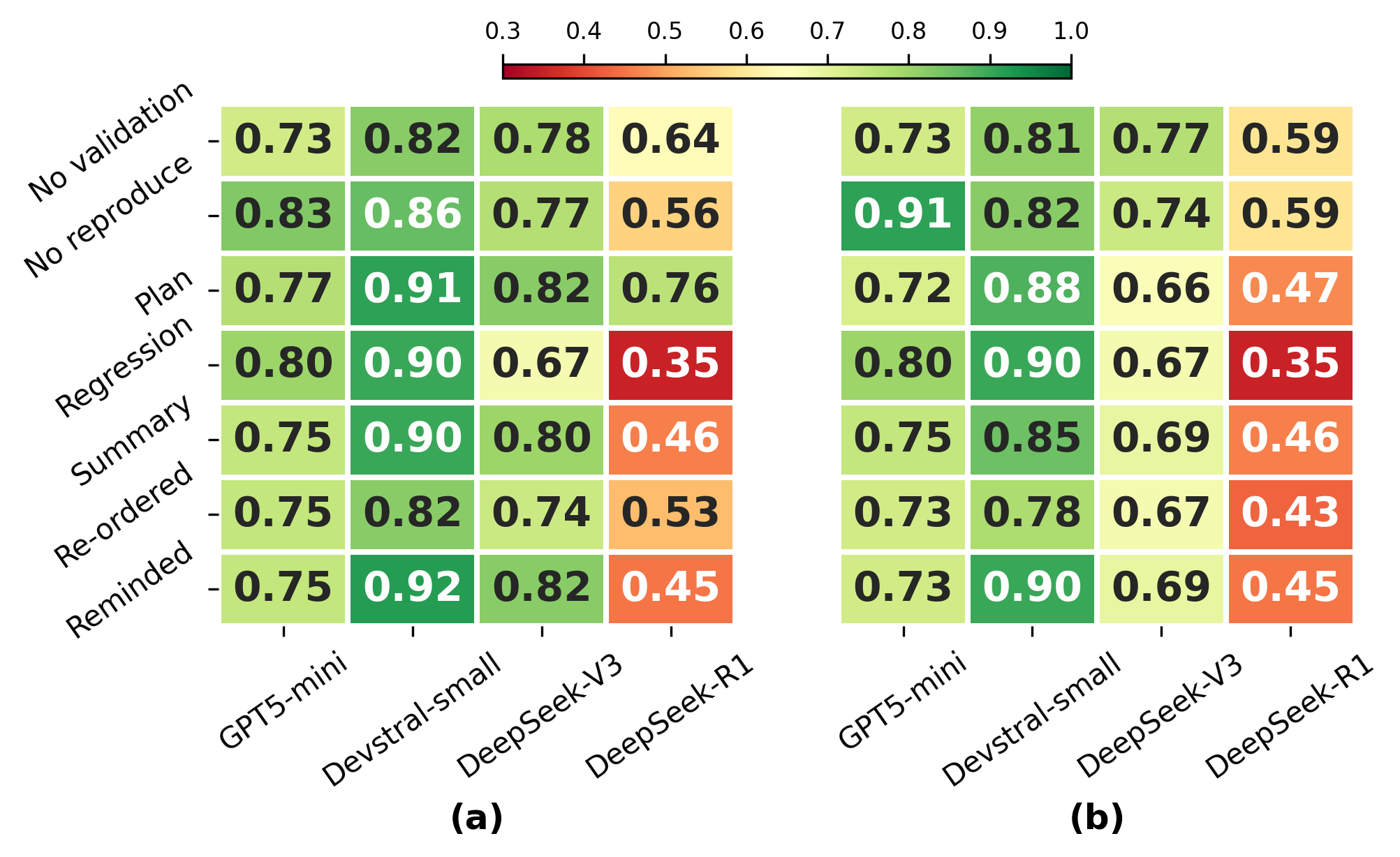}
    \vspace{-10pt}
    \caption{Compliance metrics on \swebv deterministic (a) and \swebp (b)}
    \label{fig:pro-deterministic-heatmap}
\end{figure}

\subsubsection{RQ5.1. Reordered Plan Setting}
\label{subsubsec:with-reorder}
Previous settings challenge agents to achieve high \textit{plan phase compliance} ($PPC$) and \emph{Plan Phase Fidelity} ($PPF$). Challenging \textbf{plan order compliance} ($POC$) involves reordering phases and investigating plan compliance. As in augmented plan settings, to avoid bias or noise, we reorder the phases so that the new plan remains relevant to the task. Specifically, we instruct the agent to patch the bug immediately after navigation and to postpone generating the reproduction test after patching, primarily for patch validation rather than bug localization. 

The reordered plan ($\mathcal{L}^\star(\Phi)=\colorbox[HTML]{CCC7E6}{N}\colorbox[HTML]{FFED99}{P}\colorbox[HTML]{EEC7D4}{R}\colorbox[HTML]{D9EDCC}{V}$) slightly 
impacts agents' behavior. Phase flow analysis (Figure~\ref{fig:reordered-reminded-compliance-sankey}) shows that \Vthree
often proceeds directly from navigation to patching, consistent with the modified plan. As shown in Figure~\ref{fig:upset-reordered-reminded}a, this slightly improves its success rate (from 191 resolved under the default plan to 196). In contrast, \devstral performs reproduction before patching despite the reordered instruction, resulting in a lower $POC$ as reflected in Figure~\ref{fig:heatmap-reordered}. For models that rely on reproduction, executing it before patching remains preferred, as it helps confirm the bug and improve the overall process. The reordering leads to moderately reduced success rates for \devstral, which consistently relies on reproduction, and \gptf, which adaptively incorporates reproduction as problem difficulty increases.

\noindent \textbf{Finding 11. Agents prioritize effective workflows over prescribed phase ordering.} Agents do not rigidly follow suboptimal ordering constraints; instead, they override them in favor of execution orders that better support their problem-solving process. 

\noindent \textbf{Finding 12. Delaying weakly internalized phases can reduce interference, but it does not consistently improve performance.} For models that do not naturally rely on a phase (e.g., \colorbox[HTML]{EEC7D4}{Reproduction} in \Vthree), postponing it 
can reduce 
interference with early 
steps. However, this benefit is not consistent: after accounting for nondeterminism (\S \ref{subsec-rq7-nondeterminism}), the reordered plan yields only a negligible change in success rate (38.3\% to 38\%).

\vspace{-5pt}
\subsubsection{RQ5.2. Reminded Plan Setting}
\label{subsubsec:with-reminder}
To mitigate the context window pressure (\S \ref{subsubsec:contributing-factors-to-plan-compliance}), where the initial plan becomes less influential as the trajectory length increases, we introduce a Reminded plan setting. In this variant, the Standard plan is periodically re-inserted into the context every \textit{five} steps. 
Figure~\ref{fig:heatmap-reminded} shows that periodic plan reminders improve plan compliance for \Vthree and maintain similar $PC$  for the other models. This leads to consistent improvements in success rates across models, as shown in Figure~\ref{fig:upset-reordered-reminded}b. 
The reminders prevent drifting into irrelevant sub-goals (e.g., exploring unrelated directories) 
and return focus to the repair task.

\section{Factors Beyond the Scaffold and LLM}
\label{sec:nondeterminism-contextwindow}

\subsection{RQ6. Generalization to Other Benchmarks}
\label{subsec-rq6-contamination}

In previous RQs, we observed that plan variants affect model performance and plan compliance differently, largely due to their internalized problem-solving strategies. One potential explanation is data contamination, where models may overfit to the
\swebv dataset. 
To assess generalization of the findings, we repeated RQ1--RQ5 on \swebp~\cite{swebenchpro},
which aims to
minimize overlap with LLM training corpora through licensing constraints.
To minimize moving factors, we focused on 266 Python instances of \swebp. This benchmark is more challenging and less contaminated, and the studied models 
achieve near-zero success rates in many instances.
To obtain more meaningful comparisons, we selected instances that \emph{at least one} of the LLMs could resolve under the Standard plan settings, leaving us with 31 instances. 

Figure~\ref{fig:pro-deterministic-heatmap}b shows the plan compliance ($PC$) values of studied agents under all plan settings. We observe that plan compliance drops by 13\% on average across all agents. The phase flow analysis in Figure~\ref{fig:sankey-pro} demonstrates a different trend under the Standard plan setting in \swebp compared to \swebv~(Figure~\ref{fig:compliance-sankey}), specifically for \Vthree and \Rone. This is likely because \swebp instances are more challenging
and most of the problem-solving effort goes into repetitive \colorbox[HTML]{CCC7E6}{Navigation} and \colorbox[HTML]{FFED99}{Patching}, without reaching
\colorbox[HTML]{D9EDCC}{Validation}. 
Similar to \swebv, the majority of the agents still skip \colorbox[HTML]{EEC7D4}{Reproduction}, achieving relatively high $PC$ scores on the \emph{No Reproduction} setting.

\subsection{RQ7. Impact of Nondeterminism}
\label{subsec-rq7-nondeterminism}
To mitigate any bias due to agents' inherent nondeterminism, we repeated the experiments under the Standard plan three times. 
Pairwise McNemar tests~\cite{McNemar_1947} show statistically significant differences across runs, confirming the non-determinism. We then identify instances that are consistently resolved or unresolved across Standard plan runs and evaluate the remaining plan variants on this reproducible subset (\gptf: 401, \devstral: 308, \Vthree: 323, and \Rone: 287). Figure~\ref{fig:pro-deterministic-heatmap}a reports the plan compliance $PC$, remaining nearly identical to that observed on the full benchmark. The results are consistent with our earlier findings: plan reduction, augmentation, and reordering affect models differently depending on their underlying problem-solving strategies, as reflected in the No Plan setting. Periodic plan reminders consistently improve performance by maintaining focus on the core task.

\vspace{-8pt}
\section{Related Work}
\label{sec:relatedwork}

Planning has become a central mechanism for improving the reliability of agents, especially for long-horizon and tool-using tasks. 
Several studies focus on constructing and refining plans in single- and multi-agent systems. Agent-Oriented Planning~\citep{AOP} and PMC~\cite{PMC} decompose complex tasks into structured subtasks and coordinate multiple agents to satisfy constraints, while EAGLET~\cite{si_et_al_2025} and Plan-and-Act \citep{plan-and-act} separate planning and execution into distinct LLMs to improve long-horizon reasoning. ReWoo took an extreme stance, planning all actions up-front~\cite{xu_et_al_2023}. Liu et al.~\cite{graphectory} introduce process-centric metrics, but as we saw in Finding~3 (\S \ref{sec:rq1-plan}), those alone are insufficient to understand plan compliance.

SAGE~\citep{SAGE} shows that
plans distilled from prior executions can guide future behavior and improve performance on software engineering tasks. While these approaches demonstrate the benefits of planning, they primarily evaluate success at the task level and implicitly assume that agents will follow the generated or provided plans during execution. \citet{jia2025your} assess whether a web agent’s actions align with its stated plan using LLM-based judges. However, this approach relies on costly and potentially unstable LLM scoring, limiting its scalability. In contrast, our work introduces mathematically defined plan compliance metrics that enable systematic analysis of agent behavior under controlled plan variations.


\section{Threats to Validity}
\label{sec:threats}


\noindent \textbf{External Validity.}
We evaluate four models 
and conduct experiments on two different benchmarks, 
which consist of real-world GitHub issues and differ in 
task composition, providing complementary evaluation settings. 
We use a structured plan common among several programming agents~(e.g., \MSA, \TA, \OH) as the standard plan to avoid bias
and to be representative of common practice.

\noindent \textbf{Internal Validity.}
To minimize the impact of nondeterminism of agents, we use a consistent default configuration across all experiments and repeat each experiment three times, focusing on stable behavioral patterns rather than random artifacts.

\noindent \textbf{Construct Validity.}
Our analysis relies on a phase-level abstraction of trajectories, which may omit fine-grained or project-specific actions.
However, it captures the core problem-solving stages that determine plan compliance: navigation, reproduction, patching, and validation. 
Our pipeline is built on top of peer-reviewed artifacts and is validated with well-vetted tools. We distinguish between key variations, such as newly generated tests and regression tests. Remaining low-level differences are treated as general actions and do not affect the main conclusions.
\section{Discussion}
\label{sec:discussion}
\noindent \textbf{Generalization to Other Domains.}
Although this work focuses on programming agents, our process-centric analysis is not limited to software engineering. Applying it to another domain requires defining a domain-specific plan-phase vocabulary and mapping agent actions to these phases. The computation of plan compliance metrics remains unchanged. For example, a scientific discovery agent following an Observation-Research-Hypothesis-Experiment-Analysis workflow~\cite{scientific_method} can be represented using the corresponding phase alphabet, with actions mapped through predefined rules or an LLM-based classifier. Controlled plan variations, such as phase removal and periodic reminders, can similarly be adapted to other domains, whereas plan augmentation requires domain knowledge to introduce task-relevant phases. Thus, our framework provides a general methodology for studying how instructed plans shape agent behavior and task success across domains.

\noindent \textbf{Implications for Agent Design and Training.}
Our findings suggest that plan compliance should serve as a process-level signal rather than a maximized objective. Strict compliance is neither necessary nor always beneficial: agents may productively adapt an instructed plan to the task, whereas missing phases, harmful reordering, or loss of plan awareness can hinder successful execution. During execution, compliance signals can support online monitoring and trigger targeted reminders or adaptive plan refinement when harmful deviations emerge. During post-training, they can provide process-level supervision for teaching agents to follow task-appropriate plans while retaining the flexibility to adapt, rather than memorizing fixed workflows or optimizing solely for task success. This analysis can also support long-term monitoring as models, scaffolds, and task distributions evolve.
\section{Conclusion}
\label{sec:conclusion}

This paper analyzes 21,120 trajectories to
assess plan compliance in programming agents. It introduces novel plan compliance metrics and runs agents under a
variety of plans. 
It also evaluates the impact of different system-prompt plans on the
agent's success in the issue-resolution task. We find that while plans clearly matter for task success, agents often
struggle to comply with them. This highlights the potential to further boost agent performance in future work via better plans and/or improved plan compliance.


\section{Data Availability Statement}
\label{sec:data-availability}

The artifacts of this paper are publicly available at \cite{website}.

\begin{acks}
This work is supported by NSF CCF-2238045 and IBM-Illinois Discovery Accelerator Institute (IIDAI) grants. We thank anonymous reviewers for their thoughtful comments and feedback.
\end{acks}

\balance
\bibliographystyle{ACM-Reference-Format}
\bibliography{references}

\end{document}